\RequirePackage{silence}
\documentclass[longauth]{aa}

\usepackage{graphicx}
\usepackage{natbib}
\usepackage{scalerel}
\usepackage{comment}
\usepackage{placeins}
\usepackage[nolist,nohyperlinks]{acronym}

\acrodef{AA}{azimuth or $\alpha$ angle}
\acrodef{ADC}{analogue-to-digital converter}
\acrodef{ADU}{analogue-to-digital unit}
\acrodef{AOI}{angle of incidence}
\acrodef{ASIC}{application specific integrated circuit}
\acrodef{BFE}{brighter-fatter effect}
\acrodef{CaLA}{camera-lens assembly}
\acrodef{CCD}{charge-coupled device}
\acrodef{CoLA}{corrector-lens assembly}
\acrodef{CDS}{Correlated Double Sampling}
\acrodef{CFC}{cryo-flex cable}
\acrodef{CFHT}{Canada-France-Hawaii Telescope}
\acrodef{CFRP}{carbon-fibre reinforced plastic}
\acrodef{CGH}{computer-generated hologram}
\acrodef{CME}{coronal mass ejection}
\acrodef{CNES}{Centre National d'Etude Spacial}
\acrodef{CPPM}{Centre de Physique des Particules de Marseille}
\acrodef{CPU}{central processingunit}
\acrodef{CR}{cosmic ray}
\acrodef{CTE}{coefficient of thermal expansion}
\acrodef{CME}{coronal mass ejection}
\acrodef{CTI}{charge-transfer inefficiency}
\acrodef{DCU}{Detector Control Unit}
\acrodef{DES}{Dark Energy Survey}
\acrodef{DPU}{Data Processing Unit}
\acrodef{DS}{Detector System}
\acrodef{EDS}{Euclid Deep Survey}
\acrodef{EE}{encircled energy}
\acrodef{EPER}{extended pixel-edge response}
\acrodef{ESA}{European Space Agency}
\acrodef{ESP}{Emission of Solar Protons}
\acrodef{ECSS}{European Cooperation for Space Standardization}
\acrodef{EWS}{Euclid Wide Survey}
\acrodef{FDIR}{Fault Detection, Isolation and Recovery}
\acrodef{FGS}{fine guidance sensor}
\acrodef{FOM}{figure of merit}
\acrodef{FOV}{field of view}
\acrodef{FPA}{focal-plane array}
\acrodef{FPR}{false positive rate}
\acrodef{FWA}{filter-wheel assembly}
\acrodef{FWC}{full-well capacity}
\acrodef{FWHM}{full width at half maximum}
\acrodef{GOES}{Geostationary Operational Environmental Satellites}
\acrodef{GCR}{Galactic cosmic ray}
\acrodef{GWA}{grism-wheel assembly}
\acrodef{H2RG}{HAWAII-2RG}
\acrodef{HST}{\textit{Hubble} Space Telescope}
\acrodef{IP2I}{Institut de Physique des 2 Infinis de Lyon}
\acrodef{JWST}{{\em James Webb} Space Telescope}
\acrodef{IAD}{ion-assisted deposition}
\acrodef{ICU}{Instrument Control Unit}
\acrodef{IPC}{inter-pixel capacitance}
\acrodef{ISES}{International Space Environmental Services}
\acrodef{JWST}{\textit{James Webb} Space Telescope}
\acrodef{LAM}{Laboratoire d'Astrophysique de Marseille}
\acrodef{LED}{light-emitting diode}
\acrodef{LSB}{low surface brightness}
\acrodef{MACC}{multiple accumulated}
\acrodef{MEF}{multi-extension FITS}
\acrodef{MER PF}{MER processing function}
\acrodef{MLI}{multi-layer insulation}
\acrodef{MMU}{Mass Memory Unit}
\acrodef{MPE}{Max-Planck-Institut für extraterrestrische Physik}
\acrodef{MPIA}{Max-Planck-Institut für Astronomie}
\acrodef{NA}{numerical aperture}
\acrodef{NASA}{National Aeronautic and Space Administration}
\acrodef{NIEL}{non-ionising energy loss}
\acrodef{JPL}{NASA Jet Propulsion Laboratory}
\acrodef{MZ-CGH}{multi-zonal computer-generated hologram}
\acrodef{NI-CU}{NISP calibration unit}
\acrodef{NI-OA}{near-infrared optical assembly}
\acrodef{NI-GWA}{NISP Grism Wheel Assembly}
\acrodef{NIR}{near-infrared}
\acrodef{NISP}{Near-Infrared Spectrometer and Photometer}
\acrodef{NOAA}{National Oceanic and Atmospheric Administration}
\acrodef{PA}{position angle}
\acrodef{PARMS}{plasma-assisted reactive magnetron sputtering}
\acrodef{PLM}{payload module}
\acrodef{PRNU}{pixel-response non-uniformity}
\acrodef{PTC}{photon transfer curve}
\acrodef{PTFE}{polytetrafluoroethylene}
\acrodef{PV}{performance verification}
\acrodef{PWM}{pulse-width modulation}
\acrodef{PSF}{point spread function}
\acrodef{QE}{quantum efficiency}
\acrodef{QF}{quality factor}
\acrodef{ROE}{readout electronic block unit}
\acrodef{ROI}{region of interest}
\acrodef{ROIC}{readout-integrated circuit}
\acrodef{ROS}{reference observing sequence}
\acrodef{SAA}{Solar aspect angle}
\acrodef{SCA}{sensor chip array}
\acrodef{SCE}{sensor chip electronic}
\acrodef{SCS}{sensor chip system}
\acrodef{SGS}{science ground segment}
\acrodef{SGPS}{Solar and Galactic Proton Sensor}
\acrodef{SHS}{Shack-Hartmann sensor}
\acrodef{SNR}[S/N]{signal-to-noise ratio}
\acrodef{SED}{spectral energy distribution}
\acrodef{SiC}{silicon carbide}
\acrodef{SEP}{Solar energetic particle}
\acrodef{SSN}{Sunspot number}
\acrodef{STIX}{Spectrometer/Telescope for Imaging X-rays}
\acrodef{SolO}{Solar Orbiter}
\acrodef{TP}{trap pumping}
\acrodef{TPR}{true positive rate}
\acrodef{UTR}{up-the-ramp}
\acrodef{SVM}{service module}
\acrodef{VIS}{visible imager}
\acrodef{VIS PF}{VIS processing function}
\acrodef{WD}{white dwarf}
\acrodef{WCS}{world coordinate system}
\acrodef{WFE}{wavefront error}
\acrodef{ZP}{zero point}
\acrodef{DPDD}{Data Product Description Document}
\acrodef{OTF}{on-the-fly}
\acrodef{RR}{regression reprocessing}
\acrodef{EUDF}{Euclid ultra-deep field }
\acrodef{PF}{processing function }
\acrodef{PA}{position angle}
\acrodef{RMS}{Root mean square} 
\acrodef{PE}{Processing element} 

\usepackage[table]{xcolor}

\usepackage{txfonts}
\usepackage[pdfencoding=auto,psdextra]{hyperref}
\hypersetup{
    colorlinks=true,
    linkcolor=blue,
    filecolor=magenta,
    urlcolor=blue,
    citecolor=blue
}
\newcommand{\pz}{\phantom{0}}
\newcommand{\pp}{\phantom{$-$}}

\makeatletter
\renewcommand*\aa@pageof{, page \thepage{} of \pageref*{LastPage}}
\makeatother

\usepackage[utf8]{inputenc}

\usepackage[switch, modulo]{lineno}

\renewcommand{\linenumbers}[0]{}

\usepackage{euclid}
\newcommand{\Gaia}{\textit{Gaia}}

\usepackage[dvipsnames]{xcolor}
\usepackage[normalem]{ulem}

\begin{document}

   \title{Euclid Quick Data Release (Q2) --
   The Euclid Galactic Bulge Survey\thanks{This paper is published on behalf of the Euclid Consortium.}}

%%%% Version Monday 22nd of June 2026 12:33:37 PM UT *PRELIMINARY*
%%%% Assumes the new A&A style file from Oct 2025 or later
%%%% Please do not edit the author list -- contact ECEB Bureau for changes
\newcommand{\orcid}[1]{} %% if already defined in aa.cls: comment, or use renewcommand			   
\author{J.-P.~Beaulieu\thanks{\email{beaulieu@iap.fr}}\inst{\ref{aff1},\ref{aff2}}
\and E.~Bachelet\orcid{0000-0002-6578-5078}\inst{\ref{aff3}}
\and M.~Gilles\orcid{0009-0007-5241-4116}\inst{\ref{aff4}}
\and C.~Ranc\orcid{0000-0003-2388-4534}\inst{\ref{aff4}}
\and E.~N.~Rektsini\orcid{0000-0002-1530-4870}\inst{\ref{aff1}}
\and E.~Kerins\orcid{0000-0002-1743-4468}\inst{\ref{aff5}}
\and M.~T.~Penny\orcid{0000-0001-7506-5640}\inst{\ref{aff6}}
\and H.~Verma\orcid{0000-0002-6302-251X}\inst{\ref{aff6}}
\and S.~Mottet\inst{\ref{aff4}}
\and R.~Vavrek\inst{\ref{aff7}}
\and V.~Bozza\orcid{0000-0003-4590-0136}\inst{\ref{aff8},\ref{aff9}}
\and J.-C.~Cuillandre\orcid{0000-0002-3263-8645}\inst{\ref{aff10}}
\and M.~Chang\inst{\ref{aff4}}
\and L.~R.~Bedin\orcid{0000-0003-4080-6466}\inst{\ref{aff11}}
\and K.~Kuijken\orcid{0000-0002-3827-0175}\inst{\ref{aff12}}
\and C.~Laigle\orcid{0009-0008-5926-818X}\inst{\ref{aff4}}
\and M.~Libralato\orcid{0000-0001-9673-7397}\inst{\ref{aff11}}
\and I.~McDonald\orcid{0000-0003-0356-0655}\inst{\ref{aff5}}
\and R.~Nakajima\orcid{0009-0009-1213-7040}\inst{\ref{aff13}}
\and V.~Popoff\inst{\ref{aff4}}
\and J.~Rhodes\orcid{0000-0002-4485-8549}\inst{\ref{aff14}}
\and E.~Thygesen\orcid{0000-0001-8812-0565}\inst{\ref{aff2}}
\and S.~Awan\inst{\ref{aff15}}
\and J.~E.~Davies\orcid{0000-0002-5079-9098}\inst{\ref{aff16}}
\and A.~M.~Di~Giorgio\orcid{0000-0002-4767-2360}\inst{\ref{aff17}}
\and T.~Flanet\inst{\ref{aff4}}
\and C.~Grenet\inst{\ref{aff4}}
\and O.~Herent\inst{\ref{aff4}}
\and P. Hudelot\inst{\ref{aff18}}
\and K.~Jahnke\orcid{0000-0003-3804-2137}\inst{\ref{aff16}}
\and R.~Kohley\inst{\ref{aff7}}
\and H.~J.~McCracken\orcid{0000-0002-9489-7765}\inst{\ref{aff4}}
%\and [tbc]Patricia Liebing\inst{\ref{aff19}}
\and H.~N.~Nguyen-Kim\inst{\ref{aff4}}
%\and [tbc]Mathew Page\inst{\ref{aff19}}
\and M.~Schirmer\orcid{0000-0003-2568-9994}\inst{\ref{aff16}}
\and F.~Soldano\inst{\ref{aff4}}
%\and [tbc]Pierre Ferruit\inst{\ref{aff20}}
\and B.~Altieri\orcid{0000-0003-3936-0284}\inst{\ref{aff7}}
\and S.~Andreon\orcid{0000-0002-2041-8784}\inst{\ref{aff21}}
\and N.~Auricchio\orcid{0000-0003-4444-8651}\inst{\ref{aff22}}
\and H.~Aussel\orcid{0000-0002-1371-5705}\inst{\ref{aff10}}
\and C.~Baccigalupi\orcid{0000-0002-8211-1630}\inst{\ref{aff23},\ref{aff24},\ref{aff25},\ref{aff26}}
\and M.~Baldi\orcid{0000-0003-4145-1943}\inst{\ref{aff27},\ref{aff22},\ref{aff28}}
\and A.~Balestra\orcid{0000-0002-6967-261X}\inst{\ref{aff11}}
\and S.~Bardelli\orcid{0000-0002-8900-0298}\inst{\ref{aff22}}
\and P.~Battaglia\orcid{0000-0002-7337-5909}\inst{\ref{aff22}}
\and R.~Bender\orcid{0000-0001-7179-0626}\inst{\ref{aff29},\ref{aff30}}
\and A.~Biviano\orcid{0000-0002-0857-0732}\inst{\ref{aff24},\ref{aff23}}
\and M.~Brescia\orcid{0000-0001-9506-5680}\inst{\ref{aff31},\ref{aff32}}
\and S.~Camera\orcid{0000-0003-3399-3574}\inst{\ref{aff33},\ref{aff34},\ref{aff35}}
\and V.~Capobianco\orcid{0000-0002-3309-7692}\inst{\ref{aff35}}
\and C.~Carbone\orcid{0000-0003-0125-3563}\inst{\ref{aff36}}
\and V.~F.~Cardone\inst{\ref{aff37},\ref{aff38}}
\and J.~Carretero\orcid{0000-0002-3130-0204}\inst{\ref{aff39},\ref{aff40}}
\and M.~Castellano\orcid{0000-0001-9875-8263}\inst{\ref{aff37}}
\and G.~Castignani\orcid{0000-0001-6831-0687}\inst{\ref{aff22}}
\and S.~Cavuoti\orcid{0000-0002-3787-4196}\inst{\ref{aff32},\ref{aff9}}
\and A.~Cimatti\inst{\ref{aff41}}
\and C.~Colodro-Conde\inst{\ref{aff42}}
\and G.~Congedo\orcid{0000-0003-2508-0046}\inst{\ref{aff43}}
\and C.~J.~Conselice\orcid{0000-0003-1949-7638}\inst{\ref{aff5}}
\and L.~Conversi\orcid{0000-0002-6710-8476}\inst{\ref{aff44},\ref{aff7}}
\and Y.~Copin\orcid{0000-0002-5317-7518}\inst{\ref{aff45}}
\and F.~Courbin\orcid{0000-0003-0758-6510}\inst{\ref{aff46},\ref{aff47},\ref{aff48}}
\and H.~M.~Courtois\orcid{0000-0003-0509-1776}\inst{\ref{aff49}}
\and M.~Cropper\orcid{0000-0003-4571-9468}\inst{\ref{aff15}}
\and H.~Degaudenzi\orcid{0000-0002-5887-6799}\inst{\ref{aff50}}
\and G.~De~Lucia\orcid{0000-0002-6220-9104}\inst{\ref{aff24}}
\and C.~Dolding\orcid{0009-0003-7199-6108}\inst{\ref{aff15}}
\and H.~Dole\orcid{0000-0002-9767-3839}\inst{\ref{aff51}}
\and F.~Dubath\orcid{0000-0002-6533-2810}\inst{\ref{aff50}}
\and X.~Dupac\inst{\ref{aff7}}
\and S.~Dusini\orcid{0000-0002-1128-0664}\inst{\ref{aff52}}
\and S.~Escoffier\orcid{0000-0002-2847-7498}\inst{\ref{aff53}}
\and M.~Farina\orcid{0000-0002-3089-7846}\inst{\ref{aff17}}
\and R.~Farinelli\inst{\ref{aff22}}
\and F.~Faustini\orcid{0000-0001-6274-5145}\inst{\ref{aff37}}
\and S.~Ferriol\inst{\ref{aff45}}
\and F.~Finelli\orcid{0000-0002-6694-3269}\inst{\ref{aff22},\ref{aff54}}
\and S.~Fotopoulou\orcid{0000-0002-9686-254X}\inst{\ref{aff55}}
\and N.~Fourmanoit\orcid{0009-0005-6816-6925}\inst{\ref{aff53}}
\and M.~Frailis\orcid{0000-0002-7400-2135}\inst{\ref{aff24}}
\and L.~Gabarra\orcid{0000-0002-8486-8856}\inst{\ref{aff56}}
\and S.~Galeotta\orcid{0000-0002-3748-5115}\inst{\ref{aff24}}
\and W.~Gillard\orcid{0000-0003-4744-9748}\inst{\ref{aff53}}
\and B.~Gillis\orcid{0000-0002-4478-1270}\inst{\ref{aff43}}
\and C.~Giocoli\orcid{0000-0002-9590-7961}\inst{\ref{aff22},\ref{aff28}}
\and P.~G\'omez-Alvarez\orcid{0000-0002-8594-5358}\inst{\ref{aff57},\ref{aff7}}
\and A.~Grazian\orcid{0000-0002-5688-0663}\inst{\ref{aff11}}
\and F.~Grupp\inst{\ref{aff29},\ref{aff30}}
\and L.~Guzzo\orcid{0000-0001-8264-5192}\inst{\ref{aff58},\ref{aff21},\ref{aff59}}
\and W.~G.~Hartley\orcid{0000-0002-4563-5101}\inst{\ref{aff50}}
\and S.~V.~H.~Haugan\orcid{0000-0001-9648-7260}\inst{\ref{aff60}}
\and G.~Helou\orcid{0000-0003-3367-3415}\inst{\ref{aff61},\ref{aff62}}
\and S.~Hemmati\orcid{0000-0003-2226-5395}\inst{\ref{aff62}}
\and H.~Hoekstra\orcid{0000-0002-0641-3231}\inst{\ref{aff12}}
\and W.~Holmes\orcid{0009-0007-8554-4646}\inst{\ref{aff14}}
\and I.~M.~Hook\orcid{0000-0002-2960-978X}\inst{\ref{aff63}}
\and F.~Hormuth\inst{\ref{aff64}}
\and A.~Hornstrup\orcid{0000-0002-3363-0936}\inst{\ref{aff65},\ref{aff66}}
\and M.~Jhabvala\inst{\ref{aff67}}
\and B.~Joachimi\orcid{0000-0001-7494-1303}\inst{\ref{aff68}}
\and S.~Kermiche\orcid{0000-0002-0302-5735}\inst{\ref{aff53}}
\and B.~Kubik\orcid{0009-0006-5823-4880}\inst{\ref{aff45}}
\and M.~K\"ummel\orcid{0000-0003-2791-2117}\inst{\ref{aff30}}
\and M.~Kunz\orcid{0000-0002-3052-7394}\inst{\ref{aff69}}
\and H.~Kurki-Suonio\orcid{0000-0002-4618-3063}\inst{\ref{aff70},\ref{aff71}}
\and R.~Laureijs\inst{\ref{aff72}}
\and A.~M.~C.~Le~Brun\orcid{0000-0002-0936-4594}\inst{\ref{aff73}}
\and S.~Ligori\orcid{0000-0003-4172-4606}\inst{\ref{aff35}}
\and P.~B.~Lilje\orcid{0000-0003-4324-7794}\inst{\ref{aff60}}
\and V.~Lindholm\orcid{0000-0003-2317-5471}\inst{\ref{aff70},\ref{aff71}}
\and M.~Magliocchetti\orcid{0000-0001-9158-4838}\inst{\ref{aff17}}
\and G.~Mainetti\orcid{0000-0003-2384-2377}\inst{\ref{aff74}}
\and O.~Mansutti\orcid{0000-0001-5758-4658}\inst{\ref{aff24}}
\and O.~Marggraf\orcid{0000-0001-7242-3852}\inst{\ref{aff13}}
\and M.~Martinelli\orcid{0000-0002-6943-7732}\inst{\ref{aff37},\ref{aff38}}
\and N.~Martinet\orcid{0000-0003-2786-7790}\inst{\ref{aff75}}
\and F.~Marulli\orcid{0000-0002-8850-0303}\inst{\ref{aff76},\ref{aff22},\ref{aff28}}
\and R.~J.~Massey\orcid{0000-0002-6085-3780}\inst{\ref{aff77}}
\and E.~Medinaceli\orcid{0000-0002-4040-7783}\inst{\ref{aff22}}
\and M.~Melchior\inst{\ref{aff78}}
\and M.~Meneghetti\orcid{0000-0003-1225-7084}\inst{\ref{aff22},\ref{aff28}}
\and E.~Merlin\orcid{0000-0001-6870-8900}\inst{\ref{aff11}}
\and G.~Meylan\orcid{0000-0001-6503-0209}\inst{\ref{aff79}}
\and A.~Mora\orcid{0000-0002-1922-8529}\inst{\ref{aff80}}
\and M.~Moresco\orcid{0000-0002-7616-7136}\inst{\ref{aff76},\ref{aff22}}
\and C.~Moretti\orcid{0000-0003-3314-8936}\inst{\ref{aff24},\ref{aff23},\ref{aff25}}
\and L.~Moscardini\orcid{0000-0002-3473-6716}\inst{\ref{aff76},\ref{aff22},\ref{aff28}}
\and C.~Neissner\orcid{0000-0001-8524-4968}\inst{\ref{aff81},\ref{aff40}}
\and S.-M.~Niemi\orcid{0009-0005-0247-0086}\inst{\ref{aff82}}
\and J.~W.~Nightingale\orcid{0000-0002-8987-7401}\inst{\ref{aff83}}
\and C.~Padilla\orcid{0000-0001-7951-0166}\inst{\ref{aff81}}
\and S.~Paltani\orcid{0000-0002-8108-9179}\inst{\ref{aff50}}
\and F.~Pasian\orcid{0000-0002-4869-3227}\inst{\ref{aff24}}
\and W.~J.~Percival\orcid{0000-0002-0644-5727}\inst{\ref{aff84},\ref{aff85},\ref{aff86}}
\and V.~Pettorino\orcid{0000-0002-4203-9320}\inst{\ref{aff82}}
\and G.~Polenta\orcid{0000-0003-4067-9196}\inst{\ref{aff87}}
\and M.~Poncet\inst{\ref{aff88}}
\and G.~D.~Racca\orcid{0000-0002-9883-8981}\inst{\ref{aff12},\ref{aff82}}
\and F.~Raison\orcid{0000-0002-7819-6918}\inst{\ref{aff29}}
\and A.~Renzi\orcid{0000-0001-9856-1970}\inst{\ref{aff89},\ref{aff52},\ref{aff22}}
\and G.~Riccio\inst{\ref{aff32}}
\and I.~Risso\orcid{0000-0003-2525-7761}\inst{\ref{aff90},\ref{aff91},\ref{aff21}}
\and F.~Rizzo\orcid{0000-0002-9407-585X}\inst{\ref{aff24}}
\and E.~Romelli\orcid{0000-0003-3069-9222}\inst{\ref{aff24}}
\and M.~Roncarelli\orcid{0000-0001-9587-7822}\inst{\ref{aff22}}
\and B.~Rusholme\orcid{0000-0001-7648-4142}\inst{\ref{aff62}}
\and R.~Saglia\orcid{0000-0003-0378-7032}\inst{\ref{aff30},\ref{aff29}}
\and Z.~Sakr\orcid{0000-0002-4823-3757}\inst{\ref{aff92},\ref{aff93},\ref{aff94}}
\and A.~G.~S\'anchez\orcid{0000-0003-1198-831X}\inst{\ref{aff29}}
\and D.~Sapone\orcid{0000-0001-7089-4503}\inst{\ref{aff95}}
\and P.~Schneider\orcid{0000-0001-8561-2679}\inst{\ref{aff13}}
\and T.~Schrabback\orcid{0000-0002-6987-7834}\inst{\ref{aff96}}
\and A.~Secroun\orcid{0000-0003-0505-3710}\inst{\ref{aff53}}
\and E.~Sihvola\orcid{0000-0003-1804-7715}\inst{\ref{aff97}}
\and P.~Simon\inst{\ref{aff13}}
\and C.~Sirignano\orcid{0000-0002-0995-7146}\inst{\ref{aff89},\ref{aff52}}
\and G.~Sirri\orcid{0000-0003-2626-2853}\inst{\ref{aff28}}
\and J.~Skottfelt\orcid{0000-0003-1310-8283}\inst{\ref{aff98}}
\and L.~Stanco\orcid{0000-0002-9706-5104}\inst{\ref{aff52}}
\and P.~Tallada-Cresp\'{i}\orcid{0000-0002-1336-8328}\inst{\ref{aff39},\ref{aff40}}
\and A.~N.~Taylor\inst{\ref{aff43}}
\and I.~Tereno\orcid{0000-0002-4537-6218}\inst{\ref{aff99},\ref{aff100}}
\and S.~Toft\orcid{0000-0003-3631-7176}\inst{\ref{aff101},\ref{aff102}}
\and R.~Toledo-Moreo\orcid{0000-0002-2997-4859}\inst{\ref{aff103},\ref{aff104}}
\and F.~Torradeflot\orcid{0000-0003-1160-1517}\inst{\ref{aff40},\ref{aff39}}
\and A.~Tsyganov\inst{\ref{aff105}}
\and I.~Tutusaus\orcid{0000-0002-3199-0399}\inst{\ref{aff106},\ref{aff107},\ref{aff93}}
\and J.~Valiviita\orcid{0000-0001-6225-3693}\inst{\ref{aff70},\ref{aff71}}
\and T.~Vassallo\orcid{0000-0001-6512-6358}\inst{\ref{aff24},\ref{aff108}}
\and Y.~Wang\orcid{0000-0002-4749-2984}\inst{\ref{aff62}}
\and J.~Weller\orcid{0000-0002-8282-2010}\inst{\ref{aff30},\ref{aff29}}
\and A.~Zacchei\orcid{0000-0003-0396-1192}\inst{\ref{aff24},\ref{aff23}}
\and F.~M.~Zerbi\orcid{0000-0002-9996-973X}\inst{\ref{aff21}}
\and E.~Zucca\orcid{0000-0002-5845-8132}\inst{\ref{aff22}}
\and M.~Sereno\orcid{0000-0003-0302-0325}\inst{\ref{aff22},\ref{aff28}}}
										   
%%%% please do not edit the affiliation list -- contact ECEB Bureau for changes
\institute{Institut d'Astrophysique de Paris, 98bis Boulevard Arago, 75014, Paris, France\label{aff1}
\and
School of Natural Sciences, University of Tasmania, Private Bag 37 Hobart, Tasmania 7001, Australia\label{aff2}
\and
Universit\'e de Franche-Comt\'e, Institut UTINAM, CNRS UMR6213, OSU THETA Franche-Comt\'e-Bourgogne, Observatoire de Besan\c con, BP 1615, 25010 Besan\c con Cedex, France\label{aff3}
\and
Institut d'Astrophysique de Paris, UMR 7095, CNRS, and Sorbonne Universit\'e, 98 bis boulevard Arago, 75014 Paris, France\label{aff4}
\and
Jodrell Bank Centre for Astrophysics, Department of Physics and Astronomy, University of Manchester, Oxford Road, Manchester M13 9PL, UK\label{aff5}
\and
Department of Physics \& Astronomy, Louisiana State University, 202 Nicholson Hall, Baton Rouge, LA 70803, USA\label{aff6}
\and
ESAC/ESA, Camino Bajo del Castillo, s/n., Urb. Villafranca del Castillo, 28692 Villanueva de la Ca\~nada, Madrid, Spain\label{aff7}
\and
Universita di Salerno, Dipartimento di Fisica "E.R. Caianiello", Via Giovanni Paolo II 132, I-84084 Fisciano (SA), Italy\label{aff8}
\and
INFN section of Naples, Via Cinthia 6, 80126, Napoli, Italy\label{aff9}
\and
Universit\'e Paris-Saclay, Universit\'e Paris Cit\'e, CEA, CNRS, AIM, 91191, Gif-sur-Yvette, France\label{aff10}
\and
INAF-Osservatorio Astronomico di Padova, Via dell'Osservatorio 5, 35122 Padova, Italy\label{aff11}
\and
Leiden Observatory, Leiden University, Einsteinweg 55, 2333 CC Leiden, The Netherlands\label{aff12}
\and
Universit\"at Bonn, Argelander-Institut f\"ur Astronomie, Auf dem H\"ugel 71, 53121 Bonn, Germany\label{aff13}
\and
Jet Propulsion Laboratory, California Institute of Technology, 4800 Oak Grove Drive, Pasadena, CA, 91109, USA\label{aff14}
\and
Mullard Space Science Laboratory, University College London, Holmbury St Mary, Dorking, Surrey RH5 6NT, UK\label{aff15}
\and
Max-Planck-Institut f\"ur Astronomie, K\"onigstuhl 17, 69117 Heidelberg, Germany\label{aff16}
\and
INAF-Istituto di Astrofisica e Planetologia Spaziali, via del Fosso del Cavaliere, 100, 00100 Roma, Italy\label{aff17}
\and
[tbc]Institut d'Astrophysique de Paris, 98 bis, boulevard Arago, 75014 Paris, France\label{aff18}
\and
[tbc]Mullard Space Science Laboratory, Holmbury St-Mary,  Dorking, Surrey, RH5 6NT, UK\label{aff19}
\and
[tbc]European Space Astronomy Centre, European Space Agency (ESA)  Camino bajo del Castillo, s/n  Urbanizacion Villafranca del Castillo, Villanueva de la Ca\~{n}ada, Spain\label{aff20}
\and
INAF-Osservatorio Astronomico di Brera, Via Brera 28, 20122 Milano, Italy\label{aff21}
\and
INAF-Osservatorio di Astrofisica e Scienza dello Spazio di Bologna, Via Piero Gobetti 93/3, 40129 Bologna, Italy\label{aff22}
\and
IFPU, Institute for Fundamental Physics of the Universe, via Beirut 2, 34151 Trieste, Italy\label{aff23}
\and
INAF-Osservatorio Astronomico di Trieste, Via G. B. Tiepolo 11, 34143 Trieste, Italy\label{aff24}
\and
INFN, Sezione di Trieste, Via Valerio 2, 34127 Trieste TS, Italy\label{aff25}
\and
SISSA, International School for Advanced Studies, Via Bonomea 265, 34136 Trieste TS, Italy\label{aff26}
\and
Dipartimento di Fisica e Astronomia, Universit\`a di Bologna, Via Gobetti 93/2, 40129 Bologna, Italy\label{aff27}
\and
INFN-Sezione di Bologna, Viale Berti Pichat 6/2, 40127 Bologna, Italy\label{aff28}
\and
Max Planck Institute for Extraterrestrial Physics, Giessenbachstr. 1, 85748 Garching, Germany\label{aff29}
\and
Universit\"ats-Sternwarte M\"unchen, Fakult\"at f\"ur Physik, Ludwig-Maximilians-Universit\"at M\"unchen, Scheinerstr.~1, 81679 M\"unchen, Germany\label{aff30}
\and
Department of Physics "E. Pancini", University Federico II, Via Cinthia 6, 80126, Napoli, Italy\label{aff31}
\and
INAF-Osservatorio Astronomico di Capodimonte, Via Moiariello 16, 80131 Napoli, Italy\label{aff32}
\and
Dipartimento di Fisica, Universit\`a degli Studi di Torino, Via P. Giuria 1, 10125 Torino, Italy\label{aff33}
\and
INFN-Sezione di Torino, Via P. Giuria 1, 10125 Torino, Italy\label{aff34}
\and
INAF-Osservatorio Astrofisico di Torino, Via Osservatorio 20, 10025 Pino Torinese (TO), Italy\label{aff35}
\and
INAF-IASF Milano, Via Alfonso Corti 12, 20133 Milano, Italy\label{aff36}
\and
INAF-Osservatorio Astronomico di Roma, Via Frascati 33, 00078 Monteporzio Catone, Italy\label{aff37}
\and
INFN-Sezione di Roma, Piazzale Aldo Moro, 2 - c/o Dipartimento di Fisica, Edificio G. Marconi, 00185 Roma, Italy\label{aff38}
\and
Centro de Investigaciones Energ\'eticas, Medioambientales y Tecnol\'ogicas (CIEMAT), Avenida Complutense 40, 28040 Madrid, Spain\label{aff39}
\and
Port d'Informaci\'{o} Cient\'{i}fica, Campus UAB, C. Albareda s/n, 08193 Bellaterra (Barcelona), Spain\label{aff40}
\and
Dipartimento di Fisica e Astronomia "Augusto Righi" - Alma Mater Studiorum Universit\`a di Bologna, Viale Berti Pichat 6/2, 40127 Bologna, Italy\label{aff41}
\and
Instituto de Astrof\'{\i}sica de Canarias, E-38205 La Laguna, Tenerife, Spain\label{aff42}
\and
Institute for Astronomy, University of Edinburgh, Royal Observatory, Blackford Hill, Edinburgh EH9 3HJ, UK\label{aff43}
\and
European Space Agency/ESRIN, Largo Galileo Galilei 1, 00044 Frascati, Roma, Italy\label{aff44}
\and
Universit\'e Claude Bernard Lyon 1, CNRS/IN2P3, IP2I Lyon, UMR 5822, Villeurbanne, F-69100, France\label{aff45}
\and
Institut de Ci\`{e}ncies del Cosmos (ICCUB), Universitat de Barcelona (IEEC-UB), Mart\'{i} i Franqu\`{e}s 1, 08028 Barcelona, Spain\label{aff46}
\and
Instituci\'o Catalana de Recerca i Estudis Avan\c{c}ats (ICREA), Passeig de Llu\'{\i}s Companys 23, 08010 Barcelona, Spain\label{aff47}
\and
Institut de Ciencies de l'Espai (IEEC-CSIC), Campus UAB, Carrer de Can Magrans, s/n Cerdanyola del Vall\'es, 08193 Barcelona, Spain\label{aff48}
\and
UCB Lyon 1, CNRS/IN2P3, IUF, IP2I Lyon, 4 rue Enrico Fermi, 69622 Villeurbanne, France\label{aff49}
\and
Department of Astronomy, University of Geneva, ch. d'Ecogia 16, 1290 Versoix, Switzerland\label{aff50}
\and
Universit\'e Paris-Saclay, CNRS, Institut d'astrophysique spatiale, 91405, Orsay, France\label{aff51}
\and
INFN-Padova, Via Marzolo 8, 35131 Padova, Italy\label{aff52}
\and
Aix-Marseille Universit\'e, CNRS/IN2P3, CPPM, Marseille, France\label{aff53}
\and
INFN-Bologna, Via Irnerio 46, 40126 Bologna, Italy\label{aff54}
\and
School of Physics, HH Wills Physics Laboratory, University of Bristol, Tyndall Avenue, Bristol, BS8 1TL, UK\label{aff55}
\and
Department of Physics, Oxford University, Keble Road, Oxford OX1 3RH, UK\label{aff56}
\and
FRACTAL S.L.N.E., calle Tulip\'an 2, Portal 13 1A, 28231, Las Rozas de Madrid, Spain\label{aff57}
\and
Dipartimento di Fisica "Aldo Pontremoli", Universit\`a degli Studi di Milano, Via Celoria 16, 20133 Milano, Italy\label{aff58}
\and
INFN-Sezione di Milano, Via Celoria 16, 20133 Milano, Italy\label{aff59}
\and
Institute of Theoretical Astrophysics, University of Oslo, P.O. Box 1029 Blindern, 0315 Oslo, Norway\label{aff60}
\and
California Institute of Technology, 1200 E California Blvd, Pasadena, CA 91125, USA\label{aff61}
\and
Caltech/IPAC, 1200 E. California Blvd., Pasadena, CA 91125, USA\label{aff62}
\and
Department of Physics, Lancaster University, Lancaster, LA1 4YB, UK\label{aff63}
\and
Felix Hormuth Engineering, Goethestr. 17, 69181 Leimen, Germany\label{aff64}
\and
Technical University of Denmark, Elektrovej 327, 2800 Kgs. Lyngby, Denmark\label{aff65}
\and
Cosmic Dawn Center (DAWN), Denmark\label{aff66}
\and
NASA Goddard Space Flight Center, Greenbelt, MD 20771, USA\label{aff67}
\and
Department of Physics and Astronomy, University College London, Gower Street, London WC1E 6BT, UK\label{aff68}
\and
Universit\'e de Gen\`eve, D\'epartement de Physique Th\'eorique and Centre for Astroparticle Physics, 24 quai Ernest-Ansermet, CH-1211 Gen\`eve 4, Switzerland\label{aff69}
\and
Department of Physics, P.O. Box 64, University of Helsinki, 00014 Helsinki, Finland\label{aff70}
\and
Helsinki Institute of Physics, Gustaf H{\"a}llstr{\"o}min katu 2, University of Helsinki, 00014 Helsinki, Finland\label{aff71}
\and
Kapteyn Astronomical Institute, University of Groningen, PO Box 800, 9700 AV Groningen, The Netherlands\label{aff72}
\and
Laboratoire d'etude de l'Univers et des phenomenes eXtremes, Observatoire de Paris, Universit\'e PSL, Sorbonne Universit\'e, CNRS, 92190 Meudon, France\label{aff73}
\and
Centre de Calcul de l'IN2P3/CNRS, 21 avenue Pierre de Coubertin 69627 Villeurbanne Cedex, France\label{aff74}
\and
Aix-Marseille Universit\'e, CNRS, CNES, LAM, Marseille, France\label{aff75}
\and
Dipartimento di Fisica e Astronomia "Augusto Righi" - Alma Mater Studiorum Universit\`a di Bologna, via Piero Gobetti 93/2, 40129 Bologna, Italy\label{aff76}
\and
Department of Physics, Institute for Computational Cosmology, Durham University, South Road, Durham, DH1 3LE, UK\label{aff77}
\and
University of Applied Sciences and Arts of Northwestern Switzerland, School of Engineering, 5210 Windisch, Switzerland\label{aff78}
\and
Institute of Physics, Laboratory of Astrophysics, Ecole Polytechnique F\'ed\'erale de Lausanne (EPFL), Observatoire de Sauverny, 1290 Versoix, Switzerland\label{aff79}
\and
Telespazio UK S.L. for European Space Agency (ESA), Camino bajo del Castillo, s/n, Urbanizacion Villafranca del Castillo, Villanueva de la Ca\~nada, 28692 Madrid, Spain\label{aff80}
\and
Institut de F\'{i}sica d'Altes Energies (IFAE), The Barcelona Institute of Science and Technology, Campus UAB, 08193 Bellaterra (Barcelona), Spain\label{aff81}
\and
European Space Agency/ESTEC, Keplerlaan 1, 2201 AZ Noordwijk, The Netherlands\label{aff82}
\and
School of Mathematics, Statistics and Physics, Newcastle University, Herschel Building, Newcastle-upon-Tyne, NE1 7RU, UK\label{aff83}
\and
Waterloo Centre for Astrophysics, University of Waterloo, Waterloo, Ontario N2L 3G1, Canada\label{aff84}
\and
Department of Physics and Astronomy, University of Waterloo, Waterloo, Ontario N2L 3G1, Canada\label{aff85}
\and
Perimeter Institute for Theoretical Physics, Waterloo, Ontario N2L 2Y5, Canada\label{aff86}
\and
Space Science Data Center, Italian Space Agency, via del Politecnico snc, 00133 Roma, Italy\label{aff87}
\and
Centre National d'Etudes Spatiales -- Centre spatial de Toulouse, 18 avenue Edouard Belin, 31401 Toulouse Cedex 9, France\label{aff88}
\and
Dipartimento di Fisica e Astronomia "G. Galilei", Universit\`a di Padova, Via Marzolo 8, 35131 Padova, Italy\label{aff89}
\and
Dipartimento di Fisica, Universit\`a di Genova, Via Dodecaneso 33, 16146, Genova, Italy\label{aff90}
\and
INFN-Sezione di Genova, Via Dodecaneso 33, 16146, Genova, Italy\label{aff91}
\and
Instituto de F\'isica Te\'orica UAM-CSIC, Campus de Cantoblanco, 28049 Madrid, Spain\label{aff92}
\and
Institut de Recherche en Astrophysique et Plan\'etologie (IRAP), Universit\'e de Toulouse, CNRS, UPS, CNES, 14 Av. Edouard Belin, 31400 Toulouse, France\label{aff93}
\and
Universit\'e St Joseph; Faculty of Sciences, Beirut, Lebanon\label{aff94}
\and
Departamento de F\'isica, FCFM, Universidad de Chile, Blanco Encalada 2008, Santiago, Chile\label{aff95}
\and
Universit\"at Innsbruck, Institut f\"ur Astro- und Teilchenphysik, Technikerstr. 25/8, 6020 Innsbruck, Austria\label{aff96}
\and
Department of Physics and Helsinki Institute of Physics, Gustaf H\"allstr\"omin katu 2, University of Helsinki, 00014 Helsinki, Finland\label{aff97}
\and
Centre for Electronic Imaging, Open University, Walton Hall, Milton Keynes, MK7~6AA, UK\label{aff98}
\and
Departamento de F\'isica, Faculdade de Ci\^encias, Universidade de Lisboa, Edif\'icio C8, Campo Grande, PT1749-016 Lisboa, Portugal\label{aff99}
\and
Instituto de Astrof\'isica e Ci\^encias do Espa\c{c}o, Faculdade de Ci\^encias, Universidade de Lisboa, Tapada da Ajuda, 1349-018 Lisboa, Portugal\label{aff100}
\and
Cosmic Dawn Center (DAWN)\label{aff101}
\and
Niels Bohr Institute, University of Copenhagen, Jagtvej 128, 2200 Copenhagen, Denmark\label{aff102}
\and
Universidad Polit\'ecnica de Cartagena, Departamento de Electr\'onica y Tecnolog\'ia de Computadoras,  Plaza del Hospital 1, 30202 Cartagena, Spain\label{aff103}
\and
European University of Technology EUt+, European Union\label{aff104}
\and
Centre for Information Technology, University of Groningen, P.O. Box 11044, 9700 CA Groningen, The Netherlands\label{aff105}
\and
Institute of Space Sciences (ICE, CSIC), Campus UAB, Carrer de Can Magrans, s/n, 08193 Barcelona, Spain\label{aff106}
\and
Institut d'Estudis Espacials de Catalunya (IEEC),  Edifici RDIT, Campus UPC, 08860 Castelldefels, Barcelona, Spain\label{aff107}
\and
University Observatory, LMU Faculty of Physics, Scheinerstr.~1, 81679 Munich, Germany\label{aff108}}

\abstract{
The Euclid Quick Data Release Q2 provides an unprecedented deep, wide-field, high-angular-resolution view of the inner Galactic bulge through the Euclid Galactic Bulge Survey (EGBS). Nine contiguous fields covering 4.8 deg$^2$ were observed with the VIS instrument in 23-24 March 2025. Unlike the nominal \Euclid\ survey strategy, the EGBS employed 16 dithered exposures of 400 s each, corresponding to a total integration time of 1.8 hours per field. Two calibration fields observed at similar solar aspect angles were acquired to derive high-precision point spread functions (PSFs).
The primary objective of the EGBS is exoplanet studies through gravitational microlensing, with high-angular-resolution imaging enabling lens-mass measurements to better than 10\%. The survey data can be applied immediately to previously discovered planetary microlensing events and provide a high-resolution optical reference for future NASA {\it Roman} observations of the Galactic bulge, which are expected to discover more than 1200 microlensing planets, about $10^5$ transiting planets, and hundreds of free-floating planets.
The Q2 data were processed with the \texttt{VIS-PF} pipeline.
The non-standard observing strategy required dedicated calibration products, while the extreme stellar density necessitated modifications to the astrometric calibration and cosmic-ray detection procedures. These improvements have since been incorporated into later pipeline versions.
We describe the motivation, strategy, processing, products and first results of the EGBS. Astrometric residuals with respect to the \Gaia\ DR3 catalogue are 5.5\,mas in right ascension and 4.4\,mas in declination across the survey area, while the photometric zero points are calibrated to 1.5\%. The Q2 release includes calibrated single-dither images, dedicated PSF models and calibrated photometric catalogues containing approximately 45 million detected sources per dither down to AB magnitude 26. Although the catalogues are incomplete because of extreme crowding in the Galactic bulge, these products provide a robust astrometric and photometric foundation for future exploitation of this unique data set.
}


   \titlerunning{Euclid Galactic Bulge Survey (Q2): Overview}
   \authorrunning{J.P. Beaulieu et al.}
   
   \maketitle

\section{\label{sc:Intro}Introduction}

\Euclid \citep{Laureijs11,EuclidSkyOverview} was launched in July 2023 as the second M-class mission of ESA's `Cosmic Vision' programme. The payload consists of a 1.2-m diameter Korsch telescope providing a large field of view of 0.54\,deg$^2$, observed with the VIS optical imager \citep{EuclidSkyVIS} and the NISP near-infrared instrument \citep{EuclidSkyNISP}. \Euclid is a survey mission whose primary objective is to constrain the dark energy equation of state through optical and near-infrared observations of 14\,000\,deg$^2$ of the extragalactic sky.

A number of additional science cases were identified at an early stage \citep{Laureijs11}, including the search for exoplanets via microlensing towards the Galactic bulge.
\citet{BennettRhie2002} showed that a 1\,m-class space telescope equipped with a wide-field imager can resolve and monitor faint source stars that are unresolved from the ground, extending the sensitivity of microlensing surveys to Mars-mass planets. This establishes a strong synergy between microlensing exoplanet studies and cosmological probes such as cosmic shear, which share the need for a space-based wide-field imager with high angular resolution. The microlensing science case was therefore included in early \Euclid mission concepts \citep{Beaulieu2008}, with subsequent simulations demonstrating that a four months survey could probe the mass function of cold planets down to the mass of Mars \citep{Beaulieu2013,Penny2013}. Microlensing has remained a relevant potential component of \Euclid's additional science since then \citep{Laureijs11}, ultimately precipitating the Euclid Galactic Bulge Survey (hereafter EGBS).

Euclid Quick Data Releases (QDRs) are self-contained data sets of particular scientific interest that complement and intersperse the major data releases of the Euclid mission. The EGBS is the subject of Euclid Quick Data Release 2 (Q2). The EGBS's dense stellar fields require an observing strategy distinct from that of the main cosmological surveys, leading to slightly different deliverables. 
%In this paper, we present the observing strategy, the data processing, and data products of the EGBS. 
This paper describes the science goals of the EGBS in Sect.~\ref{Sec:case}, the survey design in Sect.~\ref{Sec:design}, and its data processing in Sect.~\ref{Sec:processing}. Section~\ref{Sec:DataValidation} details the validation of the Q2 data products (calibrated images and photometry catalogues). Section~\ref{Sec:limitations} discusses some limitations that could affect their use, Sect.~\ref{Sec:opportunities} presents some opportunities, and Sect.~\ref{sec:conclusions} concludes the paper. Appendix~\ref{apdx:A} describes the contents of the Q2 data release, how to access the data, and the naming conventions of the data files.

\section{Science case}\label{Sec:case}

\subsection{Microlensing planet searches towards the Galactic bulge}

Microlensing is a rare phenomenon, as it results from the chance alignment of stars \citep{Paczynski1986}. Towards the Galactic bulge, the instantaneous probability that a star is lensed by another star is of order $10^{-6}$, requiring the monitoring of large stellar populations at high cadence. The detectability of low-mass planets depends critically on the angular size of the background star, called the source, relative to the planetary caustic: large sources dilute the microlensing signal \citep{BennettRhie1996}. Consequently, the detection of low-mass planets requires observations of source stars with small angular radii.

Ground-based surveys are limited by a typical angular resolution of around 1\,arcsec, and are therefore strongly affected by crowding and cannot efficiently monitor main-sequence stars of 0.3--0.5~$M_\odot$ in the Galactic bulge. This results in a practical detection limit for planets of 2--3~$M_\oplus$. As of 2026, about 200 cold exoplanets have been discovered by microlensing, mostly with masses of a few $M_\oplus$. Pushing the sensitivity down to Mars-mass planets and below therefore requires a space-based observatory capable of separating faint, unblended source stars in the Galactic Bulge.

\subsection{EGBS as a planet-mass measurement campaign}

A major limitation of microlensing studies has been the difficulty in determining the physical masses and distances of the lens systems. Light-curve modelling yields precise measurements of the planet-to-host mass ratio, $q$, and the projected separation, $d$, in units of the Einstein radius, but converting these into physical quantities typically requires Bayesian inference based on Galactic models, leading to large posterior uncertainties. Three independent observational constraints can substantially reduce this degeneracy.

First, many planetary microlensing events exhibit sharp features associated with caustic crossings, enabling the measurement of the time taken to cross the finite angular size of the source star. With the microlensing timescale and an estimate of the angular size of the source from photometry~\citep{Yoo2004}, this yields the angular Einstein radius, $\theta_\mathrm{E}$, typically with a precision of 10\% or better, providing a strong constraint on the lens mass and distance~\citep[e.g.,][]{Nemiroff1994, An2002}.

A second constraint arises from the measurement of the microlensing parallax, which reflects the change in the observer--lens--source geometry during the event, either due to the observer's orbital motion or the physical separation of observers in space and/or on the ground~\cite[e.g.,][]{Dong2009,Udalski2015,Specht2023}. Parallax measurements can reach precision of the order of 10\%, but are often subject to geometrical degeneracies between solutions with different lens--source relative proper motions that can lead to factor of around 2 ambiguities in mass and distance~\citep[e.g.,][]{Park2004}.

The third kind of constraint is high-angular-resolution imaging, which either directly resolves the source and lens, simultaneously measuring the lens--source relative proper motion, angular Einstein ring radius, and lens flux~\citep{Bennett2006, Bennett.2015.hst169, Bhattacharya.2020, Terry2022a,Rektsini24, Vandorou2025b}, or constrains just the lens flux when it is too close to the source or too faint~\citep{Dong2009, Ranc.2015, Beaulieu2018a}. Such measurements have been obtained %over the period 2012--2026 
using Keck adaptive optics \citep{Bennett2020a} and the \HST \citep[HST,][]{Bennett2024a}. When all three constraints are available, together with the lens--source relative proper motion, physical masses and distances can be determined to 10\% or better.

In a substantial fraction of cases, some or all measurements of observables that could determine the mass and distance to the lens fail to provide meaningful constraints~\citep[e.g.,][]{Penny2016,Terry2026MassMeasurement}. In these cases, the available constraints can be combined with a Galactic model-based prior to estimate a posterior distribution for the mass and distance to the lens~\citep[e.g.,][]{Bachelet2024}.

Over the past three decades, ground-based microlensing surveys (mostly OGLE, MOA, and KMTNet) have discovered thousands of microlensing events in the Galactic bulge~\citep{Kim2018,Mroz2019,Koshimoto2023}, including hundreds involving exoplanetary systems, a significant fraction of which fall within the EGBS footprint. A catalogue of $7801$ microlensing events detected up to 2023 within the EGBS footprint have been identified, including $51$ published planetary systems \citep{Bozza2025}. This historical sample represents an immediate science return from the EGBS, enabling mass measurements for many already known planetary systems.

\subsection{EGBS as a precursor to the \textit{Nancy Grace Roman} Space Telescope survey of the Galactic bulge}

In parallel with the development of \Euclid, the 2010 US Decadal Survey led to the development of the {\itshape WFIRST} mission, now called the {\it Nancy Grace Roman} Space Telescope (hereafter {\it Roman}). {\it Roman} will carry out a dedicated Galactic Bulge Time Domain Survey (GBTDS), aimed at measuring exoplanet demographics beyond 1\,au via microlensing. The mission is based on a 2.4-m telescope at L2, equipped with the 300\,megapixel Wide Field Instrument (WFI), covering 0.48--2.3\,$\mu$m, with a plate scale of 0.11\,arcsec\,pixel$^{-1}$. The current baseline consists of $6 \times 72$-day campaigns targeting 1.7\,deg$^2$ fields towards the region of highest microlensing optical depth near $(l, b) = (1\degf1, -1\degf2)$, with a cadence of 12~minutes in a broad filter (F146), complemented by lower-cadence observations in additional bands \citep{ROTAC2025}.

These fields remain only partially explored by ground-based surveys due to strong extinction \citep[e.g.,][]{Mroz2019,Nunota2025}, but are expected to exhibit very high microlensing event rates \citep{Kerins2009,Specht2020,Huston2026}. The {\it Roman} survey is predicted to detect around 30\,000 microlensing events, including more than 1200 bound exoplanets down to Mars mass \citep{Penny2019} and more than 200 free-floating planets down to Earth mass \citep{Johnson2020}, providing precise measurements of planet-to-host mass ratios and projected separations from the first observing season.

Given the limited observational constraints, the placement of the {\it Roman} GBTDS fields was chosen based on the predictions of population synthesis Galactic models~\citep{Kluter2025,ROTAC2025}. The EGBS fields were chosen to span a larger area than the GBTDS in order to test the model's star count predictions, which are an important component in predicting microlensing event rates~\citep{Penny2013,Penny2019,Poleski2016}, and could potentially provide impetus for a re-evaluation of the GBTDS field choices. The addition of bulge stellar luminosity function data over a wide field in the Galactic bulge will also enable refinement of Galactic models that, amongst other things, are used to infer the mass and distance of microlenses that are not amenable to observational measurement.

\citet{Bachelet2022} demonstrated that a single precursor epoch of \Euclid imaging of the Galactic bulge, covering both the {\it Roman} survey fields and previously monitored ground-based fields, would enable the measurement of relative proper motions for at least 30\% of microlensing events and provide direct constraints on lens fluxes for 42\%, providing masses of the planet-hosting stars to be known at the 10\% level. The EGBS provides such a precursor epoch, ahead of the {\it Roman} exoplanet survey expected to begin in 2027. Thus, EGBS will enable early mass determinations for planets discovered by {\it Roman} from its first observing season.

\begin{figure*}[htbp!]
\centering
\includegraphics[angle=0,width=19cm]{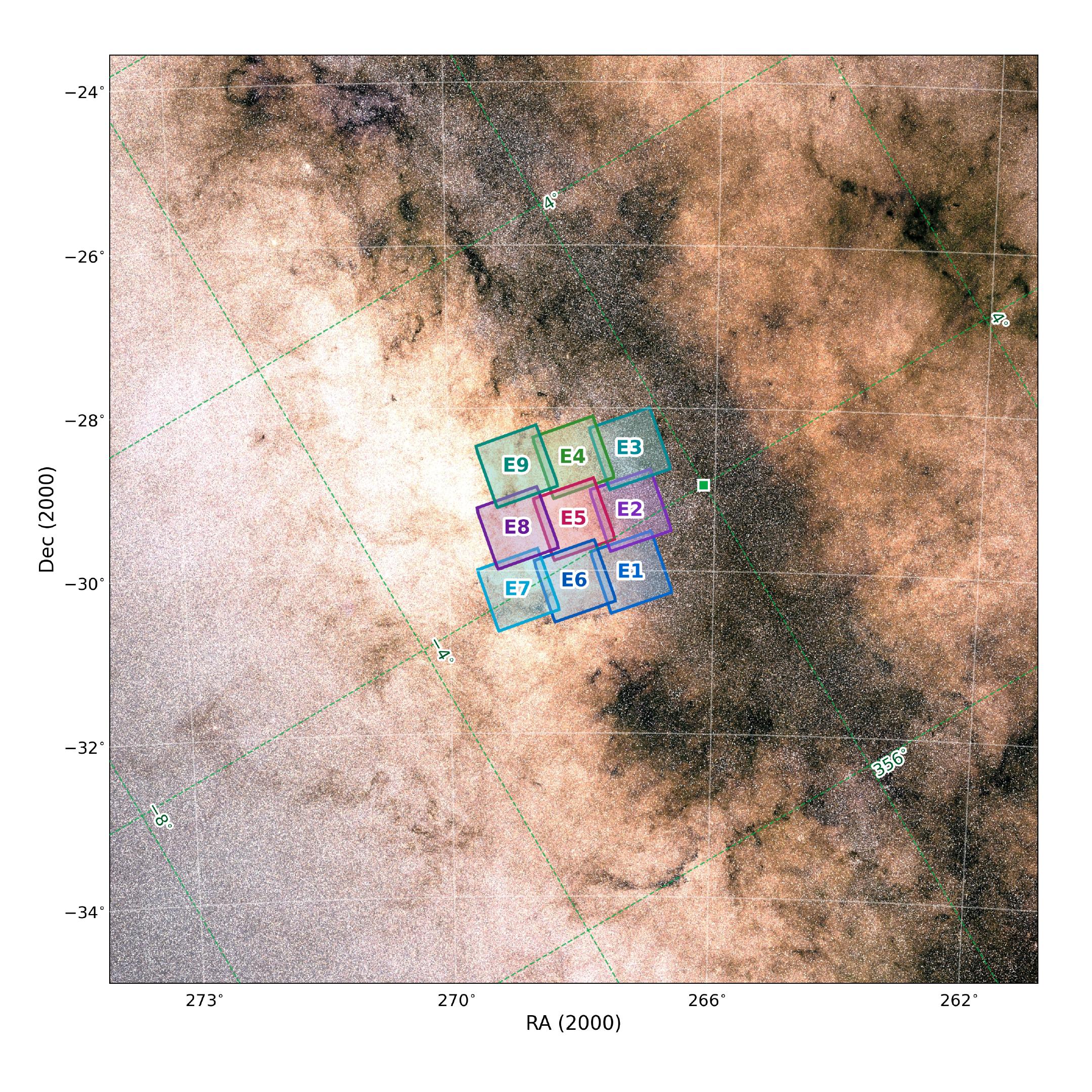}
\vskip -1.0cm
\caption{Outline of the nine EGBS fields. The background visible-light {\it Gaia} image of the Milky Way has been degraded to a resolution of 30\arcsec, and shows both a number of very bright stars, and strong and variable extinction. }
\label{fig:mosaic9fields}
\end{figure*}

%\vskip -1.0cm
\begin{figure}[htbp!]
\centering
\includegraphics[width=9cm]{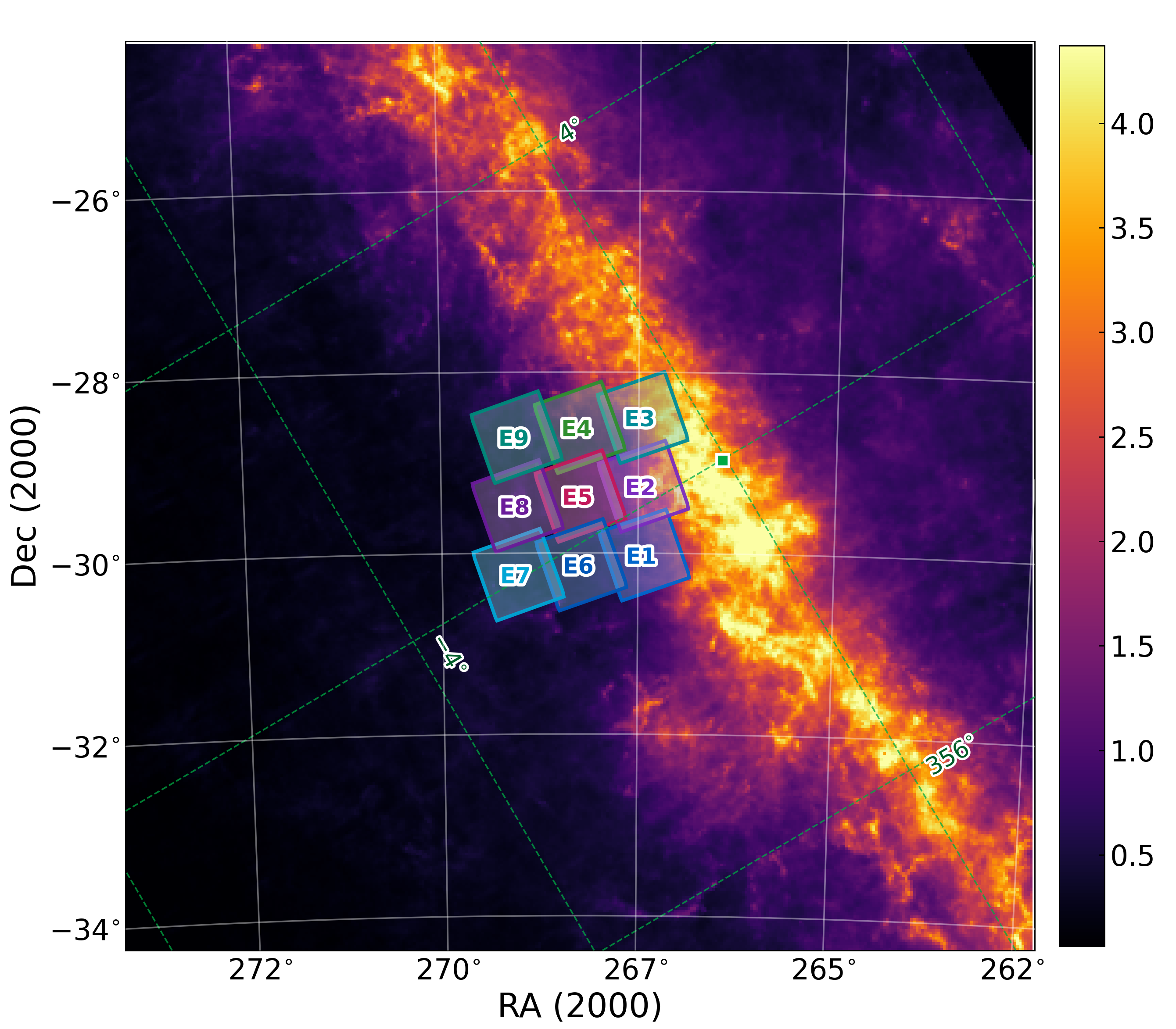}
\vskip 0.0cm
\caption{Extinction in the $K$ band towards the centre of our Galaxy as determined by \citet{Surot2020ReddeningMap}. The nine EGBS fields are shown and labelled.
}
\label{fig:Surot_ext_map}
\end{figure}

\begin{figure}[htbp!]
\centering
\includegraphics[angle=0,width=1.0\hsize]{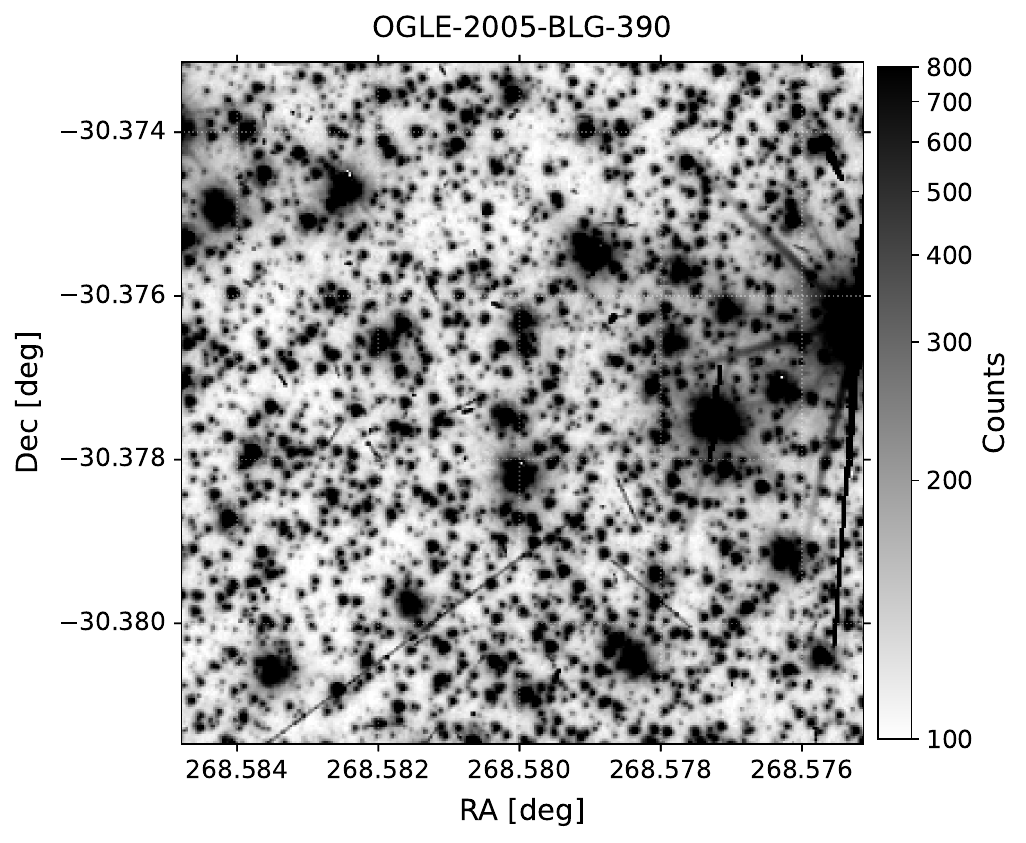}
\caption{Extract of a region from a single EGBS dither, centred on the planetary microlensing event OGLE-2005-BLG-390. }
\label{fig:100arcsec-390}
\end{figure}

\begin{table}
\caption{The nine Galactic bulge fields observed as part of the EGBS. Observations were obtained on 23–24 March 2025 ($\mathrm{MJD-OBS}\sim 60757.40$--$60758.50$), the associated calibration fields were observed during the same period.}
\smallskip
\setlength{\tabcolsep}{3pt} % change inter-olumn spacing
\begin{tabular}{lcccccc}
  \hline\hline
\noalign{\vskip 2pt}
Field  	& RA [deg] 	& Dec [deg] & Dithers &  $A_{\IE}$ & \pz$l$ & \pz$b$\\
         & (J2000)  & (J2000)   & & &[deg] &[deg]\\
  \hline
  \noalign{\vskip 2pt}
E1 	&267.425 &	$-$30.019 & \pz\pz\pz0--15 & 4.61 & 359.533 & $-$1.320 \\
E2 	&267.441 &	$-$29.259 & \pz\pz16--31 & 7.58 & \pz\pz0.192 & $-$0.942 \\
E3 	&267.456 &	$-$28.499 & \pz\pz32--47 & 8.67 &\pz\pz0.851 & $-$0.563 \\
E4 	&268.248 &	$-$28.610 & \pz\pz48--63 & 3.32 & \pz\pz1.111 & $-$1.217 \\
E5 	&268.237 &	$-$29.369 & \pz\pz64--79 & 2.51& \pz\pz0.452 & $-$1.595 \\
E6 	&268.227 &	$-$30.129 & \pz\pz80--95 & 1.69& 359.792 & $-$1.973 \\
E7 	&269.030 &	$-$30.236 & \pz96--111 & 1.50&  \pz\pz0.050 & $-$2.626 \\
E8 	&269.036 &	$-$29.476 & 112--127 & 1.93&  \pz\pz0.711 & $-$2.249 \\
E9 	&269.041 &	$-$28.716 & 128--143 & 1.85& \pz\pz1.371 & $-$1.872 \\
\hline
\end{tabular}
 \footnotesize
 \tablefoot{Each field was observed with 16 dithered exposures. The adopted naming convention assigns sequential indices to the dithers within each field. The mean \IE-band extinction, $A_{\IE}$, was estimated from a 100\,arcsec stamp centred on each field. We also provide the Galactic coordinates $(l, b)$.}
\smallskip
\label{tab:pos}
\end{table}

\vspace{24pt}
\section{EGBS design}\label{Sec:design}

\subsection{General considerations}

\Euclid was not designed to undertake studies within the crowded stellar environment of the Galactic bulge. The EGBS therefore represents a significant challenge and a big departure from the standard \Euclid observing mode, requiring careful design and testing to ensure its successful implementation. 
The principal constraints imposed upon the EGBS were: (i) the Euclid Wide Survey would not be interrupted for more than 48 hours; and (ii) sensitivity requirements would be satisfied on resumption of the Euclid Wide Survey.  Within these constraints, a pointing and survey strategy had to be devised.

Testing of possible survey designs was performed using detailed VIS and NISP simulated images of the Galactic bulge fields, together with analyses of \Euclid's imaging performance from a series of Euclid Deep Field North (EDF-N) NISP dark and science frames taken within a 48-hour time window. Of particular concern from initial simulations was the possibility of long-lived persistence effects for NISP \citep[see][]{EuclidSkyNISP} that could mean that the main survey requirements would not be met by the end of the 48 hour interruption. Ultimately, NISP persistence was characterised more accurately using the EDF-N images that showed NISP was able to recover to required levels within 12 hours of the final EGBS exposure, even for hyper-saturated observations. However, this still provided a challenge to design a VIS+NISP survey that could meet signal-to-noise requirements whilst being comfortably contained within the permitted time envelope.

\subsection{EGBS implementation}

Ultimately, observations were carried out exclusively using the VIS instrument \citep{EuclidSkyVIS}. This ensured that observations retained the highest possible spatial resolution that is critical for proper motion determinations. Nonetheless, NISP observations would have provided valuable information, in particular for mitigating the colour dependence of the PSF within the dense EGBS stellar fields, which are subject to strong and spatially variable extinction.

\begin{figure}[htbp!]
\centering
\hspace*{-0.5cm}
\includegraphics[angle=0,width=9cm]{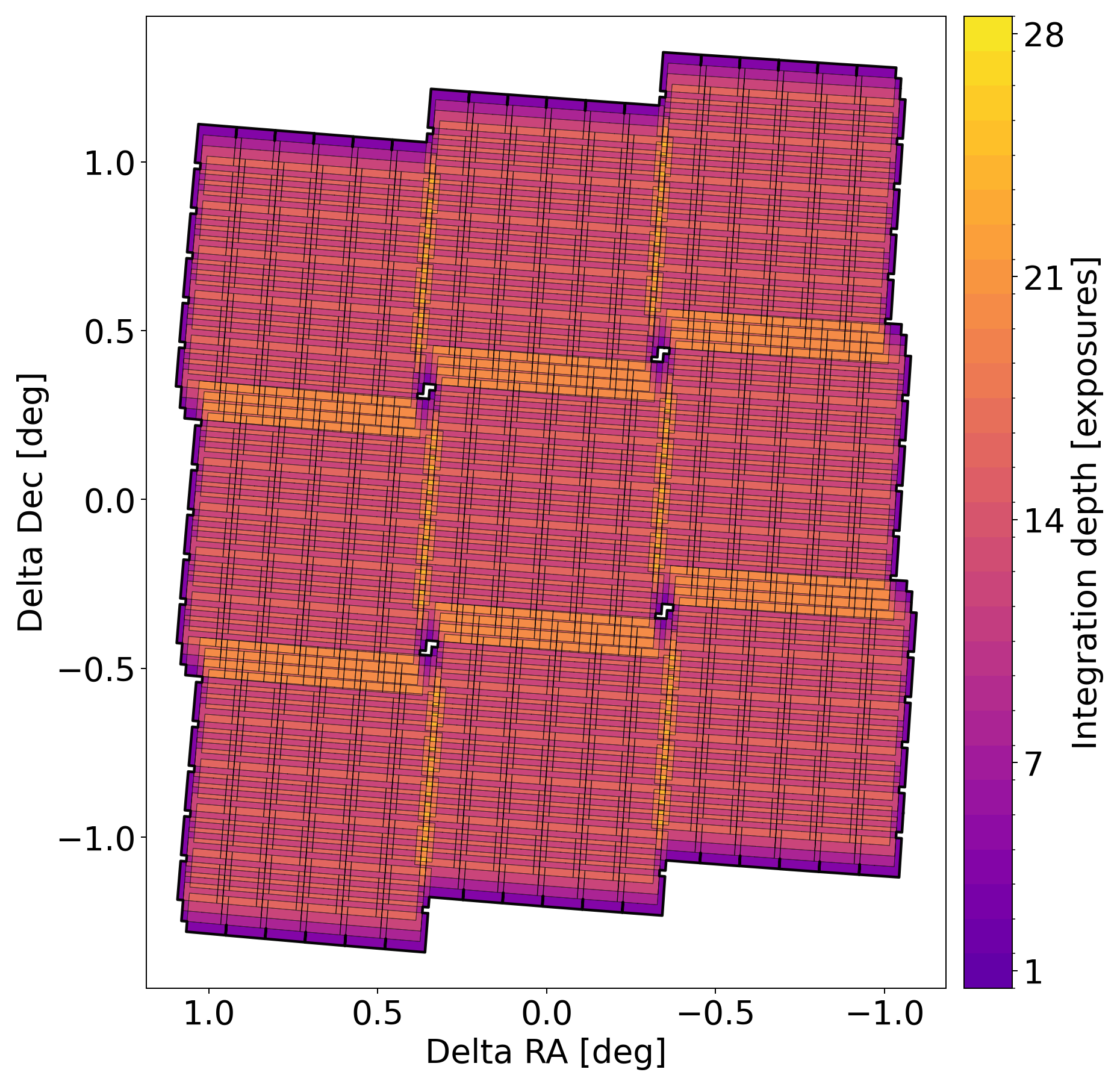}
\caption{Integration map of the EGBS fields. The colour scale indicates the number of VIS exposures contributing to each position.}
\label{fig:overlap_dither}
\end{figure}

The final VIS-only observing strategy comprises nine target fields, covering approximately 4.5~deg$^2$, with a strongly variable density and extinction (Figs.~\ref{fig:mosaic9fields} and \ref{fig:Surot_ext_map}). Each field was observed through a sequence of 16 dithered exposures, each with an integration time of 400\,s (Fig.~\ref{fig:overlap_dither}). Figure~\ref{fig:100arcsec-390} illustrates the extreme stellar density of a typical EGBS field in a single dither.

Additionally, a less crowded stellar field at $({\rm RA},{\rm Dec})\,=\,(265\degf715,+48\degf075)$ was targeted for PSF self-calibration, again with 16 dithers, taken both before and after the main EGBS fields. 
The calibration fields were observed at solar aspect angles similar to those of the EGBS fields in order to reproduce the same thermal environment experienced by the telescope. This is important because small thermally induced changes in the optical system can affect the shape and stability of the VIS point spread function. Matching the observing conditions therefore ensures that the derived PSF models are representative of the science observations. The data from these fields were used to construct a PSF model that is provided as part of Q2 data release.

\begin{figure}
\centering
\includegraphics[width=\columnwidth]{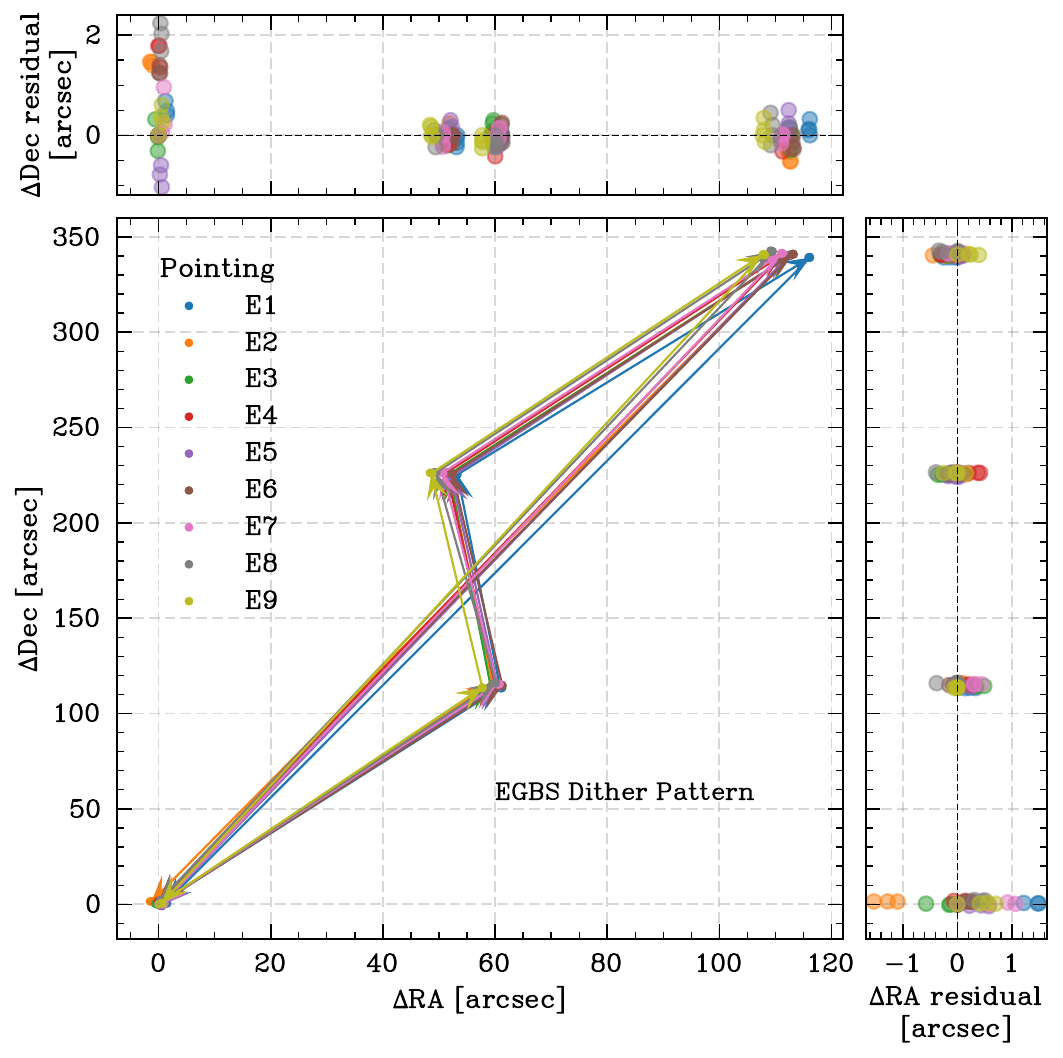}
\caption{Dither pattern for the nine EGBS pointings. The central panel shows the $\Delta$RA and $\Delta$Dec offsets, with arrows indicating the dither progression, relative to the first dither of each pointing. Only first four of the 16 dithers are displayed per pointing as a 4-step reference pattern, which has been repeated four times during EGBS. The top and right marginal panels show residual offsets of all 16 dithers with respect to the 4-step reference pattern, highlighting deviations in $\Delta$Dec and $\Delta$RA, respectively.}
\label{fig:dither_pattern}
\end{figure}

The EGBS observing sequence (hereafter EOS) for individual pointings differs from the standard Euclid Wide Survey reference observing sequence (ROS; \citealt{Scaramella-EP1,EuclidSkyOverview}). Each EOS consists of 16 dithers, whose relative offsets are given in Table~\ref{tab:ditherpos} with visualisation in Fig.~\ref{fig:dither_pattern} and illustrated in Fig.~\ref{fig:mosaic9fields}. At each dither position, VIS acquires a 400\,s nominal EGBS science exposure. This is repeated for each of the nine science fields shown in Fig.~\ref{fig:mosaic9fields}, as well as for two visits of the PSF self-calibration field.

The choice of 400\,s exposures over the nominal \Euclid VIS exposure used for the standard ROS was made to ensure that the EGBS images could be compressed within the envelope of the onboard compression timeout, which if reached causes a partial loss of data. To test compression times, simulated VIS images generated from a Besan{\c c}on Galactic population synthesis model \citep{Awiphan2016} using the {\tt gulls} image simulator~\citep{Penny2013} were injected into in-flight VIS dark frames and then compressed through the a ground-based version of the onboard software. These tests indicated compression times for the high stellar density Galactic bulge fields that were around 60\% longer than for standard \Euclid exposures within the Euclid Wide Survey. It was deemed that 400\,s was the maximum exposure time that would allow compression of the full image before timeout, although this was comfortably above the 300\,s assumed in the simulations of \cite{Bachelet2022}.

\subsection{Sixteen-dither strategy for high-S/N PSF reconstruction}

A key feature of the EGBS observing strategy is the use of 16 dithers per field, repeating four times the standard four-dither pattern of the \Euclid ROS. This effectively provides 16 independent samplings of each field, as originally proposed by \citet{Bachelet2022}. 

\begin{figure}
\centering
\includegraphics[width=\columnwidth]{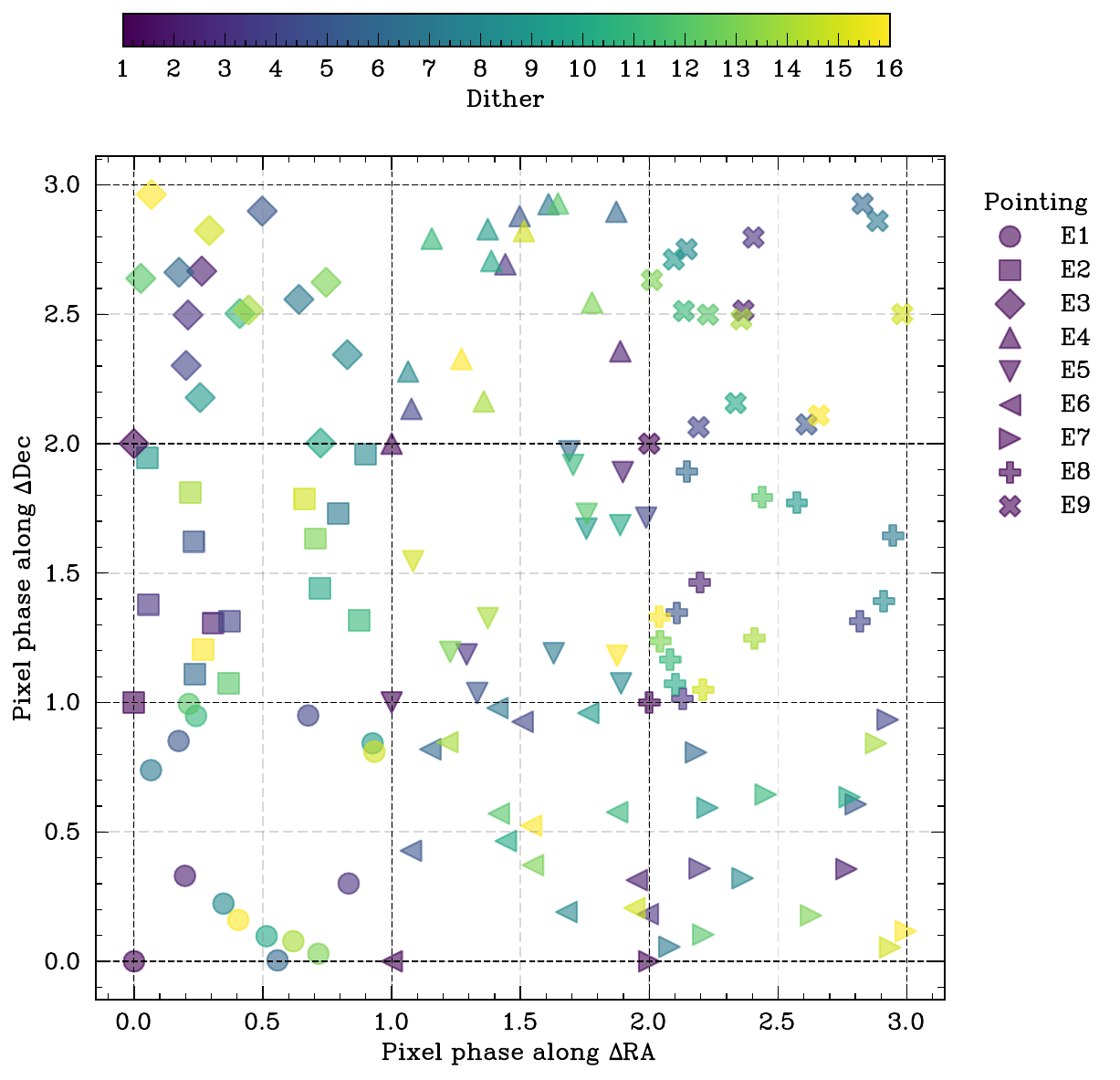}
\caption{Pixel-phase distribution of the EGBS dither sequence for nine pointings. Different pointings are separated into a 3$\times$3 layout by adding unit offset for visual clarity. This layout is shown by the dashed black grid lines indicating the unit pixel-phase boundaries for each pointing.}
\label{fig:pix_phase}
\end{figure}

This strategy significantly improves the reconstruction of the PSF \citep{Anderson2000}, while also enabling deeper imaging through stacking. In particular, multiple dithered observations enable accurate sampling of the PSF at different pixel phases. The pixel phase distribution is shown in Fig. \ref{fig:pix_phase}. Although the high stellar density in the EGBS fields could, in principle, support precise PSF reconstruction, spatial variations of the PSF across the field of view limit this approach (as we discuss below). Dithering therefore provides a critical complement. The accumulation of multiple dithers also mitigates the impact of detector systematics, such as bad pixels and cosmic rays. 

These data differ significantly from those of the Euclid Wide Survey. The nominal \texttt{VIS-PF} pipeline has been designed for high-Galactic-latitude fields characterised by low source densities \citep{_Henry2025}. In contrast, the extremely high source density of the EGBS fields, exceeding that of the EWS fields by more than two orders of magnitude, requires a careful assessment of the performance of the \texttt{VIS-PF} processing in close collaboration with the relevant science working groups.

%\vfill
%\eject

\section{Processing Q2 data}\label{Sec:processing}

We now briefly summarise the main data-processing steps, with particular emphasis on the modifications and updates applied to the \texttt{VIS-PF} pipeline~\citep{_Henry2025}. We first present the results at the scale of the CCD mosaic for a single field, in order to highlight variations introduced by the optical system, the focal plane, and the individual CCDs. We then present a global overview of the nine fields, which span a wide range of extinction though remain extremely crowded.

\subsection{Preprocessing}\label{processing}

The VIS focal plane consists of 36 back-illuminated charge-coupled devices (CCDs), each 40~$\mu$m thick and with a format of $4132 \times 4096$\,pixels, corresponding to a pixel scale of \ang{;;0.1} and sky coverage of $0.54\,\mathrm{deg}^2$. Observations are taken in a single broad band filter, \IE, covering $0.55$--$0.9\,\mathrm{\mu m}$. This filter is approximately equivalent to a combined \textit{r}+\textit{i}+$\textit{z}$\ band.
Each CCD is read out through four independent nodes, resulting in a natural subdivision into four quadrants. Consequently, a single exposure of a field comprises 144 quadrants. The layout and naming conventions for the CCDs and quadrants are described in figure~1 of \citet{_Henry2025}.

The raw VIS frames are processed by the standard pipeline \texttt{VIS-PF}, tuned for EGBS data. It is essentially the same pipeline that was used for both Q1 and DR1 data. This pipeline applies (in order) bad-pixel masking, bias subtraction, saturation masking, electronic crosstalk correction, non-linearity correction, cosmic-ray detection, `brighter--fatter' correction, flat-fielding and gain equalisation. Finally, a sky background frame is computed. Optical ghosts are also masked. 

The very high stellar density and non-standard exposure time in the EGBS means that the pipeline requires several adjustments. The probability threshold for the \texttt{DeepCR} cosmic-ray algorithm \citep{Zhang2020} is raised because the default produces many false detections in crowded fields; the sky-background parameters of the \texttt{NoiseChisel} program (from the GNU Astronomy Utilities, version 0.16; \citealt{2015ApJS..220....1A}) are also adjusted to account for the high source density. In particular, the  \texttt{NoiseChisel} \texttt{TileSize} parameter was reduced from 31 pixels to nine pixels, with the image's mode (most frequent pixel value with a resolution of 0.10 ADU) used as a constant background value when the other attempts failed. Finally, no correction for the motion of the shutter across the focal plane (the illumination correction) was made, since the non-standard EGBS exposure time differs from the two models available in \texttt{VIS-PF}. Shutter motions introduce a small residual horizontal photometric gradient across the focal plane, but in practice this effect is not detected in our data (as discussed in Sect.~\ref{Sec:DataValidation}).
%(Fig.~\ref{fig:vis-q2-bulge-photometry-fpa}). 

\subsection{Astrometric and photometric calibration}

Astrometric and photometric calibration were performed by reference against \Gaia\ source catalogues \citep{Gaia2016,GaiaDR3Summary}. Astrometric calibration was carried out using a distortion model derived from the self-calibration field, as done in (McCracken, H. et al. 2026, in prep). 

The photometric zero point was derived using the DR1 zero-point model, a time-dependent polynomial fit to \Gaia-derived zero points that accounts for the temporal evolution of the VIS throughput, including the effects of ice contamination and subsequent decontamination events (McCracken, H. et al. 2026, in prep).
Although the EGBS observations were obtained more than one year after the second \Euclid decontamination event, the DR1 time-dependent zero-point model was adopted to ensure consistency with the nominal survey processing.

\subsection{Photometric catalogue production and PSF definition }

Photometric catalogues were extracted using \texttt{Source\-Extractor} \citep{1996A&AS..117..393B}, with the Q1 PSF, to produce a calibrated source catalogue for each exposure. For each dither, the catalogue includes both instrumental and calibrated fluxes and magnitudes measured within circular apertures of 7, 13, 26, and 50 pixels, and PSF-fitting photometry. We used the same circular apertures as in Q1. Due to the crowding within the EGBS fields, small apertures are preferred for photometric measurements. 

The standard Q1 VIS PSF model \citep{_Henry2025} was used to produce our source calibrated catalogue. 
%We used the standard Q1 VIS PSF model \citep{Q1-TP002} to produce our source calibrated catalogue.
However, to enable further processing we generated a new PSF model, to be provided with Q2, that utilised the PSF field data. It was generated using observations of a less dense field that provides more isolated stars than the EGBS field. These data were taken before and after the EGBS but at the same solar aspect angle and exposure times.
Initial assessments of the astrometry and photometry produced indicated that it was sufficient for the VIS-PF pipeline, so a second run of the pipeline using a model derived from the Q2 PSF fields was not necessary. 

\begin{figure}
\centering
\includegraphics[width=\hsize]{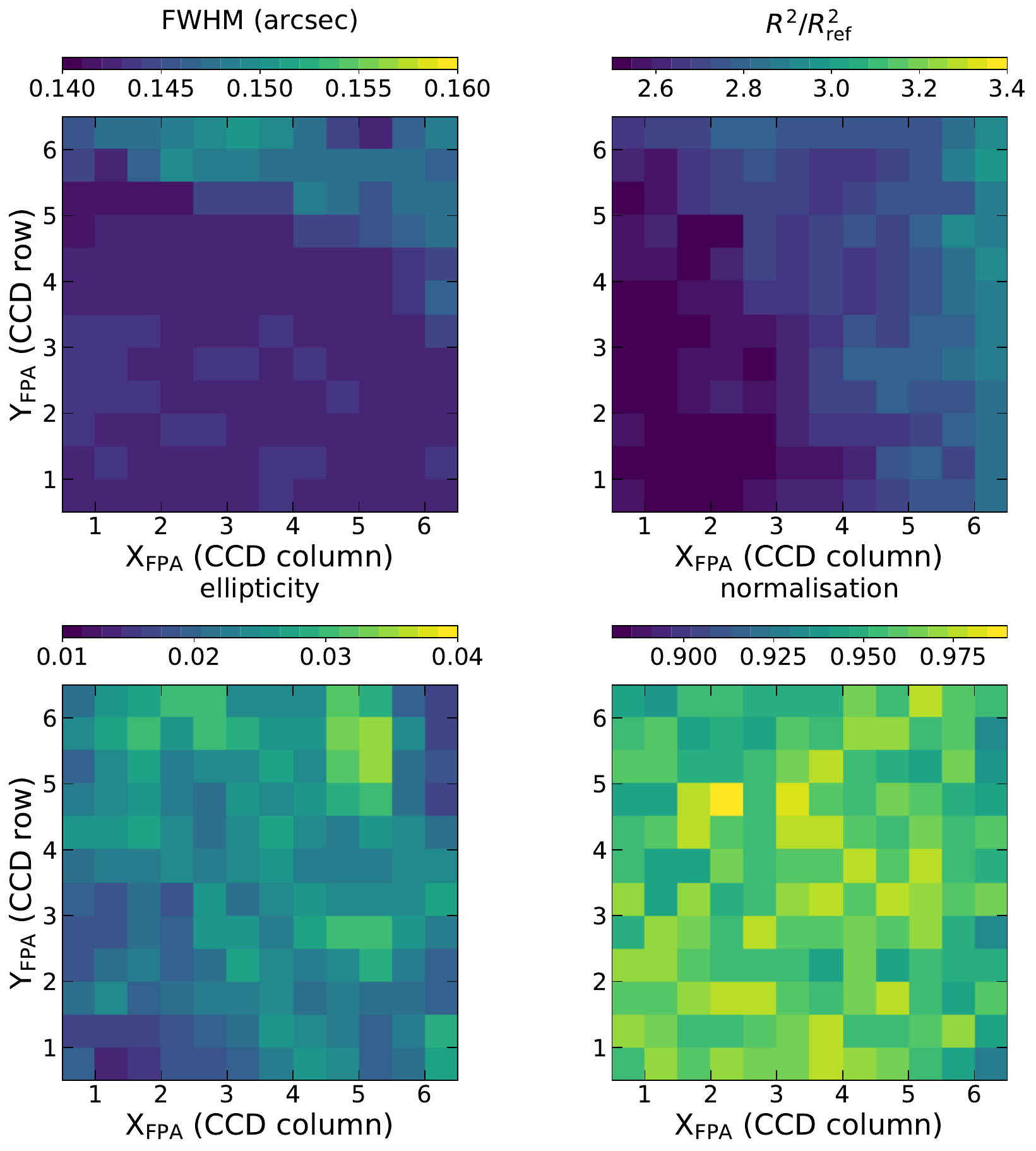}
\caption{%
Variations of FWHM, size, ellipticity, and flux normalisation factor of the VIS Q1 PSF model across the focal plane. }
\label{fig:vis-q1-psf-fpa}
\end{figure}

\begin{figure}
\centering
\includegraphics[width=\hsize]{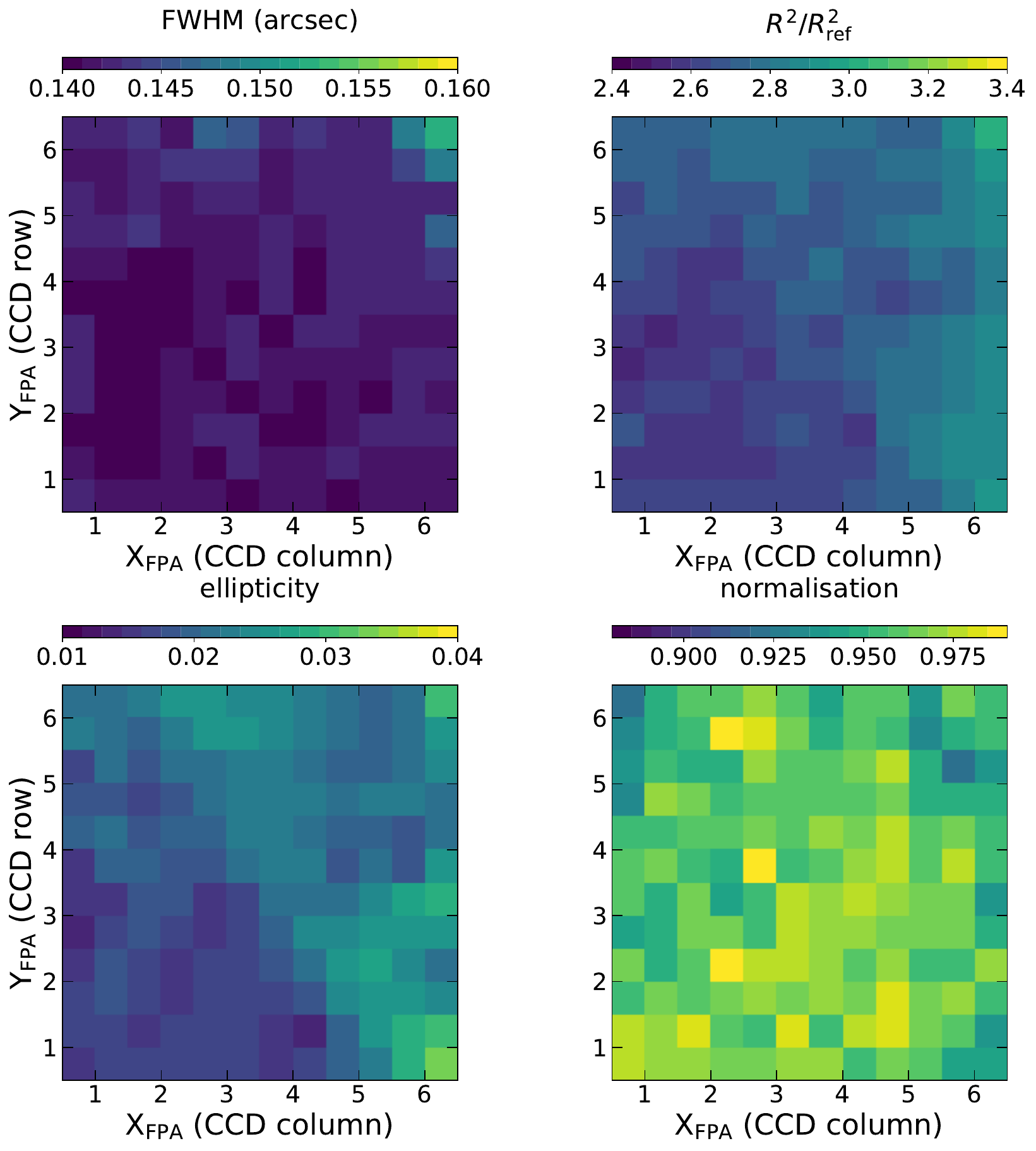}
\caption{%
Variations of FWHM, size, ellipticity, and flux normalisation factor of the VIS Q2 PSF model across the focal plane. }
\label{fig:vis-q2-psf-fpa}
\end{figure}

\begin{figure}
\centering
\includegraphics[width=\hsize]{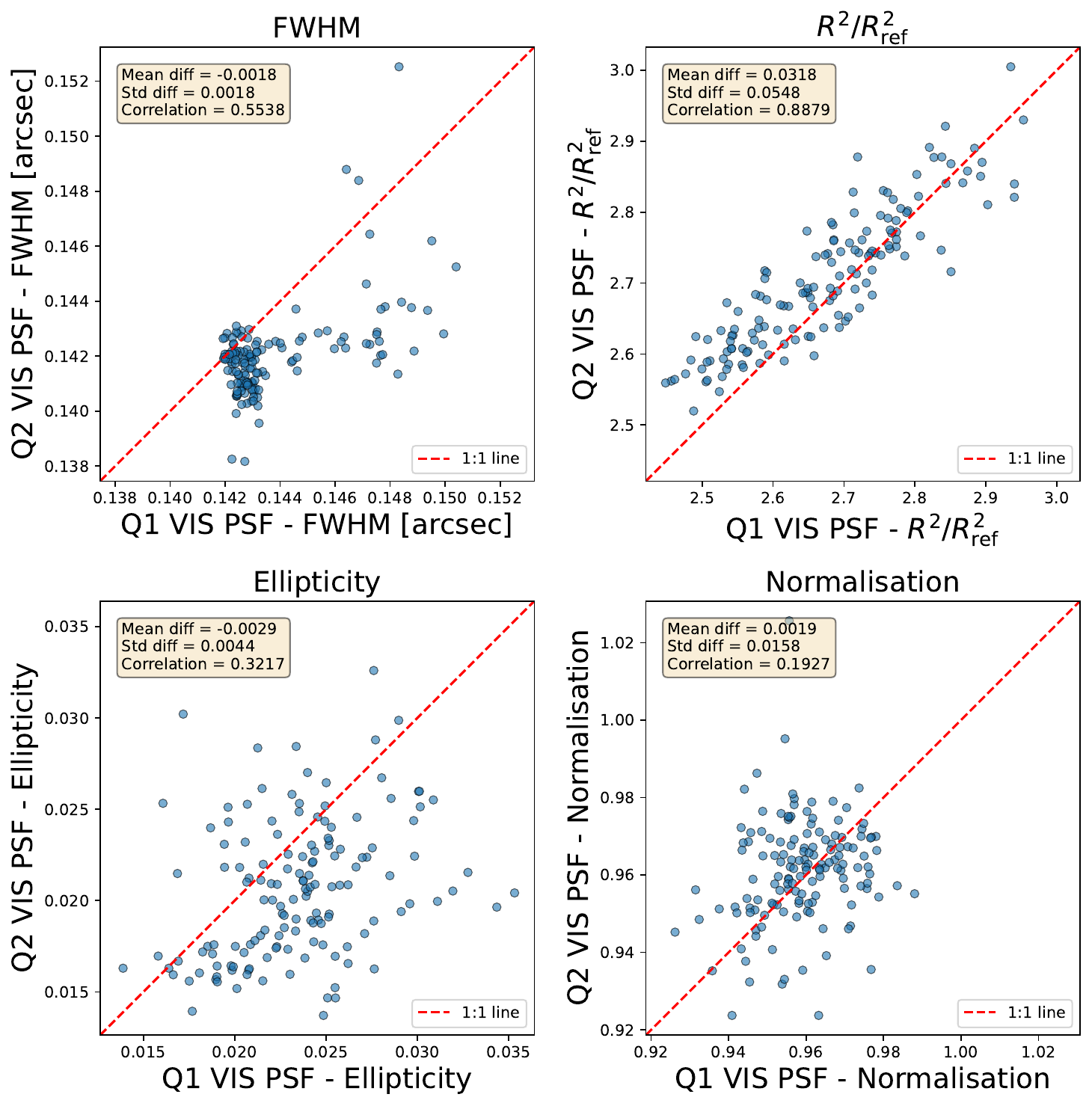}
\caption{%
Comparison of the variations of FWHM, size, ellipticity and flux normalisation factor of the Q1 and Q2 VIS PSF models across the focal plane.}
\label{fig:vis-compare-psf-fpa}
\end{figure}

In Fig.~\ref{fig:vis-q1-psf-fpa} we show the per-CCD-quadrant values of the Q1 VIS PSF metrics across the focal plane of 36 CCDs (FPA), and in Fig.~\ref{fig:vis-q2-psf-fpa} we show the same metrics for the Q2 VIS PSF model. Finally, in Fig.~\ref{fig:vis-compare-psf-fpa} we present a comparison of the two PSF models across the focal plane. We can see that the full-width at half-maximum (FWHM) of the Q2 PSF is slightly narrower and more stable than the FWHM of the Q1 PSF, while there is a strong correlation between the PSF size of both models. We define the PSF size as the sum of second orders of image moments.

For the processing of the EGBS data, the standard Q1 PSF model was adopted to ensure consistency with the other \Euclid data releases. The dedicated Q2 PSF model is nevertheless provided as part of the release, as its slightly narrower profile may offer advantages for source extraction in the extremely crowded EGBS fields and could improve the resulting photometric and astrometric precision. However, both models were generated using the PSFex tool and as a result their characteristics are affected by the tool's parameters input. A study is underway in order to understand better the effect of the parameters choices on the generated PSF models and what is the correlation between the generated PSF models and the actual stars in the field. Finally, as we show in the analysis below, the dominant limitations of the current VIS-PF catalogues arise from source-detection issues rather than from the choice of PSF model.

\section{Data product validation}\label{Sec:DataValidation}
\subsection{Astrometric and photometric calibration}

Starting from the calibrated catalogues of individual frames described above, we ran an independent validation pipeline developed in \texttt{VIS-PF} to assess the
astrometric and photometric accuracy over all observations. For each source in each calibrated frame catalogue, we compute the position on the focal plane via a per-quadrant pixel-to-mm world coordinate system (WCS) solution. Point-like sources are selected using \texttt{SourceExtractor} \texttt{FLAGS}$\,{=}\,0$, $1.0<\mathrm{FLUX\_RADIUS}<1.3$, and $18 < \IE < 22$; although, given the high source density in the EGBS, we applied an additional cut at $\IE < 20$ before the cross-match. Sources were matched against the \Gaia\ DR3 catalogue~\citep{GaiaDR3Summary} using 
a \ang{;;0.5} tolerance, with \Gaia\ positions propagated from the reference epoch ($2016.0$) to the VIS observation epoch using the \Gaia\ proper motions. Sources from different visits within $0\arcsecf2$ of one another were associated into groups corresponding to the same physical object, yielding a merged catalogue of 1\,727\,449 measurements from 207\,496 distinct sources. We applied a \Gaia\ astrometric-quality filter ($\mathrm{RUWE} < 1.2$ and $\sigma_{\mu_{\alpha^*}}, \sigma_{\mu_\delta} < 0.5\,\mathrm{mas\,yr^{-1}}$) to remove sources whose proper-motion uncertainty dominates the residual budget, leaving 366\,305 stars in the range $18.5 < \IE < 19.8$ for the astrometric validation and 266\,645 sources for the photometric validation (after a further $6\%$ outlier clipping). This catalogue provides an excellent way to assess if there are any remaining focal-plane-dependent residuals over the entire survey.

To characterise the astrometric and photometric precision, for each of the $144$ quadrants $i$ we measured the source count $n_i$, mean residual $\mu_i$, and within-quadrant dispersion $\sigma_i$. For both internally referenced and externally calibrated astrometry and photometry we computed a common source-weighted mean 
\begin{equation}
\mu_{\rm s}=\frac{\sum_i \,n_i\mu_i}{\sum_i n_i},
\label{eqn:mu_s}
\end{equation}
and two dispersions: the per-quadrant dispersion defined by
\begin{equation}
\sigma_{\rm q}^2=\frac{\sum_i n_i\,(\mu_i-\mu_{\rm s})^2}{\sum_i n_i},
\label{eqn:sigmaq}
\end{equation}
which measures spatial variation across the focal plane, and the per-source dispersion
\begin{equation}
\sigma_{\rm s}^2=\sigma_{\rm q}^2+\frac{\sum_i n_i\,\sigma_i^2}{\sum_i n_i},
\label{eqn:sigmas}
\end{equation}
which adds the within-quadrant scatter to $\sigma_{\rm q}$ to recover the dispersion of individual measurements.

\begin{figure}
\centering
\includegraphics[width=\hsize]{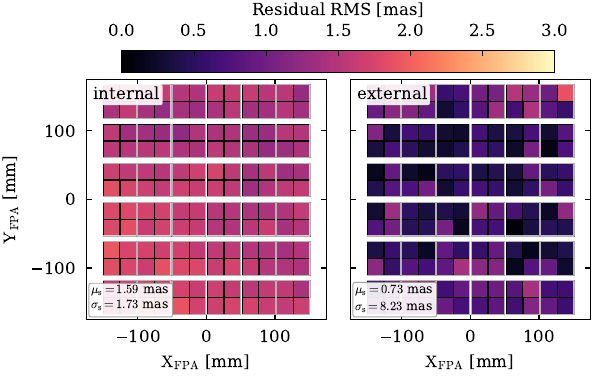}
\caption{%
Per-CCD-quadrant astrometric residual maps computed on the grouped catalogue using 366\,305 stars with
$18.5 < \IE < 19.8$ after \Gaia\ astrometric quality cuts.
\emph{Left:} Per-quadrant mean of the internal repeatability, defined as
the positional scatter of repeated observations of the same source within
each quadrant.
\emph{Right:} Magnitude of the per-quadrant mean residual vector against
\Gaia\ DR3 positions propagated to the observation epoch.
The inset gives the source-level aggregate statistics $\mu_{\rm s}$
and $\sigma_{\rm s}$, obtained by combining the per-quadrant moments
across all $144$ quadrants. }
\label{fig:vis-q2-bulge-astrometry-fpa}
\end{figure}

\begin{figure}%[h]
\centering
\includegraphics[width=\hsize]{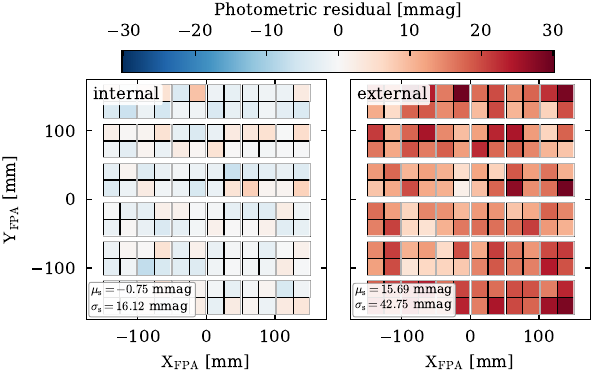}
\caption{%
Per-CCD-quadrant photometric residual maps computed on the grouped catalogue using 266\,645 sources (6\% clipped)
with $18.5 < \IE < 19.8$ and the same \Gaia\ astrometric-quality cuts as in
Fig.~\ref{fig:vis-q2-bulge-astrometry-fpa}.
\emph{Left:} Per-quadrant mean of the large-scale flat (LSF) repeatability residual,
measuring the photometric stability of repeated observations of the same source.
\emph{Right:} Per-quadrant mean difference between the VIS aperture-calibrated
magnitude \IE and the \Gaia-predicted magnitude ($G_{\rm BP}-G_{\rm RP}$ colour interpolation),
measuring absolute photometric calibration.
Both panels share a common colour scale.
The inset box gives the source-level aggregate 
statistics $\mu_{\rm s}$ and $\sigma_{\rm s}$, obtained by combining the per-quadrant moments across all $144$ quadrants.}
\label{fig:vis-q2-bulge-photometry-fpa}
\end{figure}

Figure~\ref{fig:vis-q2-bulge-astrometry-fpa} shows these per-quadrant astrometric residual maps from the grouped star catalogue, both dither-to-dither (left panel) and compared to \Gaia\ (right panel). Similarly, Fig.~\ref{fig:vis-q2-bulge-photometry-fpa} shows the same for the photometric residuals, dither-to-dither (left panel) and the external comparison with \Gaia\ (right panel); \Gaia\ magnitudes were transformed to VIS magnitudes using a variant of the transformation presented in \citet{_Henry2025}. The source-level statistics $(\mu_{\rm s},\sigma_{\rm s})$ defined above are quoted in the figure insets.

Table~\ref{tab:perfsum} summarises these per-quadrant and per-source residuals
over the survey. Overall, the photometric and astrometric performance over the
entire survey is excellent, similar to measurements in Q1 and DR1 data. We note
the higher internal and external per-source photometric residuals of
about 16\,mmag and 43\,mmag, respectively, which we attribute to
confusion in the aperture photometric measurements. These crowding errors
dominate the error budget (and are probably why the imprint of the illumination
is not visible in the focal-plane maps). Over the DR1 survey, in lower-density
fields, internal errors are approximately 2\,mmag.

\begin{table}
\centering
\small
\caption{%
Astrometric and photometric residuals for all sources with $18.5 < \IE < 19.8$, after \Gaia\ quality cuts.}
\setlength{\tabcolsep}{10pt} % change inter-olumn spacing
\begin{tabular}{lcccc}
\hline\hline
\noalign{\vskip 2pt}
 & \multicolumn{2}{c}{{\pz}Internal} & \multicolumn{2}{c}{{\pz}External} \\
   & \pz$\mu$ & \pz$\sigma$ & \pz$\mu$ & \pz$\sigma$ \\
  \hline
  \noalign{\vskip 2pt}
  \multicolumn{5}{@{}l}{{Astrometry}\ [mas]} \\
  \ \ per source     & \pp$1.59$  & \pz$1.73$  & \pz$0.73$  & \pz$8.23$  \\
  \ \ per quadrant   & \pp$1.59$  & \pz$0.15$  & \pz$0.73$  & \pz$0.37$  \\
  \multicolumn{5}{@{}l}{{Photometry}\ [mmag]} \\
  \ \ per source     & $-0.75$ & $16.12$ & $15.69$ & $42.75$ \\
  \ \ per quadrant   & $-0.75$ & \pz$2.82$  & $15.69$ & \pz$5.66$  \\
  \hline
  \label{tab:perfsum}
  \end{tabular}
   \footnotesize
 \tablefoot{Per-CCD-quadrant values represent the mean, $\mu$, and dispersion, $\sigma$, across the 144 focal plane cells.}
  \end{table}

Across the EGBS fields, we observe strong spatial variations in extinction (Fig.~\ref{fig:Surot_ext_map}), implying different mean stellar colours from one field to another. We therefore examine these effects on the scale of the nine EGBS fields individually.

\begin{figure*}[htbp!]
%\sidecaption
  \centering
\includegraphics[angle=0,width=17cm]{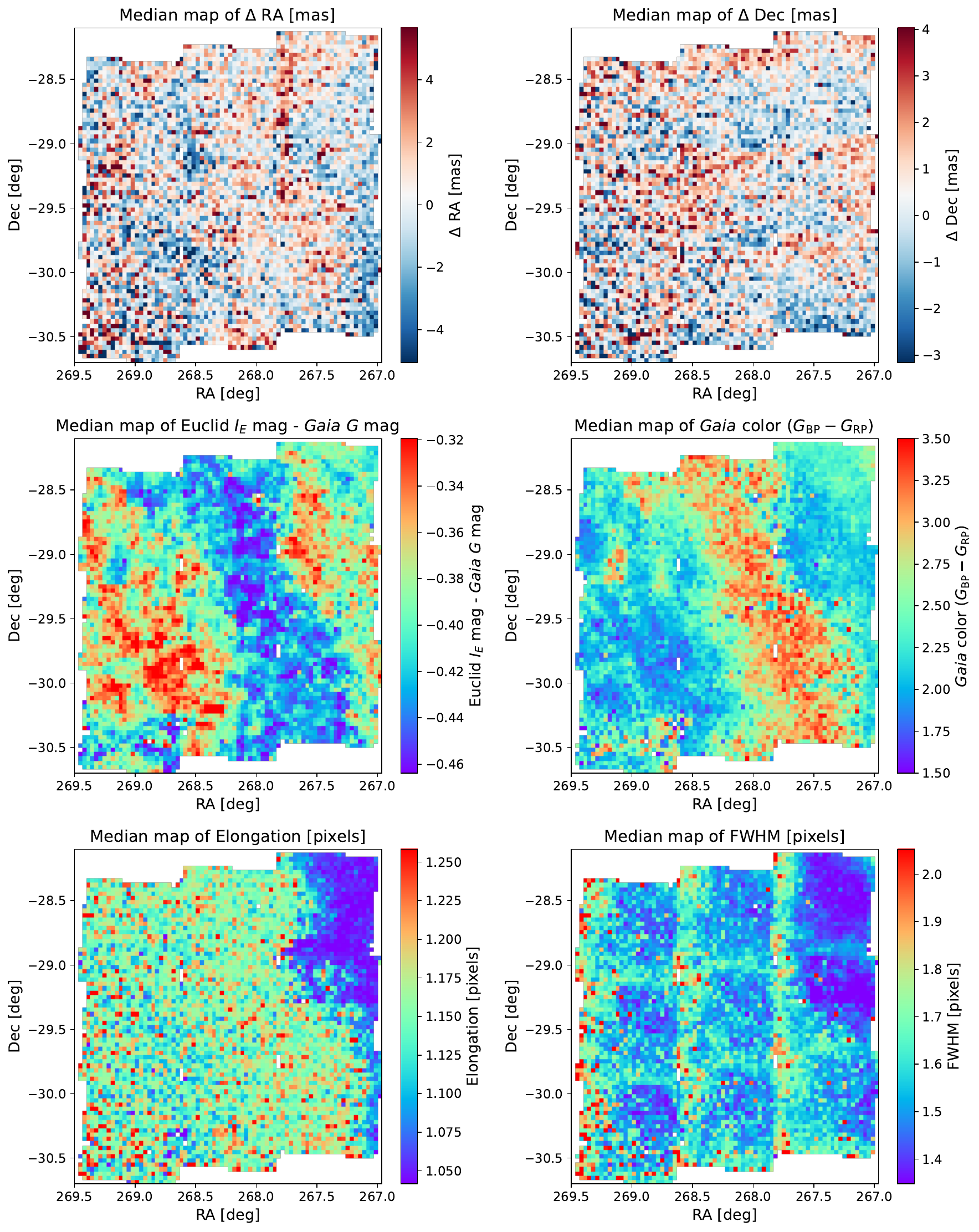}
\caption{
Maps comparing EGBS Q2 catalogue stars from the first dither to stars cross-identified with the \textit{Gaia} DR3 catalogue over the full EGBS field. The comparisons use a sample of approximately $2 \times 10^{5}$ relatively bright stars with mostly $G<19.5$. Each panel shows the median value of the following statistics computed over spatial bins of $2$\arcmin$\times2$\arcmin.
\textit{Top panels:} Residuals in milliarcsec in right ascension (\textit{left}) and declination (\textit{right}), $\Delta$RA and $\Delta$Dec, respectively. 
\textit{Middle left:} Difference between the \Euclid\ \IE\ and \textit{Gaia} $G$ magnitudes. 
\textit{Middle right:} Mean \textit{Gaia} colour, $G_{\rm BP} - G_{\rm RP}$. 
Both bands are broad, with \textit{Gaia} extending further into the red, and are defined in different magnitude systems (Vega for \textit{Gaia}, AB for \Euclid). The two colour maps are strongly anti-correlated.
\textit{Bottom panels:} The elongation and FWHM of the PSF, in pixels, for bright stars. 
 }
\label{fig:6panelastrom}
\end{figure*}

\begin{figure}[htbp!]
%\sidecaption
  \centering
\includegraphics[angle=0,width=9cm]{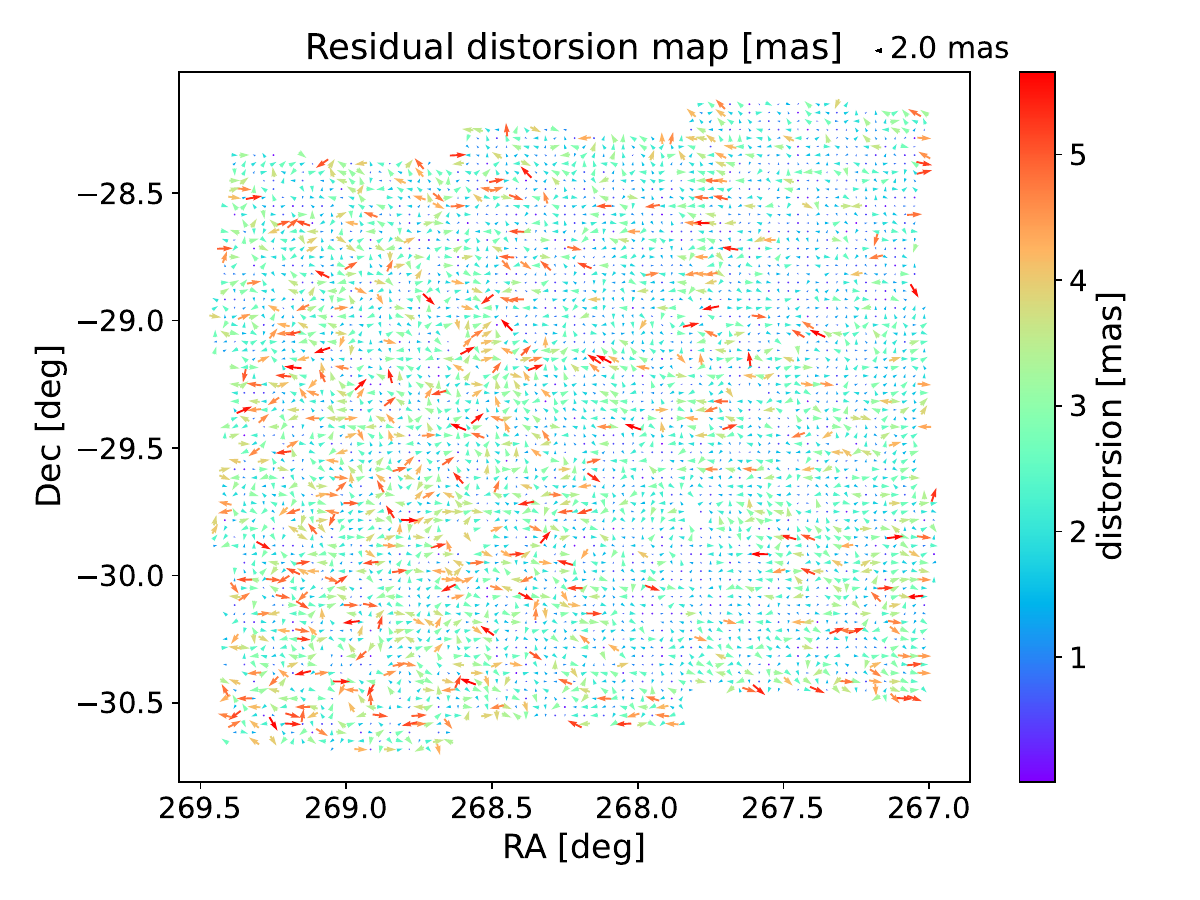}
\caption{Distortion vectors for the EGBS data, based on comparison to cross-identified \Gaia\ sources. Results have been binned by $2$\arcmin$\times2$\arcmin, as in Fig.~\ref{fig:6panelastrom}. The arrow size is proportional to the magnitude of the distortion, with an arrow outside the plot at the top showing the size of a $2.0$\,mas distortion. The colour of the arrow also indicates the size of the distortion.}
\label{fig:vectors}
\end{figure}

Figure~\ref{fig:6panelastrom} summarizes the astrometric properties for one dither in each of the nine fields, while Fig.~\ref{fig:vectors} shows the corresponding distortion vectors across all fields. The astrometry residuals are generally uniform across the nine fields. The middle panels of Fig.~\ref{fig:6panelastrom} compare the mean \Gaia\ $G$-band magnitude with the calibrated \Euclid \IE\ magnitude and the \Gaia\ colour $G_{\rm BP}-G_{\rm RP}$. Since the \Gaia\ $G$ band is significantly broader, a correlation between the two quantities is expected, and can be seen by comparing the two plots. The lower left panel of Fig.~\ref{fig:6panelastrom} shows that the mean elongation measured for bright stars is relatively uniform across the survey, with the exception of the upper-right fields, where it is lower. This may indicate a colour-dependent effect, although these fields are also characterised by lower stellar densities, which could implicate crowding. A similar pattern is seen in the map of PSF FWHM in the lower right panel of Fig.~\ref{fig:6panelastrom}, though with an additional imprint of a common intra-FPA pattern seen in each of the nine pointings. This highlights the subtlety involved in handling Q2 catalogues generated through the standard \Euclid VIS-PF processing pipeline.

\subsection{Assessment of data quality with maps of the EGBS fields}

\begin{figure}%[]
  \centering
    \hspace*{-0.5cm}
  \includegraphics[angle=0,width=10cm]{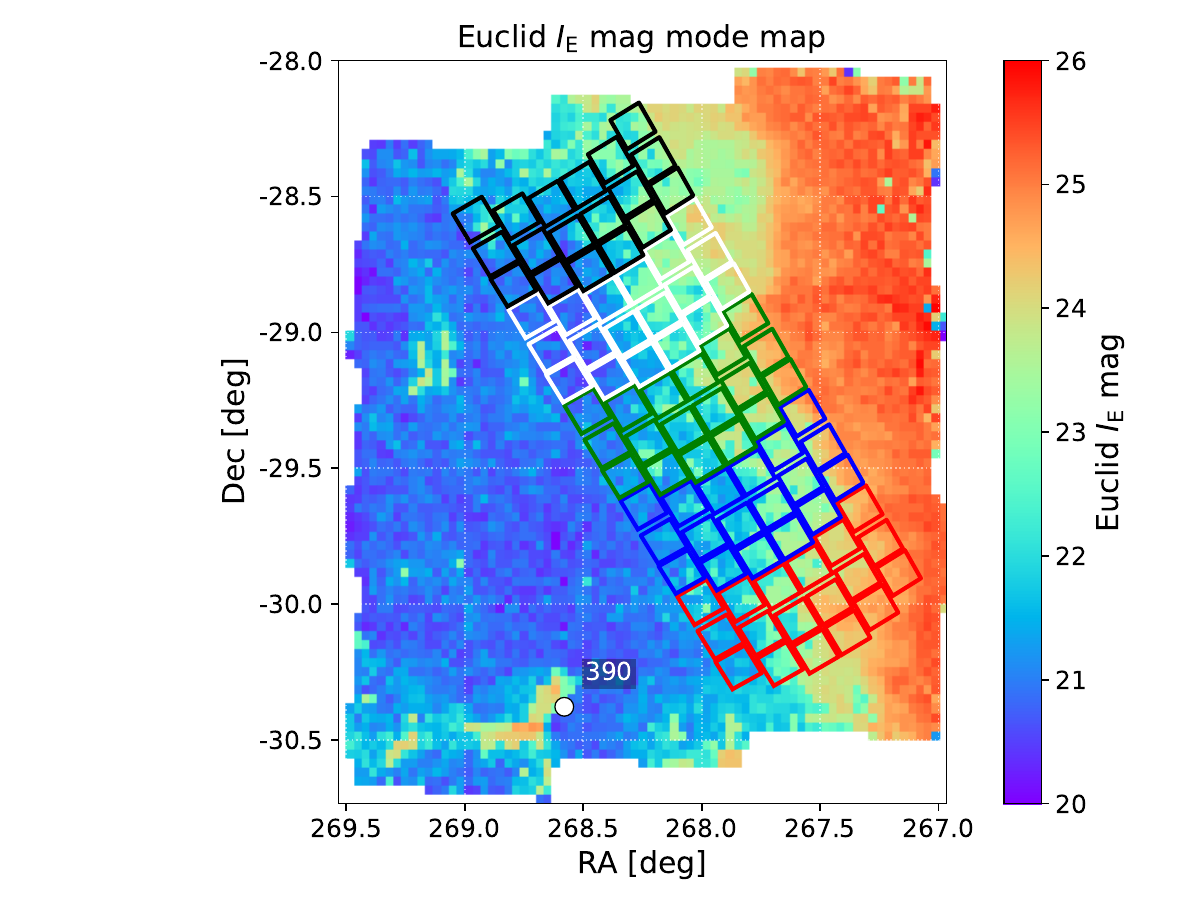}
\caption{Map of the mode of \Euclid $\IE$ magnitudes, computed at a spatial resolution of $2$\arcmin$\times2$\arcmin. 
The five pre-selected {\it Roman} microlensing fields are overplotted. The planetary microlensing event OGLE-2005-BLG-0390 mentioned in the text, is indicated.}
\label{fig:hist_mag2arcmin}
\end{figure}

For each of the nine fields, we selected the two dithers with the largest separation and merge the corresponding catalogues into a single file. When multiple detections were found within 50\,mas (0.5\,pixels) of one another, only one source was retained. 
This procedure yields a catalogue containing approximately 60 million sources from the VIS-PF processing. The multiple dither pattern increases the effective sky coverage by filling the gaps between CCDs in the VIS focal plane. In order to map the variations across the fields, we divided them into $2\arcmin\times2\arcmin$ bins. Within each spatial bin, the mode of the $\IE$ magnitude distribution was computed without applying an extinction correction. This mode varies from $\IE \approx 21$ to $\IE \approx 25$, with the faintest values occurring in highly extincted regions where the density of detectable stars is reduced (Fig.~\ref{fig:hist_mag2arcmin}).

The depth of the field is limited by the crowding and erroneous estimates of the background, which affects the detection threshold. These issues prevent the detection of the fainter stars in the densest regions, as discussed in Sect.~\ref{processing}. 

\subsubsection{Probing the limitation of Q2 processing, and mitigation}\label{ob390proc} 

To assess the level of incompleteness in the Q2 catalogues, we first consider a crowded field containing the planetary microlensing event OGLE-2005-BLG-390 \citep{Beaulieu2006}. We extracted image cubes comprising the 16 dithered VIS exposures centred on the event position.
%over a field of view of $100.5 \times 100.5$~arcsec$^2$

The 16 dithered observations obtained for each field enable the construction of deeper stacked images using a forward-modelling approach implemented in the \texttt{JAX}-based package \texttt{MIJax} (\url{https://github.com/euclid-egbs/MIJax}). This framework provides efficient estimation of complex spatial gradients, allowing the reconstruction of a higher-resolution model of the scene from multiple slightly shifted lower-resolution observations.

The first step consists of recalibrating the WCS solution for the individual dithers by extracting source catalogues and cross-matching them against \Gaia. Transformation matrices between the individual catalogues and a reference catalogue were then derived. The reference catalogue was selected such that the target of interest is located as close as possible to the centre of the field.

\begin{figure*}
\centering
\includegraphics[width=9cm]{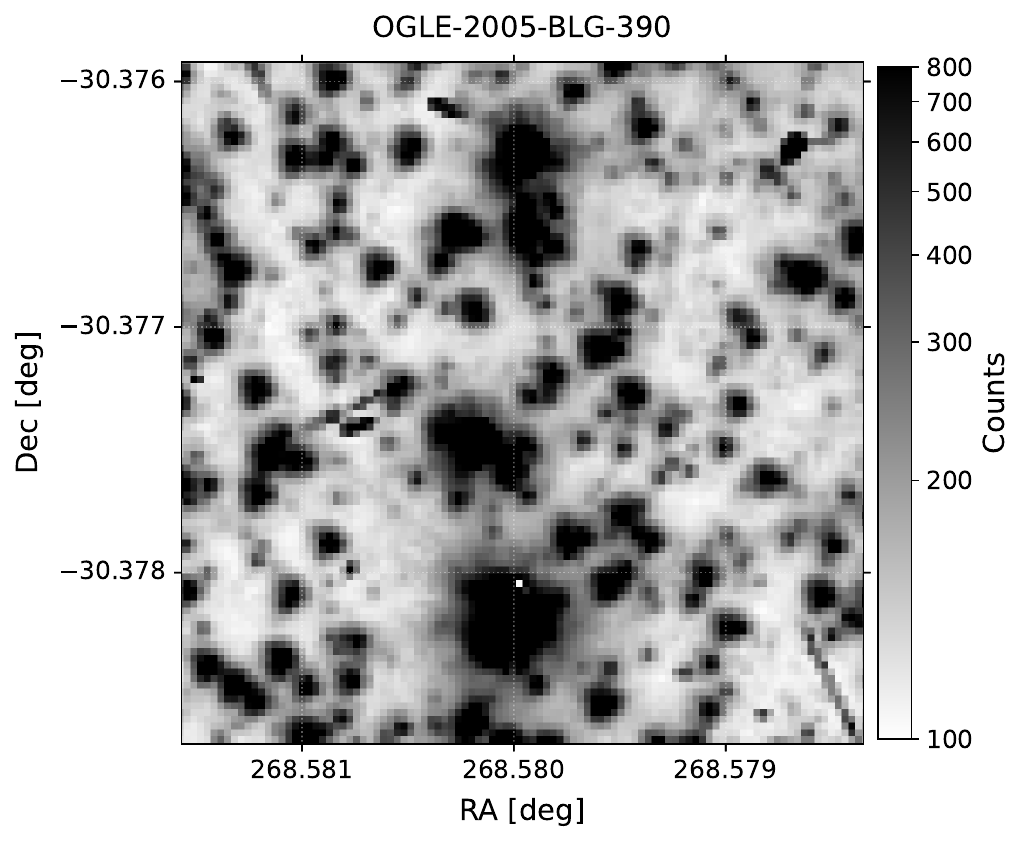}
\includegraphics[width=9cm]{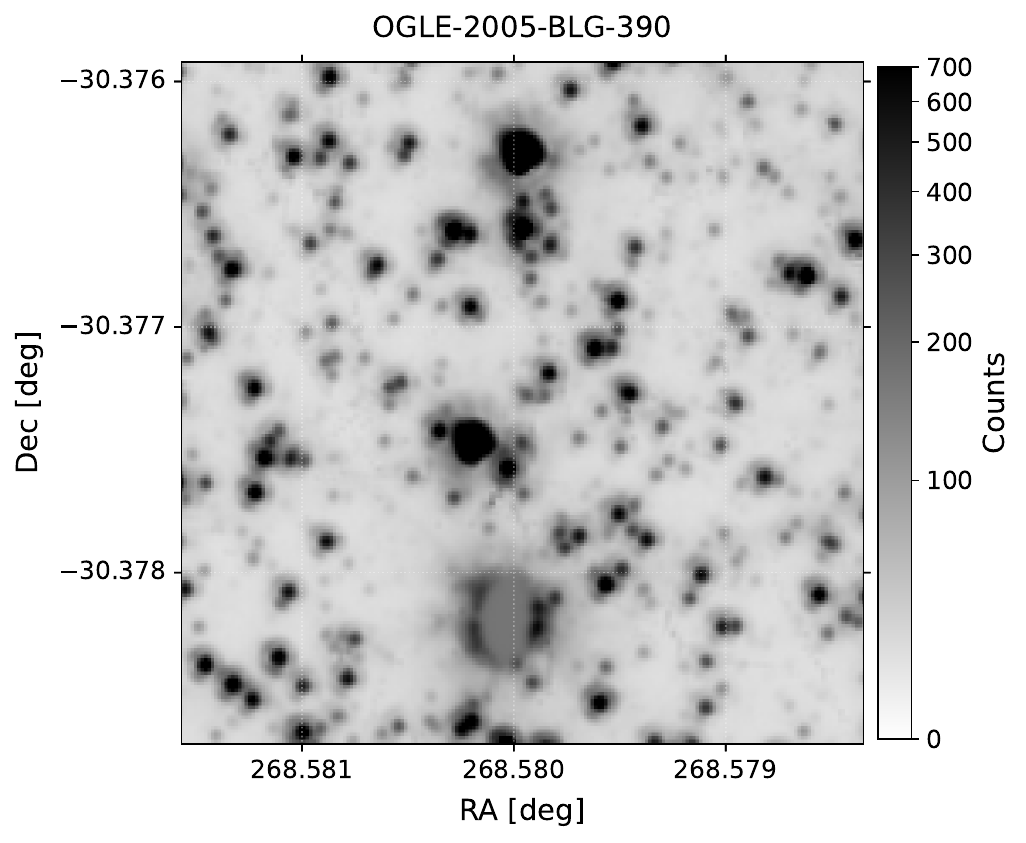}
\caption{Zoom into a $10$\arcsec$\times 10$\arcsec\,region centred on the planetary microlensing event OGLE-2005-BLG-390, extracted from the field shown in Fig.~\ref{fig:100arcsec-390}. \textit{Left:} Single \Euclid dither. \textit{Right:} Stack of 16 dithers, illustrating the gain in depth and source detectability obtained through image combination.}
\label{fig:ob05390_image}
\end{figure*}

A forward-modelling approach was subsequently applied, in which the high-resolution model was shifted and rebinned to reproduce each low-resolution observation of the field. Using the numerical gradients provided by \texttt{JAX}, the model parameters were iteratively refined until convergence was achieved. An example of a small section of such stack, super sampled image, compared to a simple dither is presented in Fig.~\ref{fig:ob05390_image}.

When the stack was obtained, the star extraction was performed once again. These sources were then used to perform PSF photometry, using a PSF model estimated internally, while an approximate zero-point was computed by comparison to the \Gaia{} catalogue. 

Finally, we computed a detection efficiency using a Monte Carlo injection-recovery method, where thousands of artificial stars were simulated in the stacks. This allows a robust estimation of the overall completeness of the source detection and, ultimately, to reconstruct an accurate luminosity function. We note that the correction is applied only if the estimated efficiency is above 10\%, which is the typical error on the efficiency.

\begin{figure}
\centering
\includegraphics[width=\hsize]{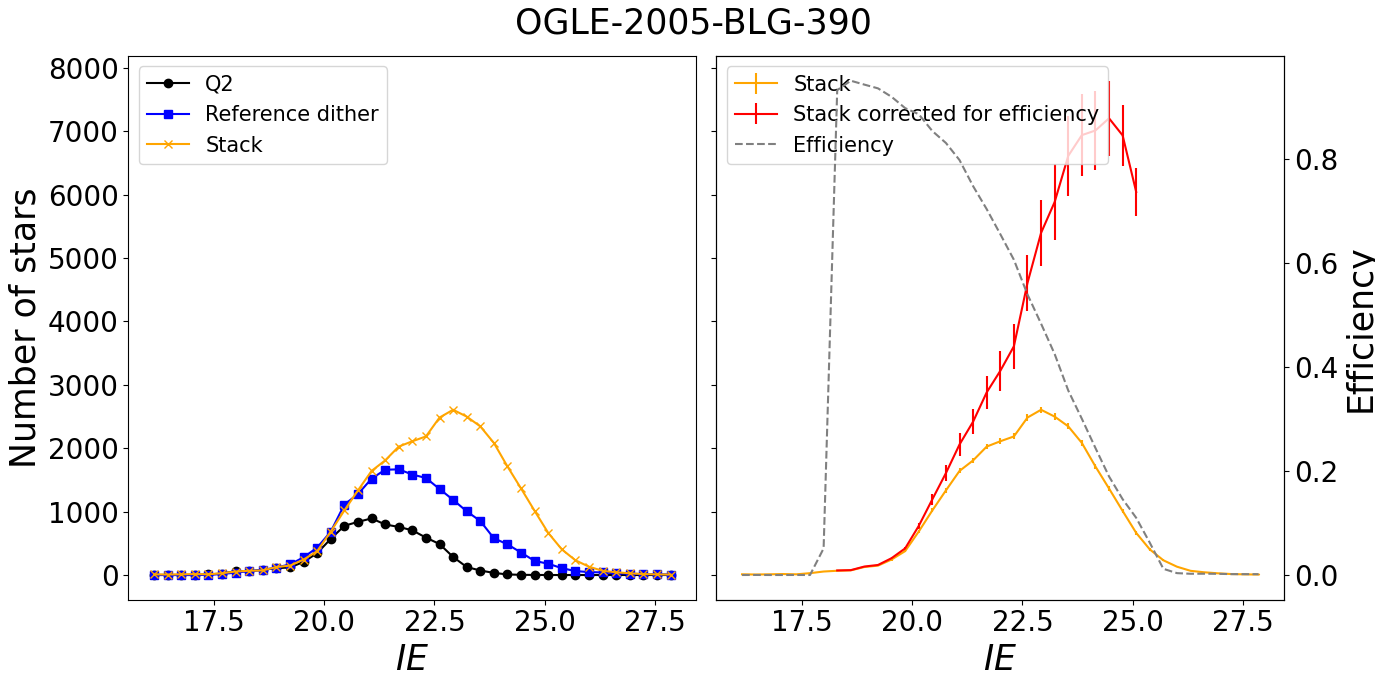}
\caption{%
\textit{Left:} Number of stars detected in a $100\arcsecf5\times 100\arcsecf5$ cutout centred around OGLE-2005-BLG-390Lb. In black are the magnitude distributions of detections available from the Q2 catalogue (7962 sources), while the blue and orange curves are detections using \texttt{DAOStarFinder}, on the reference dither (18\,792 sources) and the stack (31\,693 sources). \textit{Right:} The red curve shows distribution of counts in the stack corrected for the detection efficiency (shown as the grey-dashed curve).}
\label{fig:ob05390_completness}
\end{figure}

For the OGLE-2005-BLG-390 field, the standard Q2 catalogue contains only 7962 detected sources ($0.8\,\mathrm{stars\,arcsec^{-2}}$). Using a source-detection pipeline based on \texttt{DAOStarFinder} \citep{1987PASP...99..191S}, we detect 18\,792 stars in the reference dither ($1.9\,\mathrm{stars\,arcsec^{-2}}$) and 31\,693 stars in the stacked image ($3.1\,\mathrm{stars\,arcsec^{-2}}$). This comparison illustrates the limitations of \texttt{SourceExtractor} in extremely crowded fields (Figs.~\ref{fig:ob05390_image} and \ref{fig:ob05390_completness}).

We repeated this analysis for a 100\,arcsec stamp centred on each of the nine EGBS fields. The coordinates and mean \IE-band extinctions are listed in Table~\ref{tab:pos}. Most fields have extinctions in the range $A_{\IE}=1.5$--3.32, while three fields exhibit substantially higher extinction ($A_{\IE}=4.61$, 7.58, and 8.67).

The results for the individual fields are presented in Fig.~\ref{fig:E1-E9_completness}. For single dithers, the Q2 VIS-PF catalogues are highly incomplete, with the level of incompleteness varying significantly from field to field. This behaviour reflects a combination of crowding, background estimation, and source-detection threshold effects. Comparison with catalogues generated from individual dithers and from stacked images shows that the limitations of the Q2 catalogues cannot be explained solely by stellar density or extinction.

We therefore turn to the catalogues derived from the stacked images. Figure~\ref{fig:comp9} presents the observed \IE\ magnitude distributions per square arcminute and the corresponding cumulative distributions for the nine fields. The lower panels show the same distributions after correction for the mean extinction in each field. 

The highly extincted fields E2 and E3 exhibit markedly different magnitude distributions, indicating that a large fraction of the bulge population is obscured by dust and that the detected stars are predominantly foreground objects. In contrast, the remaining fields display similar magnitude and cumulative distributions, suggesting that the stacked-image analysis yields a more homogeneous characterisation of the underlying stellar populations. After applying a mean extinction correction and correcting for detection efficiency, the magnitude distributions and luminosity functions of the different fields become remarkably similar. This convergence indicates that much of the field-to-field variation seen in the standard VIS-PF catalogues arises from incompleteness and source-detection limitations rather than intrinsic differences in the stellar populations.

\section{Limitations of Q2 data}\label{Sec:limitations}

\subsection{PSF modelling and chromatic effects}

For the Q2 photometric and astrometric processing, we initially adopted the standard VIS PSF model used for the Q1 data release. In addition, a dedicated PSF model was derived from calibration observations obtained with the same solar aspect angle and exposure time as the EGBS science observations. This model is included in the Q2 release and is expected to provide a more accurate representation of the survey data. However, it does not include any dependence on source colour.

A possible approach to investigate chromatic effects in the VIS PSF would be to generate theoretical PSFs~\citep[e.g.,][section~7.6.4]{Paykari-EP6,EuclidSkyOverview} for different stellar spectral types and compare them with empirical PSFs measured in the EGBS fields. In practice, this is complicated by the highly variable extinction across the survey area, ranging from $A_K\,{\approx}\,0.5$ to $A_K\,{\approx}\,4$. As a consequence, the observed stellar population is significantly redder than that of the calibration fields, preventing a straightforward determination of the colour dependence of the Q2 PSF.

\FloatBarrier

\begin{figure*}[ht]
\centering
\includegraphics[width=9cm]{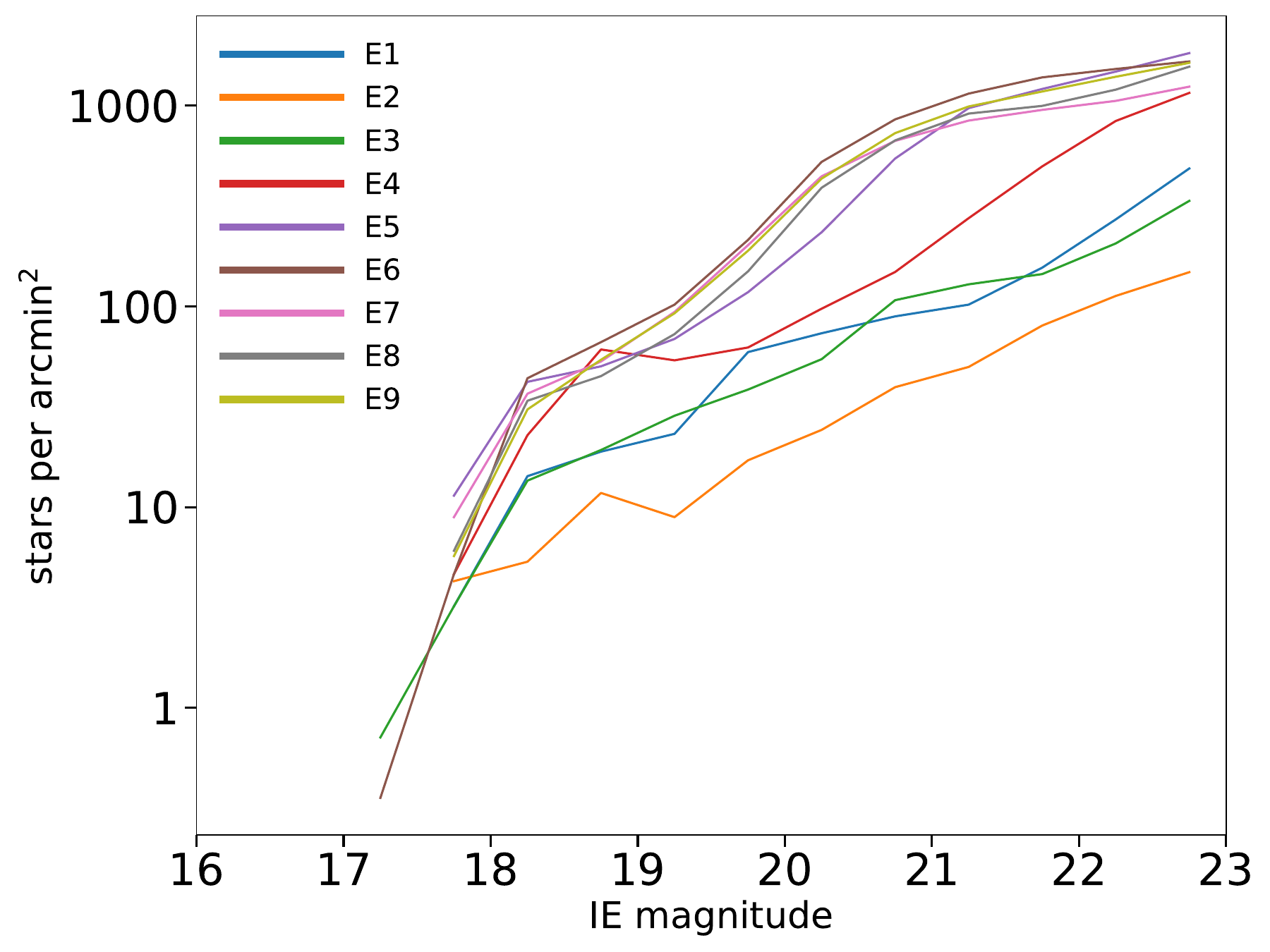}
\includegraphics[width=9cm]{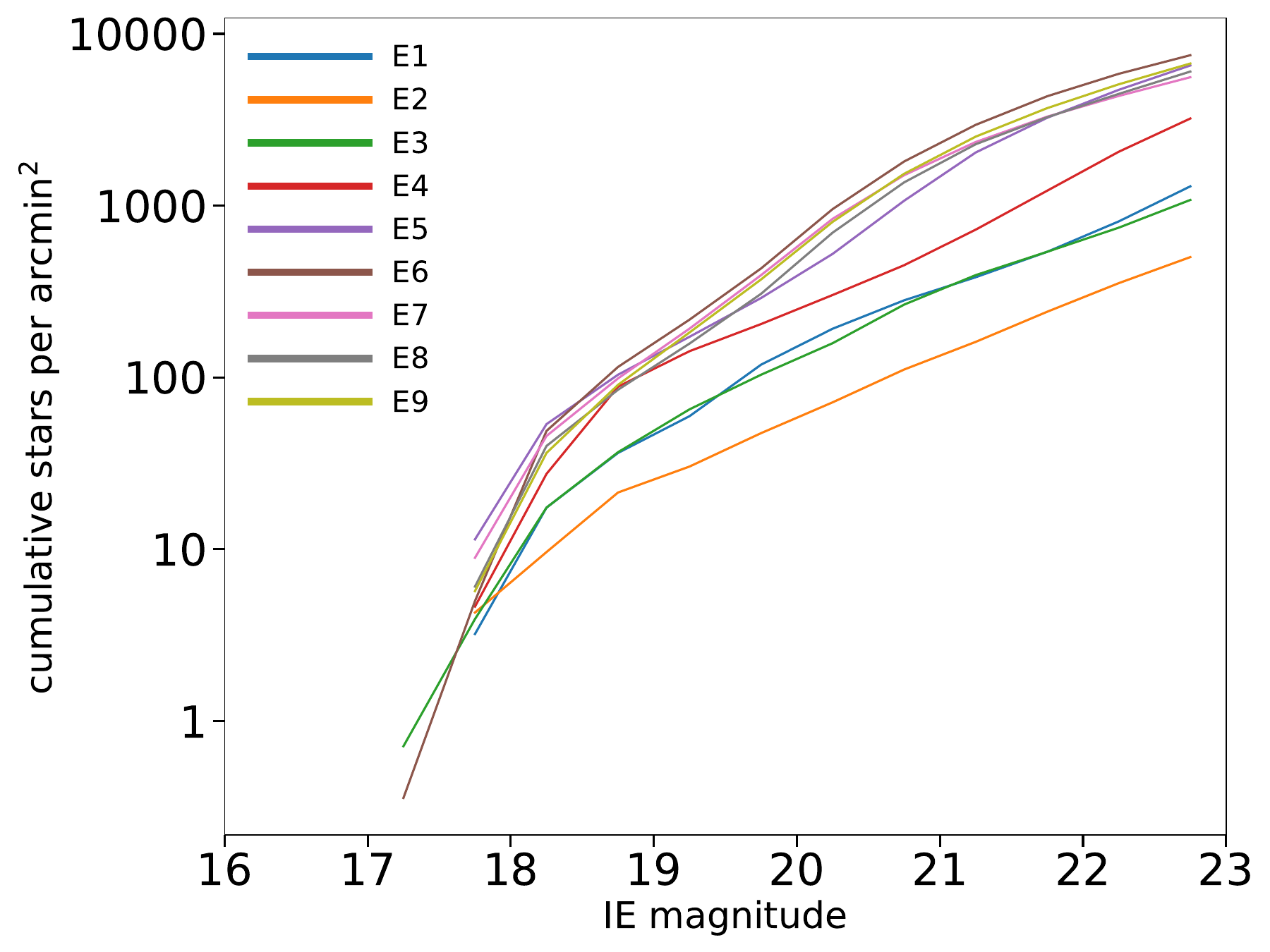}
\includegraphics[width=9cm]{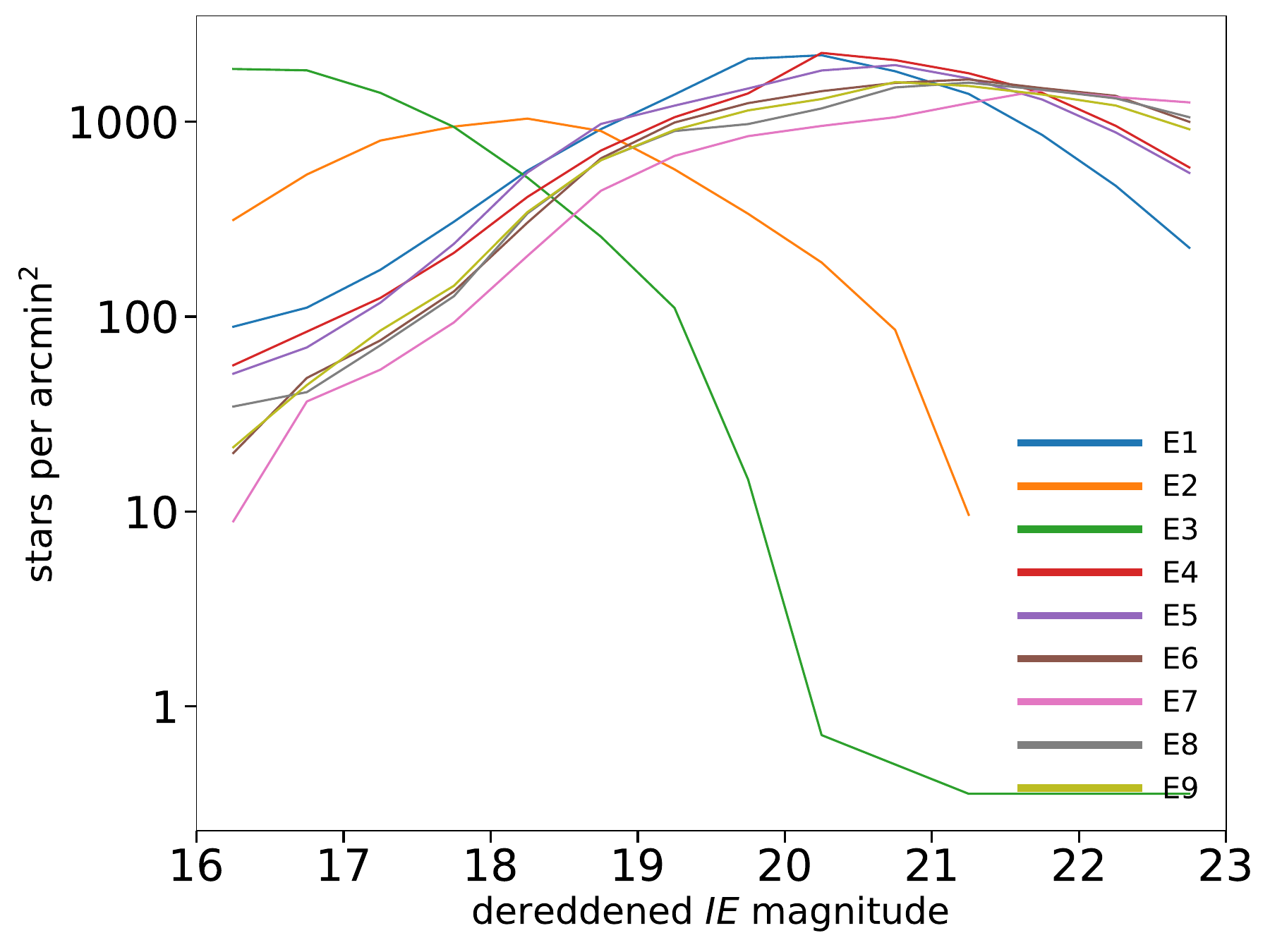}
\includegraphics[width=9cm]{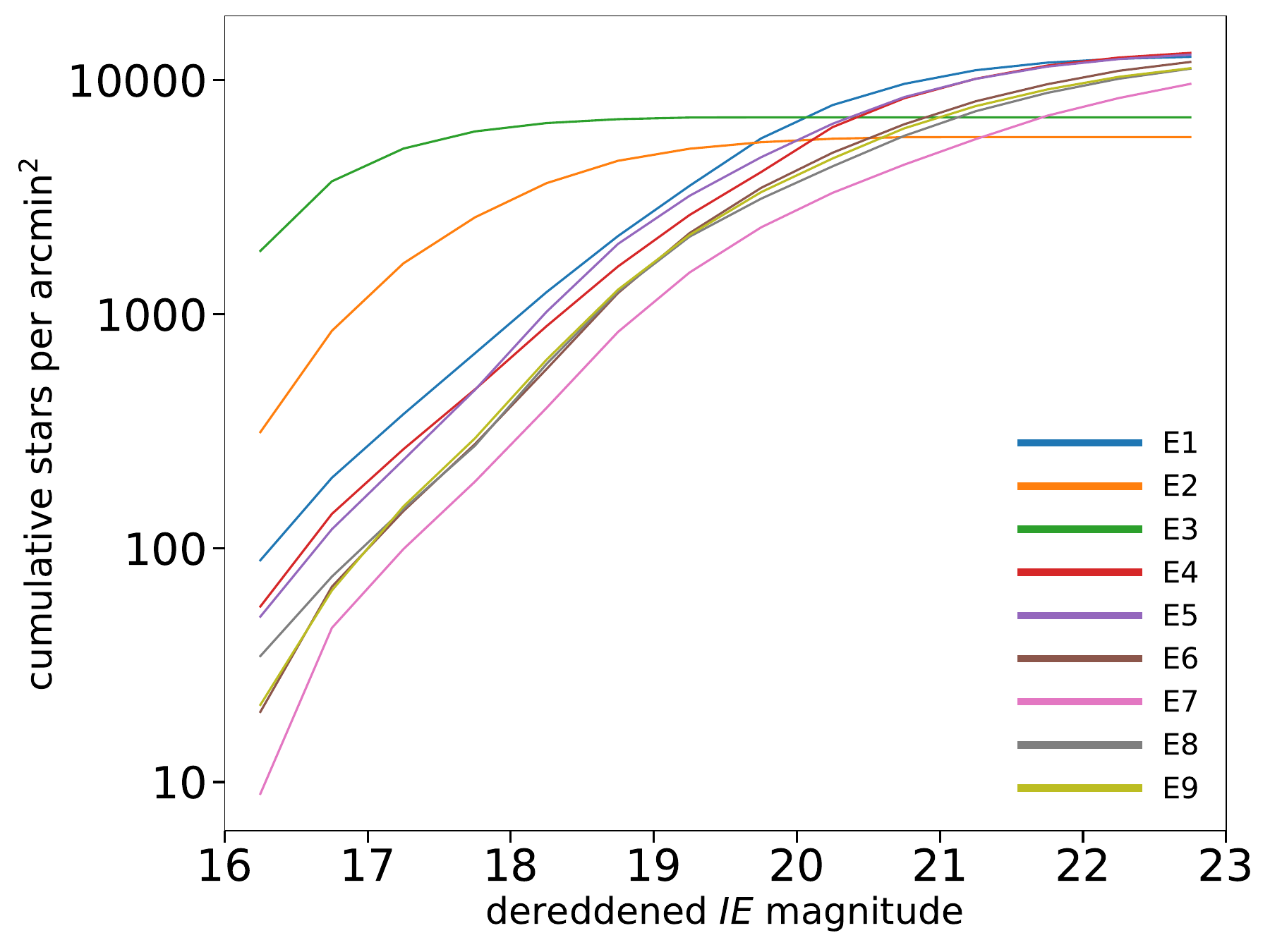}
\includegraphics[width=9cm]{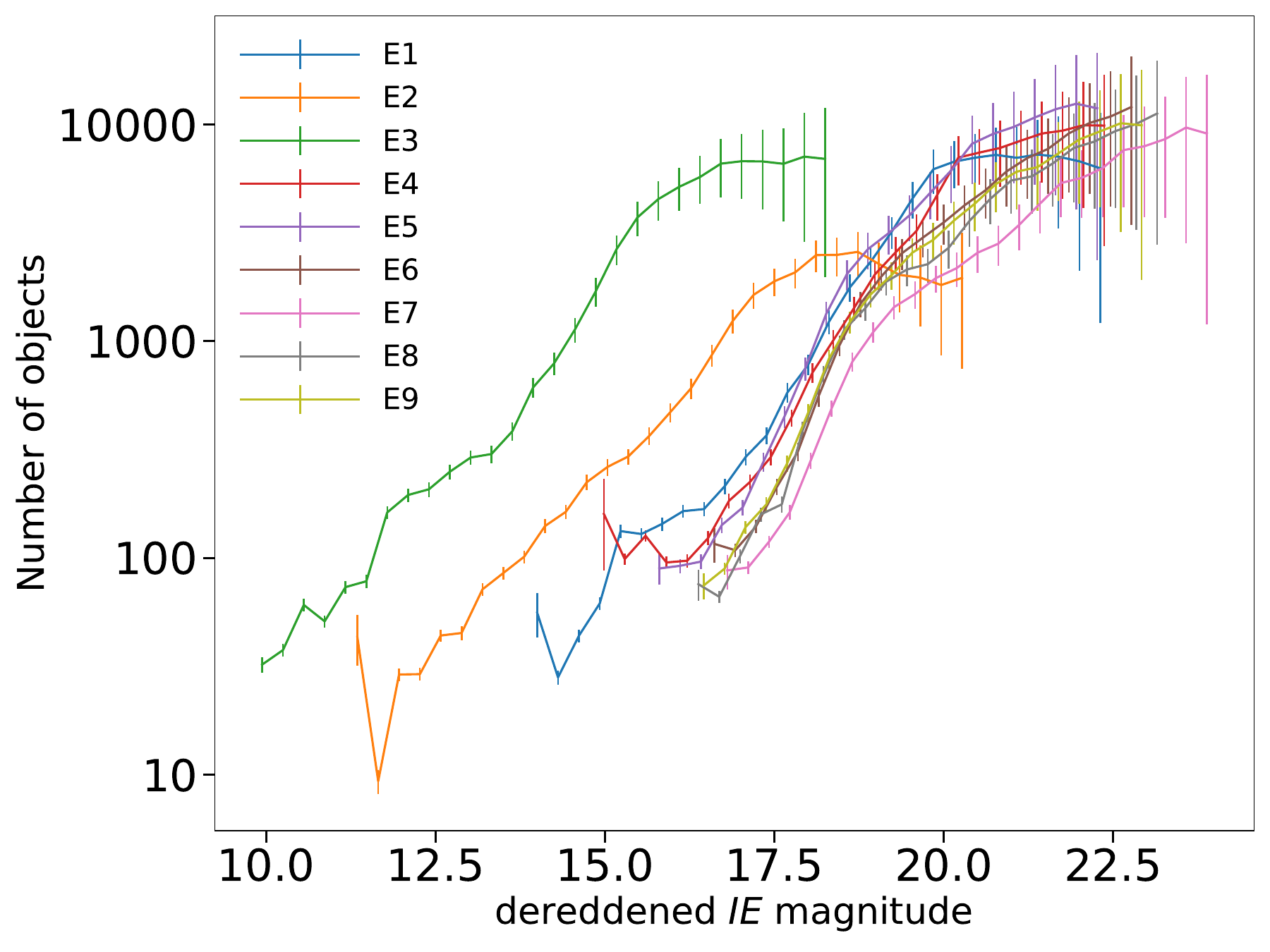}
\caption{Comparison of the \IE\ magnitude distributions in the central regions of the nine EGBS fields, derived from stacked Q2 images and calibrated using the VIS-PF catalogues. No completeness correction has been applied. The distributions illustrate the effects of extinction and crowding: the highly extincted fields E2 and E3 ($A_{\IE}=7.58$ and 8.67, respectively) contain fewer detected stars and reach fainter magnitudes, thanks to lower crowding. However, the observed populations are dominated by foreground stars, Galactic Bulge stars being subjected to large extinction. In contrast, fields E5--E9 exhibit similar magnitude distributions, consistent with comparable stellar populations and levels of crowding. \textit{Top left:} Histogram of \IE\ magnitudes. \textit{Top right:} Cumulative \IE\ magnitude distribution. \textit{Middle panels:} Same distributions after applying a mean extinction correction for each field.
\textit{Lower panel:} Distribution of star counts as a function of magnitude derived from the stacked images after correction for detection efficiency and mean extinction.}
\label{fig:comp9}
\end{figure*}
\FloatBarrier

Future analyses could address this limitation by combining \Euclid observations with external photometric catalogues, such as those obtained with CFHT or VISTA (Thygesen et al., in prep.), in order to characterise the local colour distribution of stars across the field. An alternative empirical approach is to derive colour-dependent PSF models from stars cross-matched with the HST Wide-field Survey of the Galactic Bulge \citep{Terry2026MassMeasurement} and with \Gaia\ DR3 using \texttt{PSFEx}. The HST survey covers a large fraction of the planned {\it Roman} microlensing fields, all of which lie within the EGBS footprint. Observations obtained in the $F606W$ and $F814W$ filters provide colour measurements that are largely independent of extinction corrections. Details of this analysis, the resulting PSF products, and the impact of PSF chromaticity on the photometric and astrometric precision of the calibrated source catalogue will be presented in a forthcoming paper (Rektsini et al., in prep.).

\subsection{Catalogue completeness in crowded fields}\label{completeness}

A major challenge in the analysis of the EGBS data is the extreme crowding of the Galactic bulge fields. Under these conditions, the standard VIS-PF pipeline produces source catalogues that are incomplete and spatially inhomogeneous. While the astrometric solutions and photometric zero points remain well constrained, the resulting catalogues are primarily suitable for calibration purposes rather than for statistical studies of the stellar population.

The incompleteness patterns illustrated in Figs.~\ref{fig:ob05390_completness}, \ref{fig:comp9}, and \ref{fig:E1-E9_completness} arise from a combination of effects related to background modelling, source-detection thresholds, PSF undersampling, and the extreme stellar density of the Galactic bulge. We would like to emphasise two different effects which are a direct consequence of the extreme crowding of these fields. First, Fig.~\ref{fig:comp9} shows that using our processing of the stacked images, the results are deeper, and much more uniform than the VIS-PF processing. Second, our completeness determination process is promising, before extending it to the full EGBS field. 

Users should therefore exercise caution when employing the standard Q2 catalogues for population studies, completeness analyses, or investigations requiring spatially uniform source detection.
Nevertheless, the dedicated processing described in Sect.~\ref{ob390proc} demonstrates that substantially deeper and more homogeneous catalogues can be derived from the Q2 image products, highlighting the considerable scientific potential of the EGBS data set.

\subsection{Prospects for Survey-wide Point-Source Photometry and HST Synergies}
\label{sec:future_prospects}

Building upon the analyses described in previous sections, we are exploring an additional, complementary reduction framework as a prospect for future analysis. As one of several parallel approaches currently being developed to process the full EGBS fields, this methodology will be specifically optimised for high-precision point-source astrometry and photometry in crowded, undersampled images.

The future data reduction will follow specialised methodologies designed to overcome the challenges of severe stellar crowding, as described in 
\citet{Libralato24} and Griggio et al.\ (2026). 
The core of this photometric extraction will rely on a multiple-pass software package, \texttt{KS2}, which was originally developed for HST and recently adapted to the specific instrumentation of \textit{Euclid}. This process will begin by deriving effective-PSF models from the brightest and most isolated sources to account for temporal and spatial variations. The definitive photometry will then be extracted by performing simultaneous source detection across all available images. Crucially for crowded environments like the Galactic bulge, the pipeline will utilise iterative neighbour subtraction to significantly increase the detection efficiency for faint sources that would otherwise remain buried in the noise.

As a practical application of what will be achieved on a survey-wide scale, we plan to apply this optimised pipeline to portions of the \textit{Euclid} fields that overlap with target fields from the extensive HST wide-field survey of the Galactic bulge presented in \citet{2026ApJ..1003L...1T}. This future synergy will showcase the power of combining the deep optical baselines and exquisite spatial resolution of HST with the wide-field, high-precision capabilities of \textit{Euclid}, ultimately unlocking the full potential of the survey for stellar population and kinematic studies.

\section{Opportunities from Q2 data}\label{Sec:opportunities}

\subsection{Historical microlensing events}

One of the primary science drivers of the EGBS is the analysis of individual microlensing events, both from historical ground-based surveys \citep{Bozza2025} and from the future {\it Roman} microlensing survey \citep[][Kerins et al., in prep.]{Penny2019}. However, the EGBS also provides an unprecedented combination of angular resolution, sky coverage, and depth in the inner Galactic bulge, opening a broad range of scientific applications beyond microlensing, particularly when combined with existing and future surveys.

\subsection{Stellar populations}
With more than 45 million detected sources in each dither, the EGBS enables detailed studies of the stellar populations of the Galactic bulge.
In particular, it provides a direct measurement of the stellar luminosity function across a wide range of environments and extinction levels, enabling stringent tests of Galactic population models (Verma et al., in prep.). 

\subsection{Proper motions of bulge stars}

\begin{figure}
\includegraphics[width=9 TRUECM]{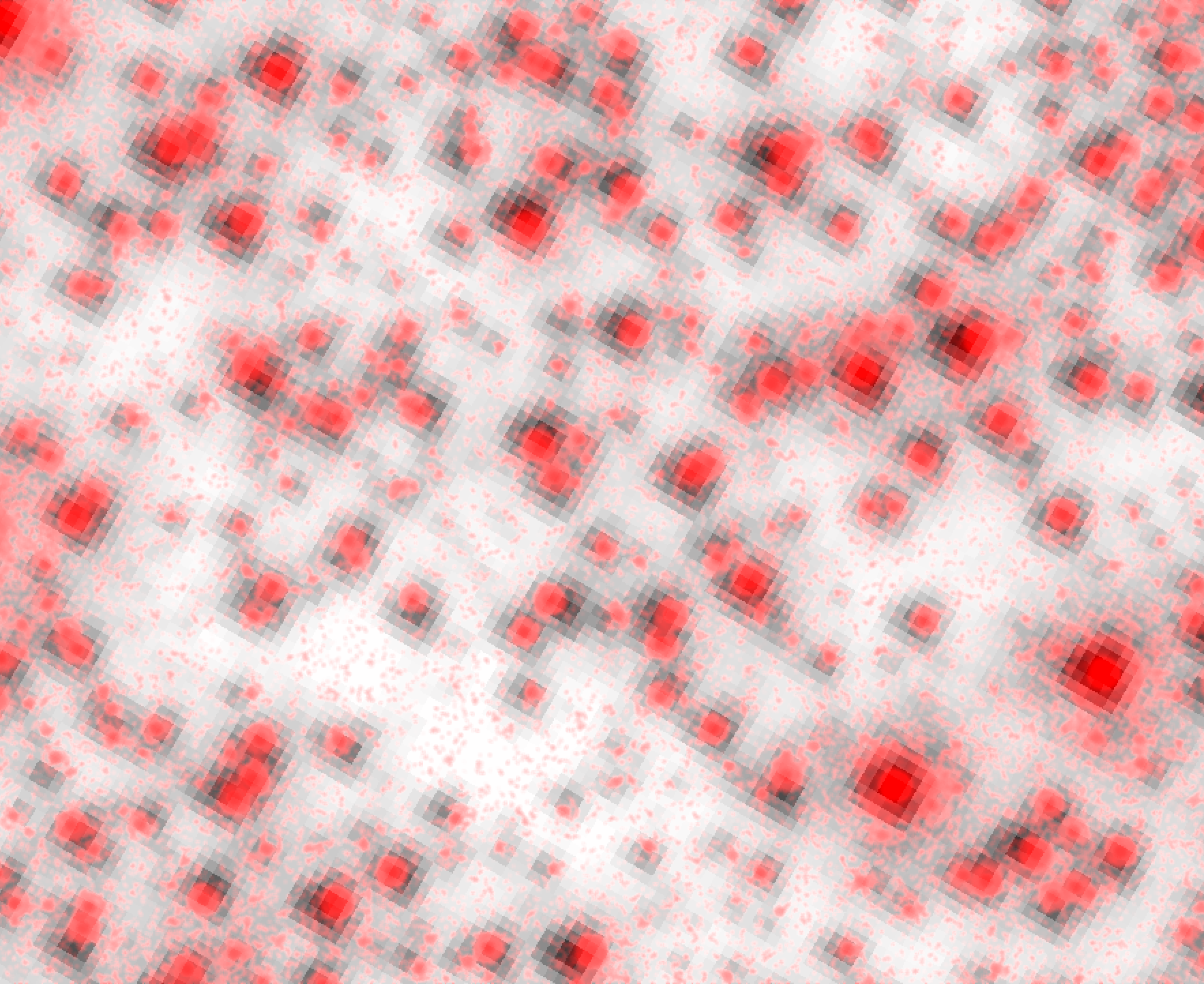}
\caption{
  \label{fig:bulgepm}
Overlay of a VIS EGBS image (greyscale) and an ACS image (red; F775W filter, programme 9805) covering an $11\arcsec\times9\arcsec$ field centred at Galactic coordinates $(l,b)=(0\degf7050,-2\degf3654)$. The two observations were obtained approximately 22 years apart. The resulting positional offsets, of order one VIS pixel, are readily apparent and reflect the proper motions of stars in the Galactic bulge and foreground disk.}
\end{figure}

Combined with future \Euclid observations or with the {\it Roman} Galactic Plane Survey \citep{GalacticPlaneSurveyDefinitionCommittee2025}, the EGBS will enable precise proper-motion measurements, that can be used for investigations of the kinematics and dynamical structure of the Galactic bulge and, their connection to stellar population properties such as age and chemical composition. They can also be used to separate foreground disk stars from bona fide bulge members (see, e.g., \citealt{sotoetal:2014,clarkeetal:2019}).

The EGBS combines a wide field of view with high angular resolution, providing the data quality required to measure proper motions for stars far fainter than those accessible to {\it Gaia}, whose performance in this region is limited by severe source confusion.

Proper-motion measurements require observations at multiple epochs. The EGBS overlaps with a significant number of HST pointings, some of which were obtained more than 22 years ago. Stellar positions can be measured in the EGBS images with a precision better than $3\,\mathrm{mas}$, enabling accurate proper-motion measurements over these long baselines. Typical bulge stars are expected to move by more than one VIS pixel (100\,mas), because a transverse velocity of $200\,\mathrm{km\,s^{-1}}$ corresponds to a proper motion of $5.2\,\mathrm{mas\,yr^{-1}}$ at a distance of 8\,kpc. This effect is illustrated in Fig.~\ref{fig:bulgepm}, which combines a portion of an ACS image obtained in 2003 with the overlapping EGBS image.

\subsection{Faint stellar populations}

Beyond stellar-population and kinematic studies, the EGBS is equally well suited to investigations of rare and intrinsically faint populations. In combination with potential future NISP observations and/or future {\it Roman} near-infrared imaging, it will enable the identification of ultracool dwarfs, brown dwarfs, and potentially free-floating planets in a region that has remained largely inaccessible owing to extreme crowding. Multi-wavelength observations will furthermore permit the construction of spectral energy distributions for large samples of sources, including dust-enshrouded objects associated with both stellar birth and the late stages of stellar evolution \citep[e.g.,][]{McDonald2017,McDonald2025b}. Such objects are often difficult to identify and characterise in crowded Galactic fields without high-angular-resolution imaging \citep{McDonald2025a}.

The EGBS includes many previously known planetary nebulae, and likely provides the highest resolution imaging of them to date. The data can act as a first epoch to measure approximately 1\,mas\,yr$^{-1}$ expansion rates~\citep{Jacob2013} and provide opportunities to test models of stellar evolution in the Galactic bulge \citep[e.g.,][]{Tan2023}.

\subsection{Time-domain astronomy}

The 16 individual dithers in each field of the EGBS were taken sequentially, so there is only a limited time baseline of about two hours for any given object in the field. The observations have a cadence of less than 10 minutes. Objects near the edges of two or more fields such that they appear in more than one may have a longer baseline, but slower, less even cadence. Many stars exhibit variability on sub-hour timescales, and with 45~million sources, it is reasonable to expect a large number of variable sources could be detected in the EGBS data. However, the short baseline will likely make identification of the type of variability challenging.

\section{Conclusions}\label{sec:conclusions}

The Euclid Q2 provides the first public release of data from the EGBS, an unprecedented deep, wide-field, and high-angular-resolution survey of the inner Galactic bulge. Using the VIS instrument, \Euclid observed nine contiguous fields covering 4.8\,deg$^2$ with a total integration time of 1.8\,h per field, delivering imaging of more than 45 million detected sources in one dither, as processed by VIS-PF, in one of the most crowded regions of the sky.

The EGBS was designed primarily to support exoplanet studies through gravitational microlensing. Its high angular resolution enables the measurement of lens fluxes for previously discovered planetary systems and provides a crucial precursor data set for the forthcoming NASA {\it Roman} microlensing survey. At the same time, the survey offers a unique resource for a broad range of Galactic science topics, including studies of stellar populations, proper motions, compact objects, planetary nebulae, and other rare populations in the inner Milky Way.

The non-standard observing strategy of the EGBS required dedicated calibration products and modifications to several elements of the VIS processing chain. The resulting astrometric and photometric calibrations achieve accuracies comparable to those obtained for standard \Euclid observations, demonstrating the robustness of the \Euclid data-processing system under observing conditions that differ significantly from those of the nominal survey. The challenges posed by these exceptionally crowded fields have already motivated improvements to the VIS-PF pipeline, several of which have been incorporated into later versions and will contribute to the processing of future \Euclid data releases.

The source catalogues released as part of Q2 remain affected by incompleteness and spatial inhomogeneities caused by the extreme crowding and variable extinction of the Galactic bulge. Nevertheless, the image products, calibration data, and accompanying catalogues provide a solid foundation for future reprocessing efforts and scientific investigations. Dedicated analyses already demonstrate that substantially deeper and more homogeneous catalogues can be extracted from the released data.

The EGBS represents one of the deepest and highest-resolution optical surveys ever obtained of the Galactic bulge over such a large area. As additional \Euclid, HST, and {\it Roman} observations become available, the scientific value of the EGBS will continue to grow, providing a unique legacy data set for Galactic and exoplanet science.

As \Euclid, HST, and future {\it Roman} observations become available -- including the expected discovery of around $10^5$ transiting planets, $1200$ bound microlensing planets, and $1000$ free-floating planets by {\it Roman}~\citep{Wilson2023,Penny2019,Johnson2020} -- the EGBS will provide a uniquely valuable high-resolution optical reference data set, enabling detailed characterisation of planetary systems and new insights into the structure, formation, and evolution of the Milky Way.

\begin{acknowledgements}

This work was supported by the SPACE-MLENS ANR grant ANR-24-CE31-3263 and the Acad\'emie Spatiale d'Ile de France. 
% APR Clément Ranc :
Authors CR, MG, NR acknowledge financial support from the Centre national d’études spatiales (CNES), France (ROR: \url{https://ror.org/04h1h0y33}) within the framework of the \Euclid and Roman missions.
This work was made possible by utilising the CANDIDE cluster at the Institut d’Astrophysique de Paris. The cluster was funded through
grants from the PNCG, CNES, DIM-ACAV, the Euclid Consortium, and the Danish
National Research Foundation Cosmic Dawn Center (DNRF140). It is maintained
by Stephane Rouberol. \AckEC \AckQtwo
\end{acknowledgements}

\bibliography{Euclid, DR1,Q2-bib,manual-refs}

@ARTICLE{EuclidSkyOverview,
author = {{Euclid Collaboration: Mellier}, Y. and {Abdurro'uf} and {Acevedo~Barroso}, J.A. and others},
	title = {Euclid - I. Overview of the Euclid mission},
	DOI= "10.1051/0004-6361/202450810",
	url= "https://doi.org/10.1051/0004-6361/202450810",
	journal = {A\&A},
	year = 2025,
	volume = 697,
	pages = "A1",
}

@ARTICLE{EuclidSkyVIS,
author = {{Euclid Collaboration: Cropper}, M. and {Al-Bahlawan}, A. and {Amiaux}, J. and others},
	title = {Euclid - II. The VIS instrument},
	DOI= "10.1051/0004-6361/202450996",
	url= "https://doi.org/10.1051/0004-6361/202450996",
	journal = {A\&A},
	year = 2025,
	volume = 697,
	pages = "A2",
}

@ARTICLE{EuclidSkyNISP,
author = {{Euclid Collaboration: Jahnke}, K. and {Gillard}, W. and {Schirmer}, M. and others},
	title = {Euclid - III. The NISP Instrument},
	DOI= "10.1051/0004-6361/202450786",
	url= "https://doi.org/10.1051/0004-6361/202450786",
	journal = {A\&A},
	year = 2025,
	volume = 697,
	pages = "A3",
}

@ARTICLE{Paykari-EP6,
       author = {{Euclid Collaboration: Paykari}, P. and {Kitching}, T. and {Hoekstra}, H. and others},
        title = "{Euclid preparation. VI. Verifying the performance of cosmic shear experiments}",
      journal = {\aap},
         year = 2020,
        month = mar,
       volume = {635},
          eid = {A139},
        pages = {A139},
          doi = {10.1051/0004-6361/201936980},
archivePrefix = {arXiv},
       eprint = {1910.10521},
 primaryClass = {astro-ph.CO},
       adsurl = {https://ui.adsabs.harvard.edu/abs/2020A&A...635A.139E}
}

@ARTICLE{Scaramella-EP1,
       author = {{Euclid Collaboration: Scaramella}, R. and {Amiaux}, J. and {Mellier}, Y. and others},
        title = "{Euclid preparation. I. The Euclid Wide Survey}",
      journal = {\aap},
         year = 2022,
        month = jun,
       volume = {662},
          eid = {A112},
        pages = {A112},
          doi = {10.1051/0004-6361/202141938},
archivePrefix = {arXiv},
       eprint = {2108.01201},
 primaryClass = {astro-ph.CO},
       adsurl = {https://ui.adsabs.harvard.edu/abs/2022A&A...662A.112E}
}

@ARTICLE{Libralato24,
       author = {{Libralato}, M. and {Bedin}, L.~R. and {Griggio}, M. and others},
        title = "{Euclid: High-precision imaging astrometry and photometry from Early Release Observations: I. Internal kinematics of NGC6397 by combining Euclid and Gaia data}",
      journal = {\aap},
         year = 2024,
        month = dec,
       volume = {692},
          eid = {A96},
        pages = {A96},
          doi = {10.1051/0004-6361/202452295},
archivePrefix = {arXiv},
       eprint = {2411.02487},
 primaryClass = {astro-ph.SR},
       adsurl = {https://ui.adsabs.harvard.edu/abs/2024A&A...692A..96L}
}

@ARTICLE{Laureijs11,
       author = {{Laureijs}, R. and {Amiaux}, J. and {Arduini}, S. and {Augu{\`e}res}, J. -L. and {Brinchmann}, J. and {Cole}, R. and {Cropper}, M. and {Dabin}, C. and {Duvet}, L. and {Ealet}, A. and {Garilli}, B. and {Gondoin}, P. and {Guzzo}, L. and {Hoar}, J. and {Hoekstra}, H. and {Holmes}, R. and {Kitching}, T. and {Maciaszek}, T. and {Mellier}, Y. and {Pasian}, F. and {Percival}, W. and {Rhodes}, J. and {Saavedra Criado}, G. and {Sauvage}, M. and {Scaramella}, R. and {Valenziano}, L. and {Warren}, S. and {Bender}, R. and {Castander}, F. and {Cimatti}, A. and {Le F{\`e}vre}, O. and {Kurki-Suonio}, H. and {Levi}, M. and {Lilje}, P. and {Meylan}, G. and {Nichol}, R. and {Pedersen}, K. and {Popa}, V. and {Rebolo Lopez}, R. and {Rix}, H. -W. and {Rottgering}, H. and {Zeilinger}, W. and {Grupp}, F. and {Hudelot}, P. and {Massey}, R. and {Meneghetti}, M. and {Miller}, L. and {Paltani}, S. and {Paulin-Henriksson}, S. and {Pires}, S. and {Saxton}, C. and {Schrabback}, T. and {Seidel}, G. and {Walsh}, J. and {Aghanim}, N. and {Amendola}, L. and {Bartlett}, J. and {Baccigalupi}, C. and {Beaulieu}, J. -P. and {Benabed}, K. and {Cuby}, J. -G. and {Elbaz}, D. and {Fosalba}, P. and {Gavazzi}, G. and {Helmi}, A. and {Hook}, I. and {Irwin}, M. and {Kneib}, J. -P. and {Kunz}, M. and {Mannucci}, F. and {Moscardini}, L. and {Tao}, C. and {Teyssier}, R. and {Weller}, J. and {Zamorani}, G. and {Zapatero Osorio}, M.~R. and {Boulade}, O. and {Foumond}, J.~J. and {Di Giorgio}, A. and {Guttridge}, P. and {James}, A. and {Kemp}, M. and {Martignac}, J. and {Spencer}, A. and {Walton}, D. and {Bl{\"u}mchen}, T. and {Bonoli}, C. and {Bortoletto}, F. and {Cerna}, C. and {Corcione}, L. and {Fabron}, C. and {Jahnke}, K. and {Ligori}, S. and {Madrid}, F. and {Martin}, L. and {Morgante}, G. and {Pamplona}, T. and {Prieto}, E. and {Riva}, M. and {Toledo}, R. and {Trifoglio}, M. and {Zerbi}, F. and {Abdalla}, F. and {Douspis}, M. and {Grenet}, C. and {Borgani}, S. and {Bouwens}, R. and {Courbin}, F. and {Delouis}, J. -M. and {Dubath}, P. and {Fontana}, A. and {Frailis}, M. and {Grazian}, A. and {Koppenh{\"o}fer}, J. and {Mansutti}, O. and {Melchior}, M. and {Mignoli}, M. and {Mohr}, J. and {Neissner}, C. and {Noddle}, K. and {Poncet}, M. and {Scodeggio}, M. and {Serrano}, S. and {Shane}, N. and {Starck}, J. -L. and {Surace}, C. and {Taylor}, A. and {Verdoes-Kleijn}, G. and {Vuerli}, C. and {Williams}, O.~R. and {Zacchei}, A. and {Altieri}, B. and {Escudero Sanz}, I. and {Kohley}, R. and {Oosterbroek}, T. and {Astier}, P. and {Bacon}, D. and {Bardelli}, S. and {Baugh}, C. and {Bellagamba}, F. and {Benoist}, C. and {Bianchi}, D. and {Biviano}, A. and {Branchini}, E. and {Carbone}, C. and {Cardone}, V. and {Clements}, D. and {Colombi}, S. and {Conselice}, C. and {Cresci}, G. and {Deacon}, N. and {Dunlop}, J. and {Fedeli}, C. and {Fontanot}, F. and {Franzetti}, P. and {Giocoli}, C. and {Garcia-Bellido}, J. and {Gow}, J. and {Heavens}, A. and {Hewett}, P. and {Heymans}, C. and {Holland}, A. and {Huang}, Z. and {Ilbert}, O. and {Joachimi}, B. and {Jennins}, E. and {Kerins}, E. and {Kiessling}, A. and {Kirk}, D. and {Kotak}, R. and {Krause}, O. and {Lahav}, O. and {van Leeuwen}, F. and {Lesgourgues}, J. and {Lombardi}, M. and {Magliocchetti}, M. and {Maguire}, K. and {Majerotto}, E. and {Maoli}, R. and {Marulli}, F. and {Maurogordato}, S. and {McCracken}, H. and {McLure}, R. and {Melchiorri}, A. and {Merson}, A. and {Moresco}, M. and {Nonino}, M. and {Norberg}, P. and {Peacock}, J. and {Pello}, R. and {Penny}, M. and {Pettorino}, V. and {Di Porto}, C. and {Pozzetti}, L. and {Quercellini}, C. and {Radovich}, M. and {Rassat}, A. and {Roche}, N. and {Ronayette}, S. and {Rossetti}, E. and {Sartoris}, B. and {Schneider}, P. and {Semboloni}, E. and {Serjeant}, S. and {Simpson}, F. and {Skordis}, C. and {Smadja}, G. and {Smartt}, S. and {Spano}, P. and {Spiro}, S. and {Sullivan}, M. and {Tilquin}, A. and {Trotta}, R. and {Verde}, L. and {Wang}, Y. and {Williger}, G. and {Zhao}, G. and {Zoubian}, J. and {Zucca}, E.},
        title = "{Euclid Definition Study Report}",
      journal = {ESA/SRE(2011)12},
         year = 2011,
        month = oct,
          eid = {arXiv:1110.3193},
        pages = {arXiv:1110.3193},
          doi = {10.48550/arXiv.1110.3193},
archivePrefix = {arXiv},
       eprint = {1110.3193},
 primaryClass = {astro-ph.CO},
       adsurl = {https://ui.adsabs.harvard.edu/abs/2011arXiv1110.3193L}
}

@ARTICLE{Anderson2000,
       author = {{Anderson}, Jay and {King}, Ivan R.},
        title = "{Toward High-Precision Astrometry with WFPC2. I. Deriving an Accurate Point-Spread Function}",
      journal = {\pasp},
         year = 2000,
        month = oct,
       volume = {112},
       number = {776},
        pages = {1360-1382},
          doi = {10.1086/316632},
archivePrefix = {arXiv},
       eprint = {astro-ph/0006325},
 primaryClass = {astro-ph},
       adsurl = {https://ui.adsabs.harvard.edu/abs/2000PASP..112.1360A}
}

@ARTICLE{Udalski2015,
       author = {{Udalski}, A. and {Yee}, J.~C. and {Gould}, A. and {Carey}, S. and {Zhu}, W. and {Skowron}, J. and {Koz{\l}owski}, S. and {Poleski}, R. and {Pietrukowicz}, P. and {Pietrzy{\'n}ski}, G. and {Szyma{\'n}ski}, M.~K. and {Mr{\'o}z}, P. and {Soszy{\'n}ski}, I. and {Ulaczyk}, K. and {Wyrzykowski}, {\L}. and {Han}, C. and {Calchi Novati}, S. and {Pogge}, R.~W.},
        title = "{Spitzer as a Microlens Parallax Satellite: Mass Measurement for the OGLE-2014-BLG-0124L Planet and its Host Star}",
      journal = {\apj},
         year = 2015,
        month = feb,
       volume = {799},
       number = {2},
          eid = {237},
        pages = {237},
          doi = {10.1088/0004-637X/799/2/237},
archivePrefix = {arXiv},
       eprint = {1410.4219},
 primaryClass = {astro-ph.EP},
       adsurl = {https://ui.adsabs.harvard.edu/abs/2015ApJ...799..237U}
}

@ARTICLE{Nemiroff1994,
       author = {{Nemiroff}, Robert J. and {Wickramasinghe}, W.~A.~D.~T.},
        title = "{Finite Source Sizes and the Information Content of Macho-Type Lens Search Light Curves}",
      journal = {\apjl},
         year = 1994,
        month = mar,
       volume = {424},
        pages = {L21},
          doi = {10.1086/187265},
archivePrefix = {arXiv},
       eprint = {astro-ph/9401005},
 primaryClass = {astro-ph},
       adsurl = {https://ui.adsabs.harvard.edu/abs/1994ApJ...424L..21N}
}

@ARTICLE{An2002,
       author = {{An}, Jin H. and {Albrow}, M.~D. and {Beaulieu}, J.-P. and {Caldwell}, J.~A.~R. and {DePoy}, D.~L. and {Dominik}, M. and {Gaudi}, B.~S. and {Gould}, A. and {Greenhill}, J. and {Hill}, K. and {Kane}, S. and {Martin}, R. and {Menzies}, J. and {Pogge}, R.~W. and {Pollard}, K.~R. and {Sackett}, P.~D. and {Sahu}, K.~C. and {Vermaak}, P. and {Watson}, R. and {Williams}, A.},
        title = "{First Microlens Mass Measurement: PLANET Photometry of EROS BLG-2000-5}",
      journal = {\apj},
         year = 2002,
        month = jun,
       volume = {572},
       number = {1},
        pages = {521-539},
          doi = {10.1086/340191},
archivePrefix = {arXiv},
       eprint = {astro-ph/0110095},
 primaryClass = {astro-ph},
       adsurl = {https://ui.adsabs.harvard.edu/abs/2002ApJ...572..521A}
}

@ARTICLE{Yoo2004,
       author = {{Yoo}, Jaiyul and {DePoy}, D.~L. and {Gal-Yam}, A. and {Gaudi}, B.~S. and {Gould}, A. and {Han}, C. and {Lipkin}, Y. and {Maoz}, D. and {Ofek}, E.~O. and {Park}, B.-G. and {Pogge}, R.~W. and {Mu-Fun Collaboration} and {Udalski}, A. and {Soszy{\'n}ski}, I. and {Wyrzykowski}, {\L}. and {Kubiak}, M. and {Szyma{\'n}ski}, M. and {Pietrzy{\'n}ski}, G. and {Szewczyk}, O. and {{\.Z}ebru{\'n}}, K. and {OGLE Collaboration}},
        title = "{OGLE-2003-BLG-262: Finite-Source Effects from a Point-Mass Lens}",
      journal = {\apj},
         year = 2004,
        month = mar,
       volume = {603},
       number = {1},
        pages = {139-151},
          doi = {10.1086/381241},
archivePrefix = {arXiv},
       eprint = {astro-ph/0309302},
 primaryClass = {astro-ph},
       adsurl = {https://ui.adsabs.harvard.edu/abs/2004ApJ...603..139Y}
}

@ARTICLE{Kim2018,
       author = {{Kim}, D.-J. and {Kim}, H.-W. and {Hwang}, K.-H. and {Albrow}, M.~D. and {Chung}, S.-J. and {Gould}, A. and {Han}, C. and {Jung}, Y.~K. and {Ryu}, Y.-H. and {Shin}, I.-G. and {Yee}, J.~C. and {Zhu}, W. and {Cha}, S.-M. and {Kim}, S.-L. and {Lee}, C.-U. and {Lee}, D.-J. and {Lee}, Y. and {Park}, B.-G. and {Pogge}, R.~W. and {KMTNet Collaboration}},
        title = "{Korea Microlensing Telescope Network Microlensing Events from 2015: Event-finding Algorithm, Vetting, and Photometry}",
      journal = {\aj},
         year = 2018,
        month = feb,
       volume = {155},
       number = {2},
          eid = {76},
        pages = {76},
          doi = {10.3847/1538-3881/aaa47b},
archivePrefix = {arXiv},
       eprint = {1703.06883},
 primaryClass = {astro-ph.EP},
       adsurl = {https://ui.adsabs.harvard.edu/abs/2018AJ....155...76K}
}

@ARTICLE{GalacticPlaneSurveyDefinitionCommittee2025,
       author = {{Roman Galactic Plane Survey Definition Committee}},
        title = "{Roman Galactic Plane Survey Definition Committee Report}",
      journal = {Final version of report submitted to the Roman Observations Time Allocation Committee on Oct 1, 2025},
         year = 2025,
        month = nov,
          eid = {arXiv:2511.07494},
        pages = {arXiv:2511.07494},
          doi = {10.48550/arXiv.2511.07494},
archivePrefix = {arXiv},
       eprint = {2511.07494},
 primaryClass = {astro-ph.GA},
       adsurl = {https://ui.adsabs.harvard.edu/abs/2025arXiv251107494G}
}

@article{ROTAC2025,
       author = {{Roman Observations Time Allocation Committee} and {Core Community Survey Definition Committees}},
        title = "{Roman Observations Time Allocation Committee: Final Report and Recommendations}",
      journal = {Report released in late April},
         year = 2025,
        month = may,
          eid = {arXiv:2505.10574},
        pages = {arXiv:2505.10574},
          doi = {10.48550/arXiv.2505.10574},
archivePrefix = {arXiv},
       eprint = {2505.10574},
 primaryClass = {astro-ph.IM},
       adsurl = {https://ui.adsabs.harvard.edu/abs/2025arXiv250510574O}
}

@article{Surot2020ReddeningMap,
       author = {{Surot}, F. and {Valenti}, E. and {Gonzalez}, O.~A. and {Zoccali}, M. and {S{\"o}kmen}, E. and {Hidalgo}, S.~L. and {Minniti}, D.},
        title = "{Mapping the stellar age of the Milky Way bulge with the VVV. III. High-resolution reddening map}",
      journal = {\aap},
         year = 2020,
        month = dec,
       volume = {644},
          eid = {A140},
        pages = {A140},
          doi = {10.1051/0004-6361/202038346},
archivePrefix = {arXiv},
       eprint = {2010.02723},
 primaryClass = {astro-ph.GA},
       adsurl = {https://ui.adsabs.harvard.edu/abs/2020A&A...644A.140S}
}

@ARTICLE{Park2004,
       author = {{Park}, B.-G. and {DePoy}, D.~L. and {Gaudi}, B.~S. and {Gould}, A. and {Han}, C. and {Pogge}, R.~W. and {muFun Collaboration} and {Abe}, F. and {Bennett}, D.~P. and {Bond}, I.~A. and {Eguchi}, S. and {Furuta}, Y. and {Hearnshaw}, J.~B. and {Kamiya}, K. and {Kilmartin}, P.~M. and {Kurata}, Y. and {Masuda}, K. and {Matsubara}, Y. and {Muraki}, Y. and {Noda}, S. and {Okajima}, K. and {Rattenbury}, N.~J. and {Sako}, T. and {Sekiguchi}, T. and {Sullivan}, D.~J. and {Sumi}, T. and {Tristram}, P.~J. and {Yanagisawa}, T. and {Yock}, P.~C.~M. and {MOA Collaboration}},
        title = "{MOA 2003-BLG-37: A Bulge Jerk-Parallax Microlens Degeneracy}",
      journal = {\apj},
         year = 2004,
        month = jul,
       volume = {609},
       number = {1},
        pages = {166-172},
          doi = {10.1086/420926},
archivePrefix = {arXiv},
       eprint = {astro-ph/0401250},
 primaryClass = {astro-ph},
       adsurl = {https://ui.adsabs.harvard.edu/abs/2004ApJ...609..166P}
}

@ARTICLE{GaiaDR3Summary,
       author = {{Gaia Collaboration: Vallenari}, A. and {Brown}, A.~G.~A. and {Prusti}, T. and {de Bruijne}, J.~H.~J. and {Arenou}, F. and {Babusiaux}, C. and {Biermann}, M. and {Creevey}, O.~L. and {Ducourant}, C. and {Evans}, D.~W. and {Eyer}, L. and {Guerra}, R. and {Hutton}, A. and {Jordi}, C. and {Klioner}, S.~A. and {Lammers}, U.~L. and {Lindegren}, L. and {Luri}, X. and {Mignard}, F. and {Panem}, C. and {Pourbaix}, D. and {Randich}, S. and {Sartoretti}, P. and {Soubiran}, C. and {Tanga}, P. and {Walton}, N.~A. and {Bailer-Jones}, C.~A.~L. and {Bastian}, U. and {Drimmel}, R. and {Jansen}, F. and {Katz}, D. and {Lattanzi}, M.~G. and {van Leeuwen}, F. and {Bakker}, J. and {Cacciari}, C. and {Casta{\~n}eda}, J. and {De Angeli}, F. and {Fabricius}, C. and {Fouesneau}, M. and {Fr{\'e}mat}, Y. and {Galluccio}, L. and {Guerrier}, A. and {Heiter}, U. and {Masana}, E. and {Messineo}, R. and {Mowlavi}, N. and {Nicolas}, C. and {Nienartowicz}, K. and {Pailler}, F. and {Panuzzo}, P. and {Riclet}, F. and {Roux}, W. and {Seabroke}, G.~M. and {Sordo}, R. and {Th{\'e}venin}, F. and {Gracia-Abril}, G. and {Portell}, J. and {Teyssier}, D. and {Altmann}, M. and {Andrae}, R. and {Audard}, M. and {Bellas-Velidis}, I. and {Benson}, K. and {Berthier}, J. and {Blomme}, R. and {Burgess}, P.~W. and {Busonero}, D. and {Busso}, G. and {C{\'a}novas}, H. and {Carry}, B. and {Cellino}, A. and {Cheek}, N. and {Clementini}, G. and {Damerdji}, Y. and {Davidson}, M. and {de Teodoro}, P. and {Nu{\~n}ez Campos}, M. and {Delchambre}, L. and {Dell'Oro}, A. and {Esquej}, P. and {Fern{\'a}ndez-Hern{\'a}ndez}, J. and {Fraile}, E. and {Garabato}, D. and {Garc{\'\i}a-Lario}, P. and {Gosset}, E. and {Haigron}, R. and {Halbwachs}, J.-L. and {Hambly}, N.~C. and {Harrison}, D.~L. and {Hern{\'a}ndez}, J. and {Hestroffer}, D. and {Hodgkin}, S.~T. and {Holl}, B. and {Jan{\ss}en}, K. and {Jevardat de Fombelle}, G. and {Jordan}, S. and {Krone-Martins}, A. and {Lanzafame}, A.~C. and {L{\"o}ffler}, W. and {Marchal}, O. and {Marrese}, P.~M. and {Moitinho}, A. and {Muinonen}, K. and {Osborne}, P. and {Pancino}, E. and {Pauwels}, T. and {Recio-Blanco}, A. and {Reyl{\'e}}, C. and {Riello}, M. and {Rimoldini}, L. and {Roegiers}, T. and {Rybizki}, J. and {Sarro}, L.~M. and {Siopis}, C. and {Smith}, M. and {Sozzetti}, A. and {Utrilla}, E. and {van Leeuwen}, M. and {Abbas}, U. and {{\'A}brah{\'a}m}, P. and {Abreu Aramburu}, A. and {Aerts}, C. and {Aguado}, J.~J. and {Ajaj}, M. and {Aldea-Montero}, F. and {Altavilla}, G. and {{\'A}lvarez}, M.~A. and {Alves}, J. and {Anders}, F. and {Anderson}, R.~I. and {Anglada Varela}, E. and {Antoja}, T. and {Baines}, D. and {Baker}, S.~G. and {Balaguer-N{\'u}{\~n}ez}, L. and {Balbinot}, E. and {Balog}, Z. and {Barache}, C. and {Barbato}, D. and {Barros}, M. and {Barstow}, M.~A. and {Bartolom{\'e}}, S. and {Bassilana}, J.-L. and {Bauchet}, N. and {Becciani}, U. and {Bellazzini}, M. and {Berihuete}, A. and {Bernet}, M. and {Bertone}, S. and {Bianchi}, L. and {Binnenfeld}, A. and {Blanco-Cuaresma}, S. and {Blazere}, A. and {Boch}, T. and {Bombrun}, A. and {Bossini}, D. and {Bouquillon}, S. and {Bragaglia}, A. and {Bramante}, L. and {Breedt}, E. and {Bressan}, A. and {Brouillet}, N. and {Brugaletta}, E. and {Bucciarelli}, B. and {Burlacu}, A. and {Butkevich}, A.~G. and {Buzzi}, R. and {Caffau}, E. and {Cancelliere}, R. and {Cantat-Gaudin}, T. and {Carballo}, R. and {Carlucci}, T. and {Carnerero}, M.~I. and {Carrasco}, J.~M. and {Casamiquela}, L. and {Castellani}, M. and {Castro-Ginard}, A. and {Chaoul}, L. and {Charlot}, P. and {Chemin}, L. and {Chiaramida}, V. and {Chiavassa}, A. and {Chornay}, N. and {Comoretto}, G. and {Contursi}, G. and {Cooper}, W.~J. and {Cornez}, T. and {Cowell}, S. and {Crifo}, F. and {Cropper}, M. and {Crosta}, M. and {Crowley}, C. and {Dafonte}, C. and {Dapergolas}, A. and {David}, M. and {David}, P. and {de Laverny}, P. and {De Luise}, F. and {De March}, R.},
        title = "{Gaia Data Release 3. Summary of the content and survey properties}",
      journal = {\aap},
         year = 2023,
        month = jun,
       volume = {674},
          eid = {A1},
        pages = {A1},
          doi = {10.1051/0004-6361/202243940},
archivePrefix = {arXiv},
       eprint = {2208.00211},
 primaryClass = {astro-ph.GA},
       adsurl = {https://ui.adsabs.harvard.edu/abs/2023A&A...674A...1G}
}

@ARTICLE{Terry2026MassMeasurement,
       author = {{Terry}, Sean K. and {Bachelet}, Etienne and {Zohrabi}, Farzaneh and {Verma}, Himanshu and {Crisp}, Alison and {Huston}, Macy J. and {McGee}, Carissma and {Penny}, Matthew and {Abrams}, Natasha S. and {Albrow}, Michael D. and {Anderson}, Jay and {Bagheri}, Fatemeh and {Beaulieu}, Jean-Philippe and {Bellini}, Andrea and {Bennett}, David P. and {Bergsten}, Galen and {Bhadra}, T. Dex and {Bhattacharya}, Aparna and {Bond}, Ian A. and {Bozza}, Valerio and {Brandon}, Christopher and {Calchi Novati}, Sebastiano and {Carey}, Sean and {Christiansen}, Jessie and {DeRocco}, William and {Gaudi}, B. Scott and {Hulberg}, Jon and {Ishitani Silva}, Stela and {Jones}, Sinclaire E. and {Kerins}, Eamonn and {Khakpash}, Somayeh and {Kruszy{\'n}ska}, Katarzyna and {Lam}, Casey and {Lu}, Jessica R. and {Malpas}, Amber and {Miyazaki}, Shota and {Mr{\'o}z}, Przemek and {Murlidhar}, Arjun and {Nataf}, David and {Newman}, Marz and {Olmschenk}, Greg and {Poleski}, Radek and {Ranc}, Cl{\'e}ment and {Rattenbury}, Nicholas J. and {Rybicki}, Krzysztof and {Saggese}, Vito and {Sobeck}, Jennifer and {Stassun}, Keivan G. and {Stephan}, Alexander P. and {Street}, Rachel A. and {Sumi}, Takahiro and {Suzuki}, Daisuke and {Vandorou}, Aikaterini and {Vyas}, Meet and {Yee}, Jennifer C. and {Zang}, Weicheng and {Zhang}, Keming and {Roman Galactic Exoplanet Survey Project Infrastructure Team}},
        title = "{Predictions of the Nancy Grace Roman Space Telescope Galactic Exoplanet Survey. IV. Lens Mass and Distance Measurements}",
      journal = {\aj},
         year = 2026,
        month = apr,
       volume = {171},
       number = {4},
          eid = {212},
        pages = {212},
          doi = {10.3847/1538-3881/ae43e6},
archivePrefix = {arXiv},
       eprint = {2510.13974},
 primaryClass = {astro-ph.EP},
       adsurl = {https://ui.adsabs.harvard.edu/abs/2026AJ....171..212T}
}

@ARTICLE{Bachelet2024,
       author = {{Bachelet}, E. and {Hundertmark}, M. and {Calchi Novati}, S.},
        title = "{Estimating Microlensing Parameters from Observables and Stellar Isochrones with pyLIMASS}",
      journal = {\aj},
         year = 2024,
        month = jul,
       volume = {168},
       number = {1},
          eid = {24},
        pages = {24},
          doi = {10.3847/1538-3881/ad4862},
archivePrefix = {arXiv},
       eprint = {2405.02230},
 primaryClass = {astro-ph.IM},
       adsurl = {https://ui.adsabs.harvard.edu/abs/2024AJ....168...24B}
}

@ARTICLE{Penny2016,
       author = {{Penny}, Matthew T. and {Henderson}, Calen B. and {Clanton}, Christian},
        title = "{Is the Galactic Bulge Devoid of Planets?}",
      journal = {\apj},
         year = 2016,
        month = oct,
       volume = {830},
       number = {2},
          eid = {150},
        pages = {150},
          doi = {10.3847/0004-637X/830/2/150},
archivePrefix = {arXiv},
       eprint = {1601.02807},
 primaryClass = {astro-ph.EP},
       adsurl = {https://ui.adsabs.harvard.edu/abs/2016ApJ...830..150P}
}

@ARTICLE{Wilson2023,
       author = {{Wilson}, Robert F. and {Barclay}, Thomas and {Powell}, Brian P. and {Schlieder}, Joshua and {Hedges}, Christina and {Montet}, Benjamin T. and {Quintana}, Elisa and {Mcdonald}, Iain and {Penny}, Matthew T. and {Espinoza}, N{\'e}stor and {Kerins}, Eamonn},
        title = "{Transiting Exoplanet Yields for the Roman Galactic Bulge Time Domain Survey Predicted from Pixel-level Simulations}",
      journal = {\apjs},
         year = 2023,
        month = nov,
       volume = {269},
       number = {1},
          eid = {5},
        pages = {5},
          doi = {10.3847/1538-4365/acf3df},
archivePrefix = {arXiv},
       eprint = {2305.16204},
 primaryClass = {astro-ph.EP},
       adsurl = {https://ui.adsabs.harvard.edu/abs/2023ApJS..269....5W}
}

@ARTICLE{Kluter2025,
       author = {{Kl{\"u}ter}, Jonas and {Huston}, Macy J. and {Aronica}, Abigail and {Johnson}, Samson A. and {Penny}, Matthew T. and {Newman}, Marz and {Zohrabi}, Farzaneh and {Crisp}, Alison L. and {Chevis}, Allison},
        title = "{SYNTHPOP: A New Framework for Synthetic Milky Way Population Generation}",
      journal = {\aj},
         year = 2025,
        month = jun,
       volume = {169},
       number = {6},
          eid = {317},
        pages = {317},
          doi = {10.3847/1538-3881/adcd7a},
archivePrefix = {arXiv},
       eprint = {2411.18821},
 primaryClass = {astro-ph.IM},
       adsurl = {https://ui.adsabs.harvard.edu/abs/2025AJ....169..317K}
}

@ARTICLE{Beaulieu2008,
       author = {{Beaulieu}, J.~P. and {Kerins}, E. and {Mao}, S. and {Bennett}, D. and {Cassan}, A. and {Dieters}, S. and {Gaudi}, B.~S. and {Gould}, A. and {Batista}, V. and {Bender}, R. and {Brillant}, S. and {Cook}, K. and {Coutures}, C. and {Dominis-Prester}, D. and {Donatowicz}, J. and {Fouqu{\'e}}, P. and {Grebel}, E. and {Greenhill}, J. and {Heyrovsky}, D. and {Horne}, K. and {Kubas}, D. and {Marquette}, J.~B. and {Menzies}, J. and {Rattenbury}, N.~J. and {Ribas}, I. and {Sahu}, K. and {Tsapras}, Y. and {Udalski}, A. and {Vinter}, C.},
        title = "{Towards A Census of Earth-mass Exo-planets with Gravitational Microlensing}",
      journal = {White Paper submission to the ESA Exo-Planet Roadmap Advisory Team.},
         year = 2008,
        month = jul,
          eid = {arXiv:0808.0005},
        pages = {arXiv:0808.0005},
          doi = {10.48550/arXiv.0808.0005},
archivePrefix = {arXiv},
       eprint = {0808.0005},
 primaryClass = {astro-ph},
       adsurl = {https://ui.adsabs.harvard.edu/abs/2008arXiv0808.0005B}
}

@ARTICLE{Koshimoto2023,
       author = {{Koshimoto}, Naoki and {Sumi}, Takahiro and {Bennett}, David P. and {Bozza}, Valerio and {Mr{\'o}z}, Przemek and {Udalski}, Andrzej and {Rattenbury}, Nicholas J. and {Abe}, Fumio and {Barry}, Richard and {Bhattacharya}, Aparna and {Bond}, Ian A. and {Fujii}, Hirosane and {Fukui}, Akihiko and {Hamada}, Ryusei and {Hirao}, Yuki and {Silva}, Stela Ishitani and {Itow}, Yoshitaka and {Kirikawa}, Rintaro and {Kondo}, Iona and {Matsubara}, Yutaka and {Miyazaki}, Shota and {Muraki}, Yasushi and {Olmschenk}, Greg and {Ranc}, Cl{\'e}ment and {Satoh}, Yuki and {Suzuki}, Daisuke and {Tomoyoshi}, Mio and {Tristram}, Paul J. and {Vandorou}, Aikaterini and {Yama}, Hibiki and {Yamashita}, Kansuke},
        title = "{Terrestrial- and Neptune-mass Free-Floating Planet Candidates from the MOA-II 9 yr Galactic Bulge Survey}",
      journal = {\aj},
         year = 2023,
        month = sep,
       volume = {166},
       number = {3},
          eid = {107},
        pages = {107},
          doi = {10.3847/1538-3881/ace689},
archivePrefix = {arXiv},
       eprint = {2303.08279},
 primaryClass = {astro-ph.EP},
       adsurl = {https://ui.adsabs.harvard.edu/abs/2023AJ....166..107K}
}

@ARTICLE{Jacob2013,
       author = {{Jacob}, R. and {Sch{\"o}nberner}, D. and {Steffen}, M.},
        title = "{The evolution of planetary nebulae. VIII. True expansion rates and visibility times}",
      journal = {\aap},
         year = 2013,
        month = oct,
       volume = {558},
          eid = {A78},
        pages = {A78},
          doi = {10.1051/0004-6361/201321532},
archivePrefix = {arXiv},
       eprint = {1307.6189},
 primaryClass = {astro-ph.SR},
       adsurl = {https://ui.adsabs.harvard.edu/abs/2013A&A...558A..78J}
}

@ARTICLE{Mroz2019,
       author = {{Mr{\'o}z}, Przemek and {Udalski}, Andrzej and {Skowron}, Jan and {Szyma{\'n}ski}, Micha{\l} K. and {Soszy{\'n}ski}, Igor and {Wyrzykowski}, {\L}ukasz and {Pietrukowicz}, Pawe{\l} and {Koz{\l}owski}, Szymon and {Poleski}, Rados{\l}aw and {Ulaczyk}, Krzysztof and {Rybicki}, Krzysztof and {Iwanek}, Patryk},
        title = "{Microlensing Optical Depth and Event Rate toward the Galactic Bulge from 8 yr of OGLE-IV Observations}",
      journal = {\apjs},
         year = 2019,
        month = oct,
       volume = {244},
       number = {2},
          eid = {29},
        pages = {29},
          doi = {10.3847/1538-4365/ab426b},
archivePrefix = {arXiv},
       eprint = {1906.02210},
 primaryClass = {astro-ph.SR},
       adsurl = {https://ui.adsabs.harvard.edu/abs/2019ApJS..244...29M}
}

@ARTICLE{Nunota2025,
       author = {{Nunota}, Kansuke and {Sumi}, Takahiro and {Koshimoto}, Naoki and {Rattenbury}, Nicholas J. and {Abe}, Fumio and {Barry}, Richard and {Bennett}, David P. and {Bhattacharya}, Aparna and {Fukui}, Akihiko and {Hamada}, Ryusei and {Hamada}, Shunya and {Hamasaki}, Naoto and {Hirao}, Yuki and {Ishitani Silva}, Stela and {Itow}, Yoshitaka and {Matsubara}, Yutaka and {Miyazaki}, Shota and {Muraki}, Yasushi and {Nagai}, Tsutsumi and {Olmschenk}, Greg and {Ranc}, Clement and {Satoh}, Yuki K. and {Suzuki}, Daisuke and {Tristram}, Paul. J. and {Vandorou}, Aikaterini and {Yama}, Hibiki and {MOA Collaboration}},
        title = "{The Microlensing Event Rate and Optical Depth from MOA-II 9 Yr Survey Toward the Galactic Bulge}",
      journal = {\apj},
         year = 2025,
        month = feb,
       volume = {979},
       number = {2},
          eid = {123},
        pages = {123},
          doi = {10.3847/1538-4357/ada352},
archivePrefix = {arXiv},
       eprint = {2410.23553},
 primaryClass = {astro-ph.GA},
       adsurl = {https://ui.adsabs.harvard.edu/abs/2025ApJ...979..123N}
}

@ARTICLE{Kerins2009,
       author = {{Kerins}, E. and {Robin}, A.~C. and {Marshall}, D.~J.},
        title = "{Synthetic microlensing maps of the Galactic bulge}",
      journal = {\mnras},
         year = 2009,
        month = jun,
       volume = {396},
       number = {2},
        pages = {1202-1210},
          doi = {10.1111/j.1365-2966.2009.14791.x},
archivePrefix = {arXiv},
       eprint = {0805.4626},
 primaryClass = {astro-ph},
       adsurl = {https://ui.adsabs.harvard.edu/abs/2009MNRAS.396.1202K}
}

@ARTICLE{Awiphan2016,
       author = {{Awiphan}, S. and {Kerins}, E. and {Robin}, A.~C.},
        title = "{Besan{\c{c}}on Galactic model analysis of MOA-II microlensing: evidence for a mass deficit in the inner bulge}",
      journal = {\mnras},
         year = 2016,
        month = feb,
       volume = {456},
       number = {2},
        pages = {1666-1680},
          doi = {10.1093/mnras/stv2625},
archivePrefix = {arXiv},
       eprint = {1510.06347},
 primaryClass = {astro-ph.EP},
       adsurl = {https://ui.adsabs.harvard.edu/abs/2016MNRAS.456.1666A}
}

@ARTICLE{Specht2020,
       author = {{Specht}, David and {Kerins}, Eamonn and {Awiphan}, Supachai and {Robin}, Annie C.},
        title = "{MaB{\ensuremath{\mu}}lS-2: high-precision microlensing modelling for the large-scale survey era}",
      journal = {\mnras},
         year = 2020,
        month = oct,
       volume = {498},
       number = {2},
        pages = {2196-2218},
          doi = {10.1093/mnras/staa2375},
archivePrefix = {arXiv},
       eprint = {2005.14668},
 primaryClass = {astro-ph.EP},
       adsurl = {https://ui.adsabs.harvard.edu/abs/2020MNRAS.498.2196S}
}

@ARTICLE{Huston2026,
       author = {{Huston}, Macy J. and {Crisp}, Alison L. and {Newman}, Marz and {Patlak}, Riley and {Penny}, Matthew T. and {Kluter}, Jonas and {McGill}, Peter and {Smith}, Leigh C. and {Karkour}, Victor and {Abrams}, Natasha S. and {Gaudi}, B. Scott and {Lam}, Casey Y. and {Lu}, Jessica R. and {Calchi Novati}, Sebastiano and {Stassun}, Keivan G. and {Terry}, Sean K. and {Zohrabi}, Farzaneh},
        title = "{An Updated \textbackslashsynthpop Model for Microlensing Simulations I: Model Description, Evaluation, and Microlensing Event Rates Near the Galactic Center}",
      journal = {\apj\, submitted},
         year = 2026,
        month = mar,
          eid = {2603.12219},
        pages = {arXiv:2603.12219},
          doi = {10.48550/arXiv.2603.12219},
archivePrefix = {arXiv},
       eprint = {2603.12219},
 primaryClass = {astro-ph.GA},
       adsurl = {https://ui.adsabs.harvard.edu/abs/2026arXiv260312219H}
}

@INPROCEEDINGS{Beaulieu2013,
       author = {{Beaulieu}, Jean-Philippe and {Tisserand}, Patrick and {Batista}, Virginie},
        title = "{Space based microlensing planet searches}",
    booktitle = {European Physical Journal Web of Conferences},
         year = 2013,
       series = {European Physical Journal Web of Conferences},
       volume = {47},
        month = apr,
    publisher = {EDP},
          eid = {15001},
        pages = {15001},
          doi = {10.1051/epjconf/20134715001},
archivePrefix = {arXiv},
       eprint = {1303.6783},
 primaryClass = {astro-ph.EP},
       adsurl = {https://ui.adsabs.harvard.edu/abs/2013EPJWC..4715001B}
}

@ARTICLE{BennettRhie1996,
       author = {{Bennett}, David P. and {Rhie}, Sun Hong},
        title = "{Detecting Earth-Mass Planets with Gravitational Microlensing}",
      journal = {\apj},
         year = 1996,
        month = nov,
       volume = {472},
        pages = {660},
          doi = {10.1086/178096},
archivePrefix = {arXiv},
       eprint = {astro-ph/9603158},
 primaryClass = {astro-ph},
       adsurl = {https://ui.adsabs.harvard.edu/abs/1996ApJ...472..660B}
}

@ARTICLE{Poleski2016,
       author = {{Poleski}, Rados{\l}aw},
        title = "{Empirical microlensing event rates predicted by a phenomenological model}",
      journal = {\mnras},
         year = 2016,
        month = feb,
       volume = {455},
       number = {4},
        pages = {3656-3661},
          doi = {10.1093/mnras/stv2569},
archivePrefix = {arXiv},
       eprint = {1505.07104},
 primaryClass = {astro-ph.GA},
       adsurl = {https://ui.adsabs.harvard.edu/abs/2016MNRAS.455.3656P}
}

@ARTICLE{BennettRhie2002,
       author = {{Bennett}, David P. and {Rhie}, Sun Hong},
        title = "{Simulation of a Space-based Microlensing Survey for Terrestrial Extrasolar Planets}",
      journal = {\apj},
         year = 2002,
        month = aug,
       volume = {574},
       number = {2},
        pages = {985-1003},
          doi = {10.1086/340977},
archivePrefix = {arXiv},
       eprint = {astro-ph/0011466},
 primaryClass = {astro-ph},
       adsurl = {https://ui.adsabs.harvard.edu/abs/2002ApJ...574..985B}
}

@ARTICLE{Bachelet2022,
       author = {{Bachelet}, E. and {Specht}, D. and {Penny}, M. and {Hundertmark}, M. and {Awiphan}, S. and {Beaulieu}, J.-P. and {Dominik}, M. and {Kerins}, E. and {Maoz}, D. and {Meade}, E. and {Nucita}, A.~A. and {Poleski}, R. and {Ranc}, C. and {Rhodes}, J. and {Robin}, A.~C.},
        title = "{Euclid-Roman joint microlensing survey: Early mass measurement, free floating planets, and exomoons}",
      journal = {\aap},
         year = 2022,
        month = aug,
       volume = {664},
          eid = {A136},
        pages = {A136},
          doi = {10.1051/0004-6361/202140351},
archivePrefix = {arXiv},
       eprint = {2202.09475},
 primaryClass = {astro-ph.EP},
       adsurl = {https://ui.adsabs.harvard.edu/abs/2022A&A...664A.136B}
}

@ARTICLE{Penny2013,
       author = {{Penny}, M.~T. and {Kerins}, E. and {Rattenbury}, N. and {Beaulieu}, J.-P. and {Robin}, A.~C. and {Mao}, S. and {Batista}, V. and {Calchi Novati}, S. and {Cassan}, A. and {Fouqu{\'e}}, P. and {McDonald}, I. and {Marquette}, J.~B. and {Tisserand}, P. and {Zapatero Osorio}, M.~R.},
        title = "{ExELS: an exoplanet legacy science proposal for the ESA Euclid mission - I. Cold exoplanets}",
      journal = {\mnras},
         year = 2013,
        month = sep,
       volume = {434},
       number = {1},
        pages = {2-22},
          doi = {10.1093/mnras/stt927},
archivePrefix = {arXiv},
       eprint = {1206.5296},
 primaryClass = {astro-ph.EP},
       adsurl = {https://ui.adsabs.harvard.edu/abs/2013MNRAS.434....2P}
}

@ARTICLE{_Henry2025,
       author = {{Euclid Collaboration}: {McCracken}, H.~J. and {Benson}, K. and {Dolding}, C. and {Flanet}, T. and {Grenet}, C. and {Herent}, O. and {Hudelot}, P. and {Laigle}, C. and {Leroy}, G. and {Liebing}, P. and {Massey}, R. and {Mottet}, S. and {Nakajima}, R. and {Nguyen-Kim}, H.~N. and {Nightingale}, J.~W. and {Skottfelt}, J. and {Smith}, L.~C. and {Soldano}, F. and {Vilenius}, E. and {Wander}, M. and {von Wietersheim-Kramsta}, M. and {Akhlaghi}, M. and {Aussel}, H. and {Awan}, S. and {Azzollini}, R. and {Basset}, A. and {Candini}, G.~P. and {Casenove}, P. and {Cropper}, M. and {Hoekstra}, H. and {Israel}, H. and {Khalil}, A. and {Kuijken}, K. and {Mellier}, Y. and {Miller}, L. and {Niemi}, S.-M. and {Page}, M.~J. and {Paterson}, K. and {Schirmer}, M. and {Walton}, N.~A. and {Zacchei}, A. and {Barrios}, J.~P.~L.~G. and {Erben}, T. and {Hayes}, R. and {Kegerreis}, J.~A. and {Lagattuta}, D.~J. and {Lan{\c{c}}on}, A. and {Aghanim}, N. and {Altieri}, B. and {Amara}, A. and {Andreon}, S. and {Appleton}, P.~N. and {Auricchio}, N. and {Baccigalupi}, C. and {Baldi}, M. and {Balestra}, A. and {Bardelli}, S. and {Battaglia}, P. and {Belikov}, A.~N. and {Bender}, R. and {Bernardeau}, F. and {Biviano}, A. and {Bonchi}, A. and {Branchini}, E. and {Brescia}, M. and {Brinchmann}, J. and {Camera}, S. and {Ca{\~n}as-Herrera}, G. and {Capobianco}, V. and {Carbone}, C. and {Carretero}, J. and {Casas}, S. and {Castander}, F.~J. and {Castellano}, M. and {Castignani}, G. and {Cavuoti}, S. and {Chambers}, K.~C. and {Cimatti}, A. and {Colodro-Conde}, C. and {Congedo}, G. and {Conselice}, C.~J. and {Conversi}, L. and {Copin}, Y. and {Courbin}, F. and {Courtois}, H.~M. and {Da Silva}, A. and {da Silva}, R. and {Degaudenzi}, H. and {De Lucia}, G. and {Di Giorgio}, A.~M. and {Dinis}, J. and {Dole}, H. and {Dubath}, F. and {Dupac}, X. and {Dusini}, S. and {Ealet}, A. and {Escoffier}, S. and {Fabricius}, M. and {Farina}, M. and {Farinelli}, R. and {Ferriol}, S. and {Finelli}, F. and {Fosalba}, P. and {Fotopoulou}, S. and {Fourmanoit}, N. and {Frailis}, M. and {Franceschi}, E. and {Galeotta}, S. and {George}, K. and {Gillard}, W. and {Gillis}, B. and {Giocoli}, C. and {G{\'o}mez-Alvarez}, P. and {Gracia-Carpio}, J. and {Granett}, B.~R. and {Grazian}, A. and {Grupp}, F. and {Guzzo}, L. and {Hailey}, M. and {Haugan}, S.~V.~H. and {Hoar}, J. and {Holmes}, W. and {Hormuth}, F. and {Hornstrup}, A. and {Jahnke}, K. and {Jhabvala}, M. and {Joachimi}, B. and {Keih{\"a}nen}, E. and {Kermiche}, S. and {Kiessling}, A. and {Kilbinger}, M. and {Kubik}, B. and {K{\"u}mmel}, M. and {Kunz}, M. and {Kurki-Suonio}, H. and {Le Boulc'h}, Q. and {Le Brun}, A.~M.~C. and {Le Mignant}, D. and {Ligori}, S. and {Lilje}, P.~B. and {Lindholm}, V. and {Lloro}, I. and {Mainetti}, G. and {Maino}, D. and {Maiorano}, E. and {Mansutti}, O. and {Marcin}, S. and {Marggraf}, O. and {Martinelli}, M. and {Martinet}, N. and {Marulli}, F. and {Masters}, D.~C. and {Maurogordato}, S. and {Medinaceli}, E. and {Mei}, S. and {Melchior}, M. and {Meneghetti}, M. and {Merlin}, E. and {Meylan}, G. and {Mora}, A. and {Moresco}, M. and {Moscardini}, L. and {Neissner}, C. and {Nichol}, R.~C. and {Padilla}, C. and {Paltani}, S. and {Pasian}, F. and {Pedersen}, K. and {Percival}, W.~J. and {Pettorino}, V. and {Pires}, S. and {Polenta}, G. and {Poncet}, M. and {Popa}, L.~A. and {Pozzetti}, L. and {Racca}, G.~D. and {Raison}, F. and {Rebolo}, R. and {Renzi}, A. and {Rhodes}, J. and {Riccio}, G. and {Romelli}, E. and {Roncarelli}, M. and {Rossetti}, E. and {Rusholme}, B. and {Saglia}, R. and {Sakr}, Z. and {S{\'a}nchez}, A.~G. and {Sapone}, D. and {Sartoris}, B. and {Schewtschenko}, J.~A. and {Schneider}, P. and {Schrabback}, T. and {Secroun}, A. and {Seidel}, G. and {Seiffert}, M. and {Serrano}, S. and {Simon}, P. and {Sirignano}, C.},
        title = "{Euclid Quick Data Release (Q1): VIS processing and data products}",
      journal = {arXiv e-prints},
         year = 2025,
        month = mar,
          eid = {arXiv:2503.15303},
        pages = {arXiv:2503.15303},
          doi = {10.48550/arXiv.2503.15303},
archivePrefix = {arXiv},
       eprint = {2503.15303},
 primaryClass = {astro-ph.IM},
       adsurl = {https://ui.adsabs.harvard.edu/abs/2025arXiv250315303E}
}

@ARTICLE{Johnson2020,
       author = {{Johnson}, Samson A. and {Penny}, Matthew and {Gaudi}, B. Scott and {Kerins}, Eamonn and {Rattenbury}, Nicholas J. and {Robin}, Annie C. and {Calchi Novati}, Sebastiano and {Henderson}, Calen B.},
        title = "{Predictions of the Nancy Grace Roman Space Telescope Galactic Exoplanet Survey. II. Free-floating Planet Detection Rates}",
      journal = {\aj},
         year = 2020,
        month = sep,
       volume = {160},
       number = {3},
          eid = {123},
        pages = {123},
          doi = {10.3847/1538-3881/aba75b},
archivePrefix = {arXiv},
       eprint = {2006.10760},
 primaryClass = {astro-ph.EP},
       adsurl = {https://ui.adsabs.harvard.edu/abs/2020AJ....160..123J}
}

@ARTICLE{Penny2019,
       author = {{Penny}, Matthew T. and {Gaudi}, B. Scott and {Kerins}, Eamonn and {Rattenbury}, Nicholas J. and {Mao}, Shude and {Robin}, Annie C. and {Calchi Novati}, Sebastiano},
        title = "{Predictions of the WFIRST Microlensing Survey. I. Bound Planet Detection Rates}",
      journal = {\apjs},
         year = 2019,
        month = mar,
       volume = {241},
       number = {1},
          eid = {3},
        pages = {3},
          doi = {10.3847/1538-4365/aafb69},
archivePrefix = {arXiv},
       eprint = {1808.02490},
 primaryClass = {astro-ph.EP},
       adsurl = {https://ui.adsabs.harvard.edu/abs/2019ApJS..241....3P}
}

@ARTICLE{Bozza2025,
       author = {{Bozza}, V. and {Salmeri}, L. and {Rota}, P. and {Bachelet}, E. and {Beaulieu}, J.-P. and {Cole}, A.~A. and {Cuillandre}, J.~C. and {Kerins}, E. and {McDonald}, I. and {Mr{\'o}z}, P. and {Penny}, M. and {Ranc}, C. and {Rektsini}, N. and {Thygesen}, E. and {Verma}, H. and {Udalski}, A. and {Poleski}, R. and {Skowron}, J.},
        title = "{Historic microlensing events in the Euclid Galactic Bulge Survey}",
      journal = {\aap\, submitted},
         year = 2025,
        month = nov,
          eid = {arXiv:2511.03307},
        pages = {arXiv:2511.03307},
          doi = {10.48550/arXiv.2511.03307},
archivePrefix = {arXiv},
       eprint = {2511.03307},
 primaryClass = {astro-ph.EP},
       adsurl = {https://arxiv.org/abs/2511.03307}
}

@ARTICLE{Specht2023,
       author = {{Specht}, D. and {Poleski}, R. and {Penny}, M.~T. and {Kerins}, E. and {McDonald}, I. and {Lee}, Chung-Uk and {Udalski}, A. and {Bond}, I.~A. and {Shvartzvald}, Y. and {Zang}, Weicheng and {Street}, R.~A. and {Hogg}, D.~W. and {Gaudi}, B.~S. and {Barclay}, T. and {Barentsen}, G. and {Howell}, S.~B. and {Mullally}, F. and {Henderson}, C.~B. and {Bryson}, S.~T. and {Caldwell}, D.~A. and {Haas}, M.~R. and {Van Cleve}, J.~E. and {Larson}, K. and {McCalmont}, K. and {Peterson}, C. and {Putnam}, D. and {Ross}, S. and {Packard}, M. and {Reedy}, L. and {Albrow}, Michael D. and {Chung}, Sun-Ju and {Kil Jung}, Youn and {Gould}, Andrew and {Han}, Cheongho and {Hwang}, Kyu-Ha and {Ryu}, Yoon-Hyun and {Shin}, In-Gu and {Yang}, Hongjing and {Yee}, Jennifer C. and {Cha}, Sang-Mok and {Kim}, Dong-Jin and {Kim}, Seung-Lee and {Lee}, Dong-Joo and {Lee}, Yongseok and {Park}, Byeong-Gon and {Pogge}, Richard W. and {Szyma{\'n}ski}, M.~K. and {Soszy{\'n}ski}, I. and {Ulaczyk}, K. and {Pietrukowicz}, P. and {Koz{\l}owski}, Sz and {Skowron}, J. and {Mr{\'o}z}, P. and {Mao}, Shude and {Fouqu{\'e}}, Pascal and {Zhu}, Wei and {Abe}, F. and {Barry}, R. and {Bennett}, D.~P. and {Bhattacharya}, A. and {Fukui}, A. and {Fujii}, H. and {Hirao}, Y. and {Itow}, Y. and {Kirikawa}, R. and {Kondo}, I. and {Koshimoto}, N. and {Matsubara}, Y. and {Matsumoto}, S. and {Miyazaki}, S. and {Muraki}, Y. and {Olmschenk}, G. and {Ranc}, C. and {Okamura}, A. and {Rattenbury}, N.~J. and {Satoh}, Y. and {Sumi}, T. and {Suzuki}, D. and {Silva}, S.~I. and {Toda}, T. and {Tristram}, P.~J. and {Vandorou}, A. and {Yama}, H. and {Beichman}, C. and {Bryden}, G. and {Calchi Novati}, S.},
        title = "{Kepler K2 Campaign 9 - II. First space-based discovery of an exoplanet using microlensing}",
      journal = {\mnras},
         year = 2023,
        month = apr,
       volume = {520},
       number = {4},
        pages = {6350-6366},
          doi = {10.1093/mnras/stad212},
archivePrefix = {arXiv},
       eprint = {2203.16959},
 primaryClass = {astro-ph.EP},
       adsurl = {https://ui.adsabs.harvard.edu/abs/2023MNRAS.520.6350S}
}

@ARTICLE{McDonald2017,
       author = {{McDonald}, I. and {Zijlstra}, A.~A. and {Watson}, R.~A.},
        title = "{Fundamental parameters and infrared excesses of Tycho-Gaia stars}",
      journal = {\mnras},
         year = 2017,
        month = oct,
       volume = {471},
       number = {1},
        pages = {770-791},
          doi = {10.1093/mnras/stx1433},
archivePrefix = {arXiv},
       eprint = {1706.02208},
 primaryClass = {astro-ph.SR},
       adsurl = {https://ui.adsabs.harvard.edu/abs/2017MNRAS.471..770M}
}

@ARTICLE{McDonald2025b,
       author = {{McDonald}, I. and {Zijlstra}, A.~A. and {Cox}, N.~J. and {Bernard-Salas}, J.},
        title = "{The Gaia All-Sky Stellar Parameters Service (GASPS)}",
      journal = {Communications of the Byurakan Astrophysical Observatory},
         year = 2025,
        month = dec,
       volume = {72},
       number = {2},
        pages = {341-350},
          doi = {10.52526/25792776-25.72.2-341},
archivePrefix = {arXiv},
       eprint = {2601.03978},
 primaryClass = {astro-ph.SR},
       adsurl = {https://ui.adsabs.harvard.edu/abs/2025CoBAO..72..341M}
}

@ARTICLE{McDonald2025a,
       author = {{McDonald}, I. and {Srinivasan}, S. and {Scicluna}, P. and {Jones}, O.~C. and {Zijlstra}, A.~A. and {Wallstr{\"o}m}, S.~H.~J. and {Danilovich}, T. and {He}, J.~H. and {Marshall}, J.~P. and {van Loon}, J. Th and {Wesson}, R. and {Kemper}, F. and {Trejo-Cruz}, A. and {Greaves}, J. and {Dharmawardena}, T. and {Cami}, J. and {Kim}, Hyosun and {Kraemer}, K.~E. and {Clark}, C.~J.~R. and {Shinnaga}, H. and {Haswell}, C. and {Imai}, H. and {Wouterloot}, J.~G.~A. and {P{\'e}rez Vidal}, A.~J. and {Rau}, G.},
        title = "{The Nearby Evolved Stars Survey (NESS) V: properties of volume-limited samples of Galactic evolved stars}",
      journal = {\mnras},
         year = 2025,
        month = jul,
       volume = {541},
       number = {1},
        pages = {516-552},
          doi = {10.1093/mnras/staf978},
archivePrefix = {arXiv},
       eprint = {2506.10542},
 primaryClass = {astro-ph.SR},
       adsurl = {https://ui.adsabs.harvard.edu/abs/2025MNRAS.541..516M}
}

@ARTICLE{Tan2023,
       author = {{Tan}, Shuyu and {Parker}, Quentin A. and {Zijlstra}, Albert A. and {Ritter}, Andreas and {Rees}, Bryan},
        title = "{When the Stars Align: A 5{\ensuremath{\sigma}} Concordance of Planetary Nebulae Major Axes in the Center of Our Galaxy}",
      journal = {\apjl},
         year = 2023,
        month = jul,
       volume = {951},
       number = {2},
          eid = {L44},
        pages = {L44},
          doi = {10.3847/2041-8213/acdbcd},
archivePrefix = {arXiv},
       eprint = {2307.07140},
 primaryClass = {astro-ph.GA},
       adsurl = {https://ui.adsabs.harvard.edu/abs/2023ApJ...951L..44T}
}

@ARTICLE{Zhang2020,
       author = {{Zhang}, Keming and {Bloom}, Joshua S.},
        title = "{deepCR: Cosmic Ray Rejection with Deep Learning}",
      journal = {\apj},
         year = 2020,
        month = jan,
       volume = {889},
       number = {1},
          eid = {24},
        pages = {24},
          doi = {10.3847/1538-4357/ab3fa6},
archivePrefix = {arXiv},
       eprint = {1907.09500},
 primaryClass = {astro-ph.IM},
       adsurl = {https://ui.adsabs.harvard.edu/abs/2020ApJ...889...24Z}
}

@article{Bhattacharya.2020,
       author = {{Bhattacharya}, Aparna and {Bennett}, David P. and {Beaulieu}, Jean Philippe and {Bond}, Ian A. and {Koshimoto}, Naoki and {Lu}, Jessica R. and {Blackman}, Joshua W. and {Vandorou}, Aikaterini and {Terry}, Sean K. and {Batista}, Virginie and {Marquette}, Jean Baptiste and {Cole}, Andrew A. and {Fukui}, Akihiko and {Henderson}, Calen B. and {Ranc}, Cl{\'e}ment},
        title = "{MOA-2007-BLG-400 A Super-Jupiter-mass Planet Orbiting a Galactic Bulge K-dwarf Revealed by Keck Adaptive Optics Imaging}",
      journal = {\aj},
         year = 2021,
        month = aug,
       volume = {162},
       number = {2},
          eid = {60},
        pages = {60},
          doi = {10.3847/1538-3881/abfec5},
       adsurl = {https://ui.adsabs.harvard.edu/abs/2021AJ....162...60B}
}

@article{Ranc.2015,
	author = {Ranc, C. and Cassan, A. and Albrow, M. D. and Kubas, D. and Bond, I. A. and Batista, V. and Beaulieu, J.-P. and Bennett, D. P. and Dominik, M. and Dong, S. and Fouqu{\'e}, P. and Gould, A. and Greenhill, J. and J{\o}rgensen, U. G. and Kains, N. and Menzies, J. and Sumi, T. and Bachelet, E. and Coutures, C. and Dieters, S. and Dominis Prester, D. and Donatowicz, J. and Gaudi, B. S. and Han, C. and Hundertmark, M. and Horne, K. and Kane, S. R. and Lee, C.-U. and Marquette, J.-B. and Park, B.-G. and Pollard, K. R. and Sahu, K. C. and Street, R. and Tsapras, Y. and Wambsganss, J. and Williams, A. and Zub, M. and Abe, F. and Fukui, A. and Itow, Y. and Masuda, K. and Matsubara, Y. and Muraki, Y. and Ohnishi, K. and Rattenbury, N. and Saito, T. and Sullivan, D. J. and Sweatman, W. L. and Tristram, P. J. and Yock, P. C. M. and Yonehara, A.},
	journal = {\aap},
	month = aug,
	pages = {A125},
	title = {MOA-2007-BLG-197: Exploring the brown dwarf desert},
	volume = {580},
	year = {2015}}

@article{Terry2022a,
	author = {{Terry}, Sean K. and {Bennett}, David P. and {Bhattacharya}, Aparna and {Koshimoto}, Naoki and {Beaulieu}, Jean-Philippe and {Blackman}, Joshua W. and {Bond}, Ian A. and {Cole}, Andrew A. and {Lu}, Jessica R. and {Marquette}, Jean Baptiste and {Ranc}, Cl{\'e}ment and {Rektsini}, Natalia and {Vandorou}, Aikaterini},
	journal = {\aj},
	month = nov,
	number = {5},
	pages = {217},
	title = {{Adaptive Optics Imaging Can Break the Central Caustic Cusp Approach Degeneracy in High-magnification Microlensing Events}},
	volume = {164},
	year = 2022}

@article{Vandorou2025b,
	author = {{Vandorou}, Aikaterini and {Bennett}, David P. and {Beaulieu}, Jean-Philippe and {Bhattacharya}, Aparna and {Blackman}, Joshua W. and {Bond}, Ian A. and {Cole}, Andrew A. and {Koshimoto}, Naoki and {Ranc}, Cl{\'e}ment and {Rektsini}, Natalia E. and {Terry}, Sean K.},
	journal = {\aj},
	month = dec,
	number = {6},
	pages = {310},
	title = {{MOA-2010-BLG-328: Keck and HST Expose the Limits of Occams Razor in Microlensing}},
	volume = {170},
	year = 2025}

@article{Bennett.2015.hst169,
	author = {Bennett, D. P. and Bhattacharya, A. and Anderson, J. and Bond, I. A. and Anderson, N. and Barry, R. and Batista, V. and Beaulieu, J.-P. and DePoy, D. L. and Dong, S. and Gaudi, B. S. and Gilbert, E. and Gould, A. and Pfeifle, R. and Pogge, R. W. and Suzuki, D. and Terry, S. and Udalski, A.},
	journal = {\apj},
	month = aug,
	pages = {169},
	title = {Confirmation of the Planetary Microlensing Signal and Star and Planet Mass Determinations for Event OGLE-2005-BLG-169},
	volume = {808},
	year = {2015}}

@article{Bennett2020a,
	author = {{Bennett}, David P. and {Bhattacharya}, Aparna and {Beaulieu}, Jean-Philippe and {Blackman}, Joshua W. and {Vandorou}, Aikaterini and {Terry}, Sean K. and {Cole}, Andrew A. and {Henderson}, Calen B. and {Koshimoto}, Naoki and {Lu}, Jessica R. and {Baptiste Marquette}, Jean and {Ranc}, Cl{\'e}ment and {Udalski}, Andrzej},
	journal = {\aj},
	month = feb,
	number = {2},
	pages = {68},
	title = {{Keck Observations Confirm a Super-Jupiter Planet Orbiting M Dwarf OGLE-2005-BLG-071L}},
	volume = {159},
	year = 2020}

@article{Bennett2024a,
	author = {{Bennett}, David P. and {Bhattacharya}, Aparna and {Beaulieu}, Jean-Philippe and {Koshimoto}, Naoki and {Blackman}, Joshua W. and {Bond}, Ian A. and {Ranc}, Cl{\'e}ment and {Rektsini}, Natalia and {Terry}, Sean K. and {Vandorou}, Aikaterini and {Lu}, Jessica R. and {Marquette}, Jean Baptiste and {Olmschenk}, Greg and {Suzuki}, Daisuke},
	journal = {\aj},
	month = jul,
	number = {1},
	pages = {15},
	title = {{Keck and Hubble Observations Show that MOA-2008-BLG-379Lb is a Super-Jupiter Orbiting an M Dwarf}},
	volume = {168},
	year = 2024}

@ARTICLE{Beaulieu2006,
  author = {Beaulieu, J.-P. and Bennett, D. P. and Fouqu{\'e}, P. and
            Williams, A. and Dominik, M. and J{\o}rgensen, U. G. and
            Kubas, D. and Cassan, A. and Coutures, C. and Greenhill, J. and
            Hill, K. and Menzies, J. and Pollard, K. and Sahu, K. C. and
            Vinter, C. and Albrow, M. D. and Allan, A. and
            et al.},
  title = {Discovery of a Cool Planet of 5.5 Earth Masses Through Gravitational Microlensing},
  journal = {Nature},
  year = {2006},
  volume = {439},
  pages = {437--440},
  doi = {10.1038/nature04441}
}

@article{Beaulieu2018a,
	author = {{Beaulieu}, J.-P. and {Batista}, V. and {Bennett}, D.~P. and {Marquette}, J.-B. and {Blackman}, J.~W. and {Cole}, A.~A. and {Coutures}, C. and {Danielski}, C. and {Dominis Prester}, D. and {Donatowicz}, J. and {Fukui}, A. and {Koshimoto}, N. and {Lon{\v{c}}ari{\'c}}, K. and {Morales}, J.~C. and {Sumi}, T. and {Suzuki}, D. and {Henderson}, C. and {Shvartzvald}, Y. and {Beichman}, C.},
	journal = {\aj},
	month = feb,
	number = {2},
	pages = {78},
	title = {{Combining Spitzer Parallax and Keck II Adaptive Optics Imaging to Measure the Mass of a Solar-like Star Orbited by a Cold Gaseous Planet Discovered by Microlensing}},
	volume = {155},
	year = 2018}

@ARTICLE{Bennett2006,
       author = {{Bennett}, David P. and {Anderson}, Jay and {Bond}, Ian A. and {Udalski}, Andrzej and {Gould}, Andrew},
        title = "{Identification of the OGLE-2003-BLG-235/MOA-2003-BLG-53 Planetary Host Star}",
      journal = {\apjl},
         year = 2006,
        month = aug,
       volume = {647},
       number = {2},
        pages = {L171-L174},
          doi = {10.1086/507585},
archivePrefix = {arXiv},
       eprint = {astro-ph/0606038},
 primaryClass = {astro-ph},
       adsurl = {https://ui.adsabs.harvard.edu/abs/2006ApJ...647L.171B}
}

@ARTICLE{Dong2009,
       author = {{Dong}, Subo and {Gould}, Andrew and {Udalski}, Andrzej and {Anderson}, Jay and {Christie}, G.~W. and {Gaudi}, B.~S. and {OGLE Collaboration} and {Jaroszy{\'n}ski}, M. and {Kubiak}, M. and {Szyma{\'n}ski}, M.~K. and {Pietrzy{\'n}ski}, G. and {Soszy{\'n}ski}, I. and {Szewczyk}, O. and {Ulaczyk}, K. and {Wyrzykowski}, {\L}. and {{\ensuremath{\mu}}FUN Collaboration} and {DePoy}, D.~L. and {Fox}, D.~B. and {Gal-Yam}, A. and {Han}, C. and {L{\'e}pine}, S. and {McCormick}, J. and {Ofek}, E. and {Park}, B.-G. and {Pogge}, R.~W. and {MOA Collaboration} and {Abe}, F. and {Bennett}, D.~P. and {Bond}, I.~A. and {Britton}, T.~R. and {Gilmore}, A.~C. and {Hearnshaw}, J.~B. and {Itow}, Y. and {Kamiya}, K. and {Kilmartin}, P.~M. and {Korpela}, A. and {Masuda}, K. and {Matsubara}, Y. and {Motomura}, M. and {Muraki}, Y. and {Nakamura}, S. and {Ohnishi}, K. and {Okada}, C. and {Rattenbury}, N. and {Saito}, To. and {Sako}, T. and {Sasaki}, M. and {Sullivan}, D. and {Sumi}, T. and {Tristram}, P.~J. and {Yanagisawa}, T. and {Yock}, P.~C.~M. and {Yoshoika}, T. and {PLANET/RoboNet Collaborations} and {Albrow}, M.~D. and {Beaulieu}, J.~P. and {Brillant}, S. and {Calitz}, H. and {Cassan}, A. and {Cook}, K.~H. and {Coutures}, Ch. and {Dieters}, S. and {Dominis Prester}, D. and {Donatowicz}, J. and {Fouqu{\'e}}, P. and {Greenhill}, J. and {Hill}, K. and {Hoffman}, M. and {Horne}, K. and {J{\o}rgensen}, U.~G. and {Kane}, S. and {Kubas}, D. and {Marquette}, J.~B. and {Martin}, R. and {Meintjes}, P. and {Menzies}, J. and {Pollard}, K.~R. and {Sahu}, K.~C. and {Vinter}, C. and {Wambsganss}, J. and {Williams}, A. and {Bode}, M. and {Bramich}, D.~M. and {Burgdorf}, M. and {Snodgrass}, C. and {Steele}, I. and {Doublier}, Vanessa and {Foellmi}, Cedric},
        title = "{OGLE-2005-BLG-071Lb, the Most Massive M Dwarf Planetary Companion?}",
      journal = {\apj},
         year = 2009,
        month = apr,
       volume = {695},
       number = {2},
        pages = {970-987},
          doi = {10.1088/0004-637X/695/2/970},
archivePrefix = {arXiv},
       eprint = {0804.1354},
 primaryClass = {astro-ph},
       adsurl = {https://ui.adsabs.harvard.edu/abs/2009ApJ...695..970D}
}

@ARTICLE{Rektsini24,
       author = {{Rektsini}, Natalia E. and {Batista}, Virginie and {Ranc}, Cl{\'e}ment and {Bennett}, David P. and {Beaulieu}, Jean-Philippe and {Blackman}, Joshua W. and {Cole}, Andrew A. and {Terry}, Sean K. and {Koshimoto}, Naoki and {Bhattacharya}, Aparna and {Vandorou}, Aikaterini and {Plunkett}, Thomas J. and {Marquette}, Jean-Baptiste},
        title = "{Precise Mass Measurement of OGLE-2013-BLG-0132/MOA-2013-BLG-148: A Saturn-mass Planet Orbiting an M Dwarf}",
      journal = {\aj},
         year = 2024,
        month = apr,
       volume = {167},
       number = {4},
          eid = {145},
        pages = {145},
          doi = {10.3847/1538-3881/ad2514},
archivePrefix = {arXiv},
       eprint = {2401.17549},
 primaryClass = {astro-ph.EP},
       adsurl = {https://ui.adsabs.harvard.edu/abs/2024AJ....167..145R}
}

@ARTICLE{sotoetal:2014,
       author = {{Soto}, M. and {Zeballos}, H. and {Kuijken}, K. and {Rich}, R.~M. and {Kunder}, A. and {Astraatmadja}, T.},
        title = "{Proper motions for HST observations in three off-axis bulge fields}",
      journal = {\aap},
         year = 2014,
        month = feb,
       volume = {562},
          eid = {A41},
        pages = {A41},
          doi = {10.1051/0004-6361/201117339},
archivePrefix = {arXiv},
       eprint = {1403.2533},
 primaryClass = {astro-ph.GA},
       adsurl = {https://ui.adsabs.harvard.edu/abs/2014A&A...562A..41S}
}

@ARTICLE{clarkeetal:2019,
       author = {{Clarke}, Jonathan P. and {Wegg}, Christopher and {Gerhard}, Ortwin and {Smith}, Leigh C. and {Lucas}, Phil W. and {Wylie}, Shola M.},
        title = "{The Milky Way bar/bulge in proper motions: a 3D view from VIRAC and Gaia}",
      journal = {\mnras},
         year = 2019,
        month = nov,
       volume = {489},
       number = {3},
        pages = {3519-3538},
          doi = {10.1093/mnras/stz2382},
archivePrefix = {arXiv},
       eprint = {1903.02003},
 primaryClass = {astro-ph.GA},
       adsurl = {https://ui.adsabs.harvard.edu/abs/2019MNRAS.489.3519C}
}

@ARTICLE{1996A&AS..117..393B,
       author = {{Bertin}, E. and {Arnouts}, S.},
        title = "{SExtractor: Software for source extraction.}",
      journal = {\aaps},
         year = 1996,
        month = jun,
       volume = {117},
        pages = {393-404},
          doi = {10.1051/aas:1996164},
       adsurl = {https://ui.adsabs.harvard.edu/abs/1996A&AS..117..393B}
}

@ARTICLE{2015ApJS..220....1A,
       author = {{Akhlaghi}, Mohammad and {Ichikawa}, Takashi},
        title = "{Noise-based Detection and Segmentation of Nebulous Objects}",
      journal = {\apjs},
         year = 2015,
        month = sep,
       volume = {220},
       number = {1},
          eid = {1},
        pages = {1},
          doi = {10.1088/0067-0049/220/1/1},
archivePrefix = {arXiv},
       eprint = {1505.01664},
 primaryClass = {astro-ph.IM},
       adsurl = {https://ui.adsabs.harvard.edu/abs/2015ApJS..220....1A}
}

@ARTICLE{Paczynski1986,
       author = {{Paczynski}, B.},
        title = "{Gravitational Microlensing by the Galactic Halo}",
      journal = {\apj},
         year = 1986,
        month = may,
       volume = {304},
        pages = {1},
          doi = {10.1086/164140},
       adsurl = {https://ui.adsabs.harvard.edu/abs/1986ApJ...304....1P}
}

@ARTICLE{1987PASP...99..191S,
       author = {{Stetson}, Peter B.},
        title = "{DAOPHOT: A Computer Program for Crowded-Field Stellar Photometry}",
      journal = {\pasp},
         year = 1987,
        month = mar,
       volume = {99},
        pages = {191},
          doi = {10.1086/131977},
       adsurl = {https://ui.adsabs.harvard.edu/abs/1987PASP...99..191S}
}

@ARTICLE{2026ApJ..1003L...1T,
       author = {{Terry}, Sean K. and {Anderson}, Jay and {Beichman}, Charles A. and {Bennett}, David P. and {Bhattacharya}, Aparna and {Beaulieu}, Jean-Philippe and {Gaudi}, B. Scott and {Green}, Joel and {Huston}, Macy J. and {Lu}, Jessica R. and {Lucas}, Ray A. and {Nataf}, David M. and {Penny}, Matthew T. and {Rektsini}, Natalia E. and {Rodriguez Sanchez-Vahamonde}, Carolina and {Vandorou}, Aikaterini},
        title = "{An HST Wide-field Survey of the Galactic Bulge: Overview, Strategy, and First Results}",
      journal = {\apjl},
         year = 2026,
        month = may,
       volume = {1003},
       number = {1},
          eid = {L1},
        pages = {L1},
          doi = {10.3847/2041-8213/ae53e8},
archivePrefix = {arXiv},
       eprint = {2605.06778},
 primaryClass = {astro-ph.GA},
       adsurl = {https://ui.adsabs.harvard.edu/abs/2026ApJ..1003L...1T}
}

\begin{appendix}
  \onecolumn 

\section{Data products\label{apdx:A}}

A complete description of the data products provided here can be found in DR1 VIS Euclid Data Products Description Document (DPDD), \url{http://st-dm.pages.euclid-sgs.uk/data-product-doc/dmdr1/visdpd/visindex.html}. The VIS data products provided in Q2 include the VIS Calibrated Quad Frame product and the VIS PSF model product. 

\subsection{Provided data products}
The Q2 data release includes 16 calibrated dithers, each of 400\,s exposure, of the nine Galactic bulge fields. The relative pointing positions are given in Table~\ref{tab:ditherpos}. We also provide in digital form a file summarizing the names and epoch of all the science and calibration files. 
In addition, there are 16 400-s dithers of the PSF reference frames that were used to construct the Q2 VIS PSF model and the Q2 VIS PSF model generated using \texttt{PSFEx}. Finally, we provide the calibrated source catalogues produced using the standard Q1 PSF model.

The VIS Calibrated quad frame product is provided as a FITS file containing three extensions for each of the 144 quadrants: the science image, the associated modelled background, and the weight map. Pixel values are stored as 32-bit floating-point numbers. Astrometric and photometric calibration parameters are provided in the FITS headers for each quadrant.  The Q2 PSF model is also provided for each quadrant. This model should be considered a first-order approximation and may be refined in subsequent analyses. It includes a spatial dependence across each quadrant. 

We also provide \texttt{EuclidExtractor}, an extraction tool that creates cutout frames centred on the desired right ascension, declination (or Galactic) coordinates. The tool generates cutouts, with the chosen size, of the calibrated scientific frames, the background, flag and S/R maps in addition to the PSF model for each dither per quadrant or pixel position. 

\begin{table*}%[h]
\centering
\caption{Offsets $\Delta {\rm RA [arcsec]}, \Delta {\rm Dec [arcsec]}$ for the different dithers relative to the first dither.}
\scalebox{0.75}{
\setlength{\tabcolsep}{3pt} % change inter-olumn spacing
\begin{tabular}{llllllllll}
\hline\hline
\noalign{\vskip 3pt}
 Dither & Pointing E1 & Pointing E2 & Pointing E3 & Pointing E4 & Pointing E5 & Pointing E6 & Pointing E7 & Pointing E8 & Pointing E9 \\
\noalign{\vskip 3pt}
\hline
\noalign{\vskip 2pt}
0 & (0.00, 0.00) & (0.00, 0.00) & (0.00, 0.00) & (0.00, 0.00) & (0.00, 0.00) & (0.00, 0.00) & (0.00, 0.00) & (0.00, 0.00) & (0.00, 0.00) \\
1 & (60.92, 113.43) & (59.53, 115.73) & (59.23, 114.17) & (59.99, 115.14) & (60.49, 114.09) & (61.30, 114.73) & (60.38, 115.34) & (60.12, 116.15) & (57.74, 113.65) \\
2 & (53.28, 225.33) & (51.81, 225.44) & (51.82, 225.45) & (51.34, 226.37) & (52.23, 224.22) & (52.40, 225.82) & (50.72, 226.34) & (49.51, 226.40) & (48.64, 225.98) \\
3 & (116.17, 339.09) & (112.84, 340.63) & (112.82, 340.43) & (111.31, 340.71) & (112.30, 339.77) & (113.15, 341.09) & (111.19, 341.39) & (109.48, 342.33) & (107.72, 340.61) \\
4 & (1.22, 0.69) & (-1.28, 1.46) & (-0.15, -0.31) & (0.15, 1.79) & (0.43, -0.60) & (0.11, 1.24) & (0.58, 0.06) & (0.31, 2.23) & (0.56, 0.61) \\
5 & (61.26, 113.40) & (59.62, 115.71) & (59.72, 114.47) & (60.19, 115.19) & (60.47, 114.00) & (61.21, 114.98) & (60.82, 115.48) & (60.11, 115.99) & (57.68, 113.39) \\
6 & (53.11, 225.27) & (51.88, 225.67) & (51.46, 225.26) & (51.76, 226.19) & (52.06, 224.52) & (52.24, 225.80) & (50.71, 226.11) & (49.39, 226.16) & (48.39, 226.19) \\
7 & (116.03, 339.22) & (112.39, 340.30) & (112.68, 340.13) & (111.01, 340.63) & (112.39, 340.01) & (113.27, 340.82) & (111.04, 341.23) & (109.49, 342.54) & (108.11, 340.48) \\
8 & (1.49, 0.48) & (-1.09, 1.39) & (-0.57, 0.32) & (-0.06, 1.78) & (0.58, -1.03) & (0.18, 1.40) & (0.92, 0.96) & (0.46, 1.68) & (0.41, 0.37) \\
9 & (61.15, 113.51) & (59.57, 115.64) & (59.57, 114.40) & (60.24, 114.97) & (60.49, 113.87) & (61.14, 114.95) & (60.68, 115.46) & (60.11, 115.91) & (57.73, 113.52) \\
10 & (53.12, 225.09) & (51.69, 225.43) & (51.54, 225.25) & (51.72, 226.28) & (52.27, 224.39) & (52.29, 225.66) & (50.65, 226.16) & (49.41, 226.42) & (48.61, 226.15) \\
11 & (115.92, 339.20) & (112.64, 340.11) & (112.60, 340.56) & (111.26, 340.39) & (112.38, 340.27) & (113.04, 340.96) & (111.02, 341.31) & (109.14, 342.78) & (107.92, 340.95) \\
12 & (1.47, 0.40) & (-1.53, 1.46) & (-0.13, -0.04) & (0.28, 1.35) & (0.22, -0.78) & (0.25, 1.24) & (1.06, 0.22) & (0.50, 2.02) & (0.70, 0.26) \\
13 & (61.06, 113.31) & (59.62, 115.78) & (59.54, 114.25) & (60.04, 114.72) & (60.74, 114.03) & (61.32, 114.58) & (60.69, 115.28) & (59.74, 115.92) & (57.74, 113.55) \\
14 & (53.19, 225.18) & (51.97, 225.48) & (51.73, 225.58) & (51.55, 226.18) & (52.21, 224.25) & (52.49, 225.82) & (50.69, 226.41) & (49.12, 226.50) & (48.80, 225.95) \\
15 & (116.14, 339.42) & (112.73, 340.12) & (112.51, 340.60) & (111.03, 340.63) & (112.29, 339.92) & (113.15, 340.85) & (111.30, 341.41) & (109.20, 342.13) & (107.97, 340.71) \\
\hline
\end{tabular}
}
\label{tab:ditherpos}
\end{table*}

\subsection{Data access}
The public \Euclid science archive at the ESAC Science Data Centre (ESDC) opened on 24 June 2026, offering the Q2 data products online. The data themselves and information about the release is available at \url{https://www.cosmos.esa.int/web/euclid/q2-data-release}.

\subsection{File naming conventions}

For each field, each dither, and each exposure we have in the first folder five files: 

\texttt{EUC\_VIS\_SWL-DET-\{obsid\}-\{dither\}-\{expnum\}\_\_*.fits}

\qquad
  VIS calibrated frame that includes the science image, the map of flags and
  the noise map;

\texttt{EUC\_VIS\_SWL-BKG-\{obsid\}-\{dither\}-\{expnum\}\_\_*.fits}

\qquad
  background model of the corresponding calibrated frame, generated by 
  \texttt{NoiseChisel};
  
\texttt{EUC\_VIS\_SWL-WGT-\{obsid\}-\{dither\}-\{expnum\}\_\_*.fits}

\qquad
  VIS calibrated exposure weight map. It is the master PRNU image with all   invalid pixels set to 0;
  
\texttt{EUC\_VIS\_SWL-CAT-\{obsid\}-\{dither\}-\{expnum\}\_\_*.fits}

\qquad
  photometry catalogue of the frame obtained via \texttt{SourceExtractor};
  
\texttt{EUC\_VIS\_SWL-CAT-\{obsid\}-\{dither\}-\{expnum\}\_\_*.xml}

\qquad
  metadata associated with the VIS calibrated photometry catalogue.
\\

\noindent In the following filenames \texttt{\{obsid\}}, \texttt{\{dither\}}, and \texttt{\{expnum\}} are defined as follows.
\begin{itemize}
\item 
\texttt{\{obsid\}} is the observation ID, it is either \texttt{067068},\texttt{067070}, or \texttt{067071}.

\noindent \texttt{067068} corresponds to the first $16 \times 400$-s exposures of the PSF calibration field acquired between 2025-03-23T07:56:00 and 2025-03-23T10:06:20.

\noindent \texttt{067070} corresponds to the $144 \times 400$-s exposures of the scientific field.

\noindent \texttt{067071} corresponds to the second $16 \times 400$-s exposures of the PSF calibration field acquired between 2025-03-24T08:58:00 and 2025-03-24T11:08:20.

\item
\texttt{\{dither\}} is the dither number, a sequential number lying between 0 and 143, or between 0 and 15 for the calibration field.
Each field was observed 16 times with different dither positions, as summarised in Table~\ref{tab:pos}. The relative position of the dithers for the different frames and different fields are given in Table~\ref{tab:ditherpos} and the corresponding pixel phase distribution is shown in Fig.~\ref{fig:pix_phase}.

\item
\texttt{\{expnum\}} is the exposure sequential number inside the dither; it is always 1.
\end{itemize}

\section{Processing the centre of nine fields\label{apdx:B}}
For completeness, we present the corresponding results for all nine $100'' \times 100''$ stamps analysed in Sect.~\ref{completeness}, using the same format as Fig.~\ref{fig:ob05390_completness}. The stamps are centred on the EGBS fields, whose coordinates and mean \IE-band extinctions are given in Table~\ref{tab:pos}.

\begin{figure*}
\centering
\includegraphics[width=9cm]{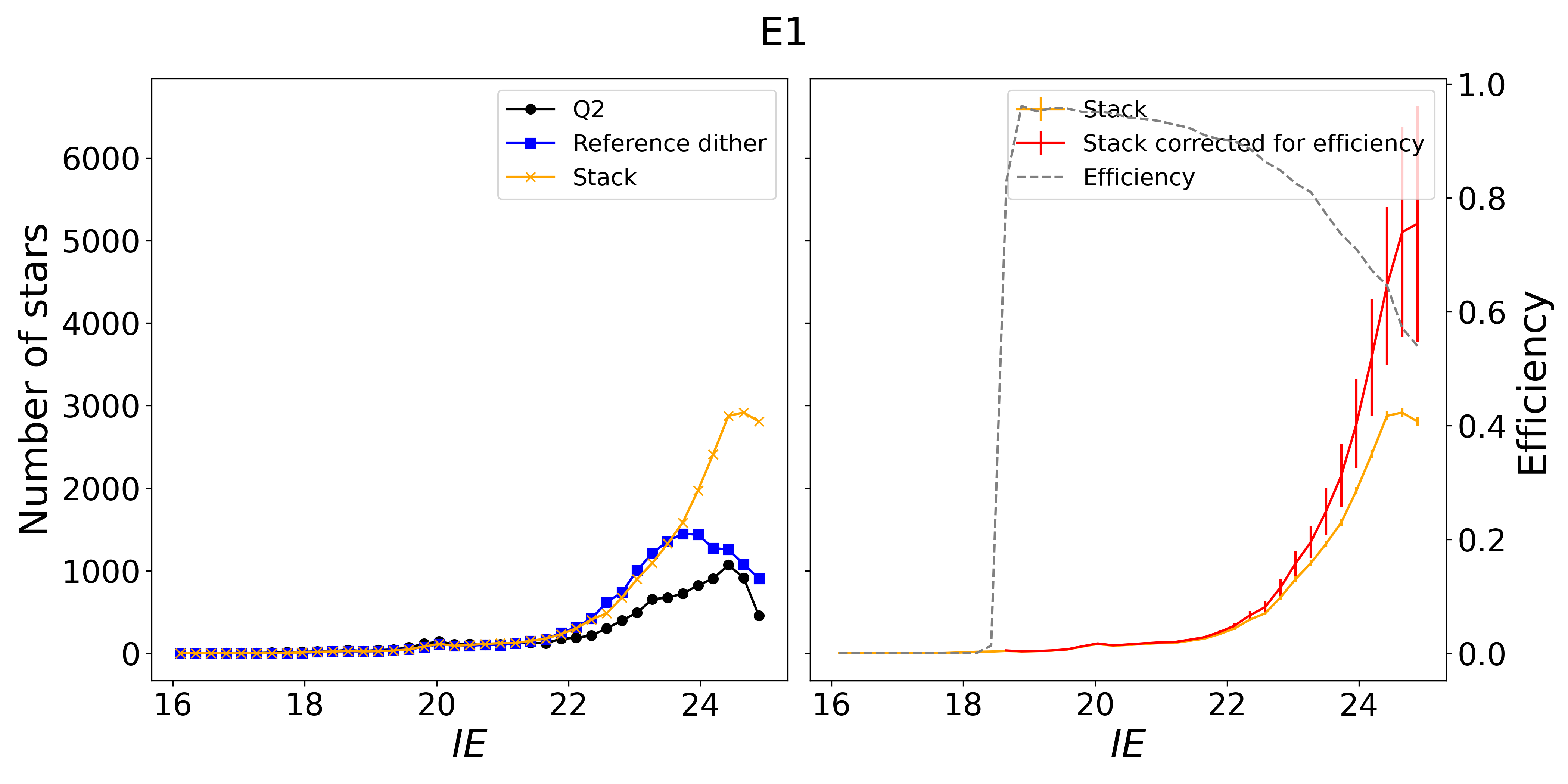}
\includegraphics[width=9cm]{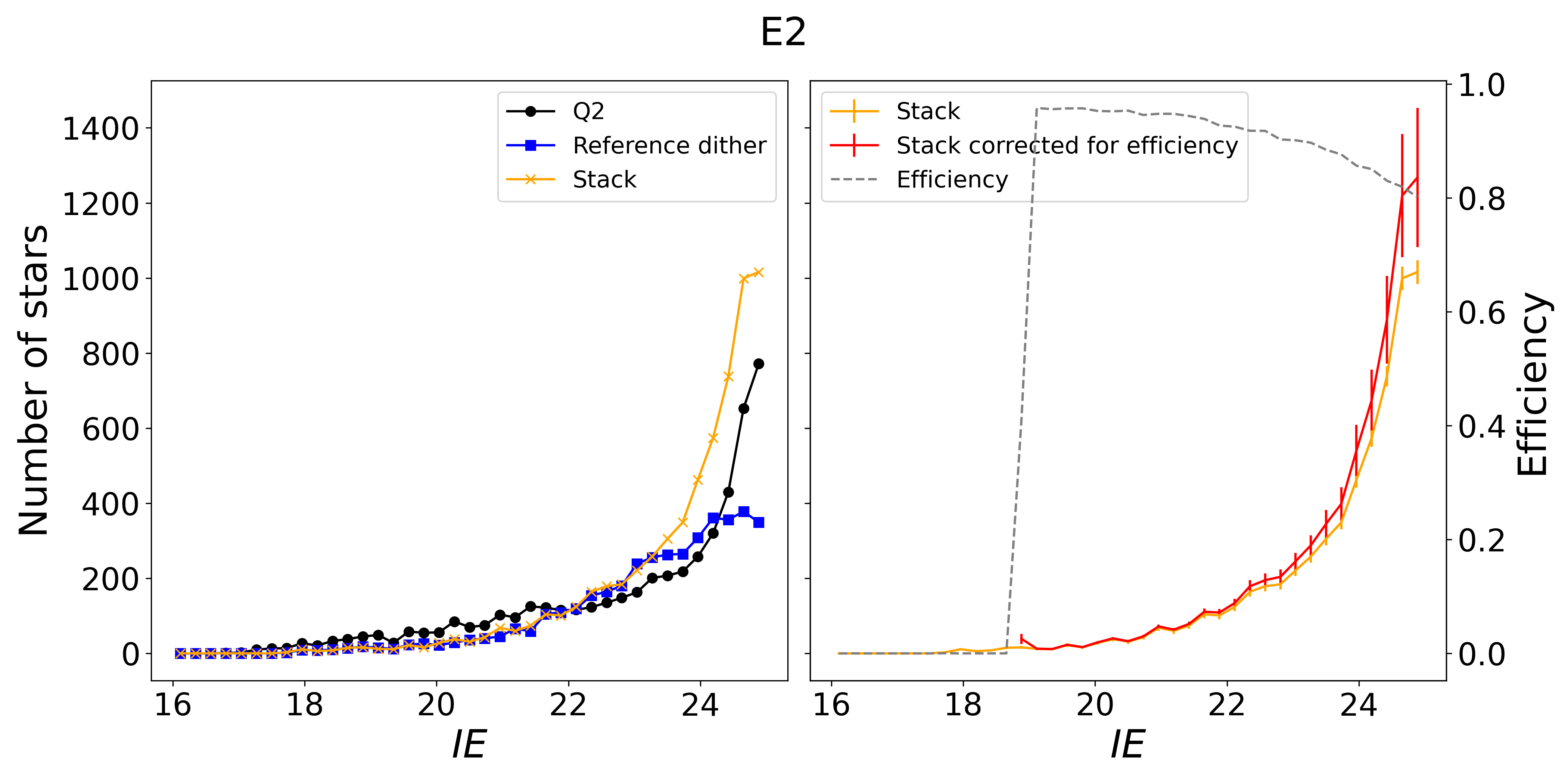}
\includegraphics[width=9cm]{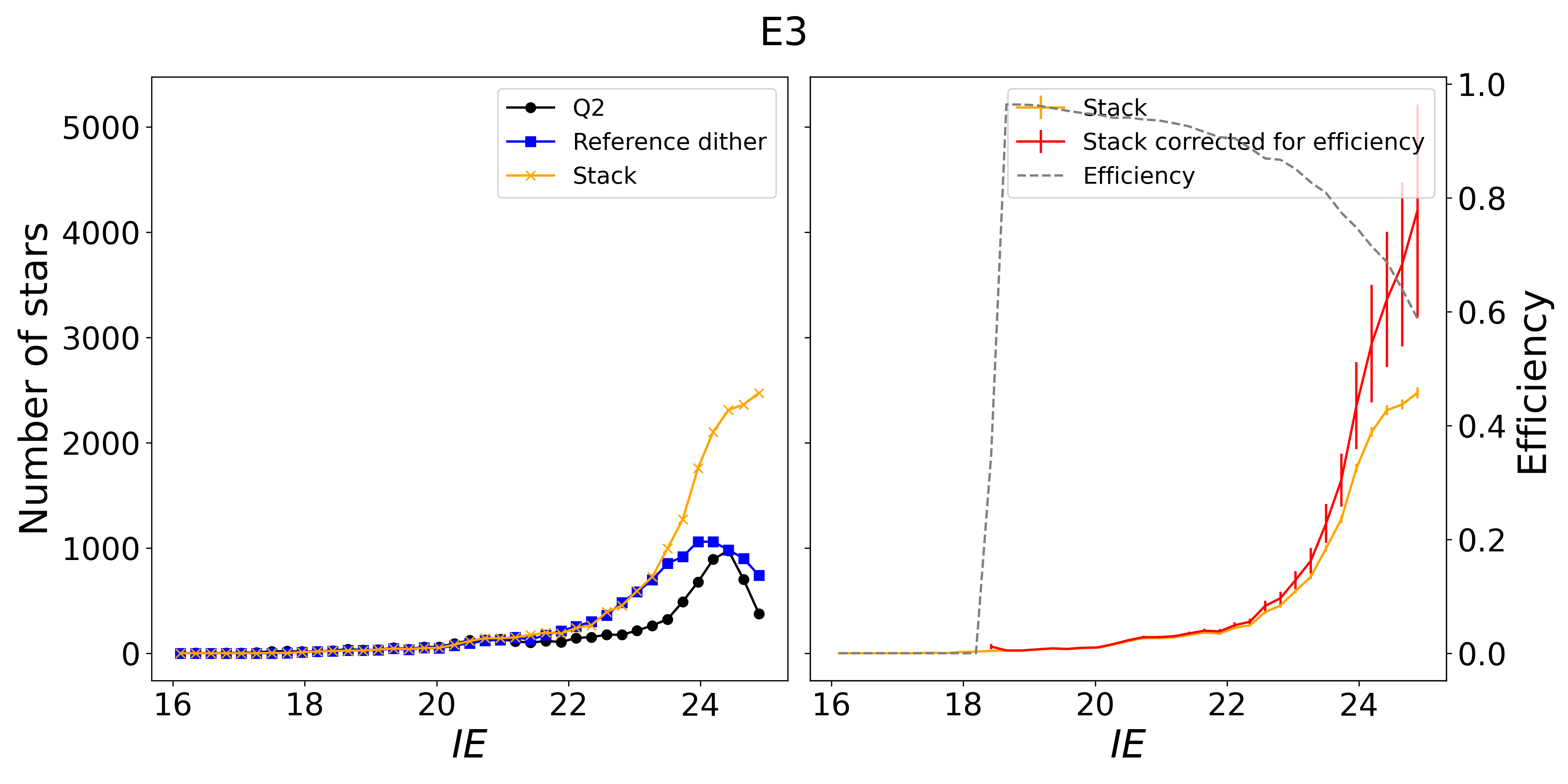}
\includegraphics[width=9cm]{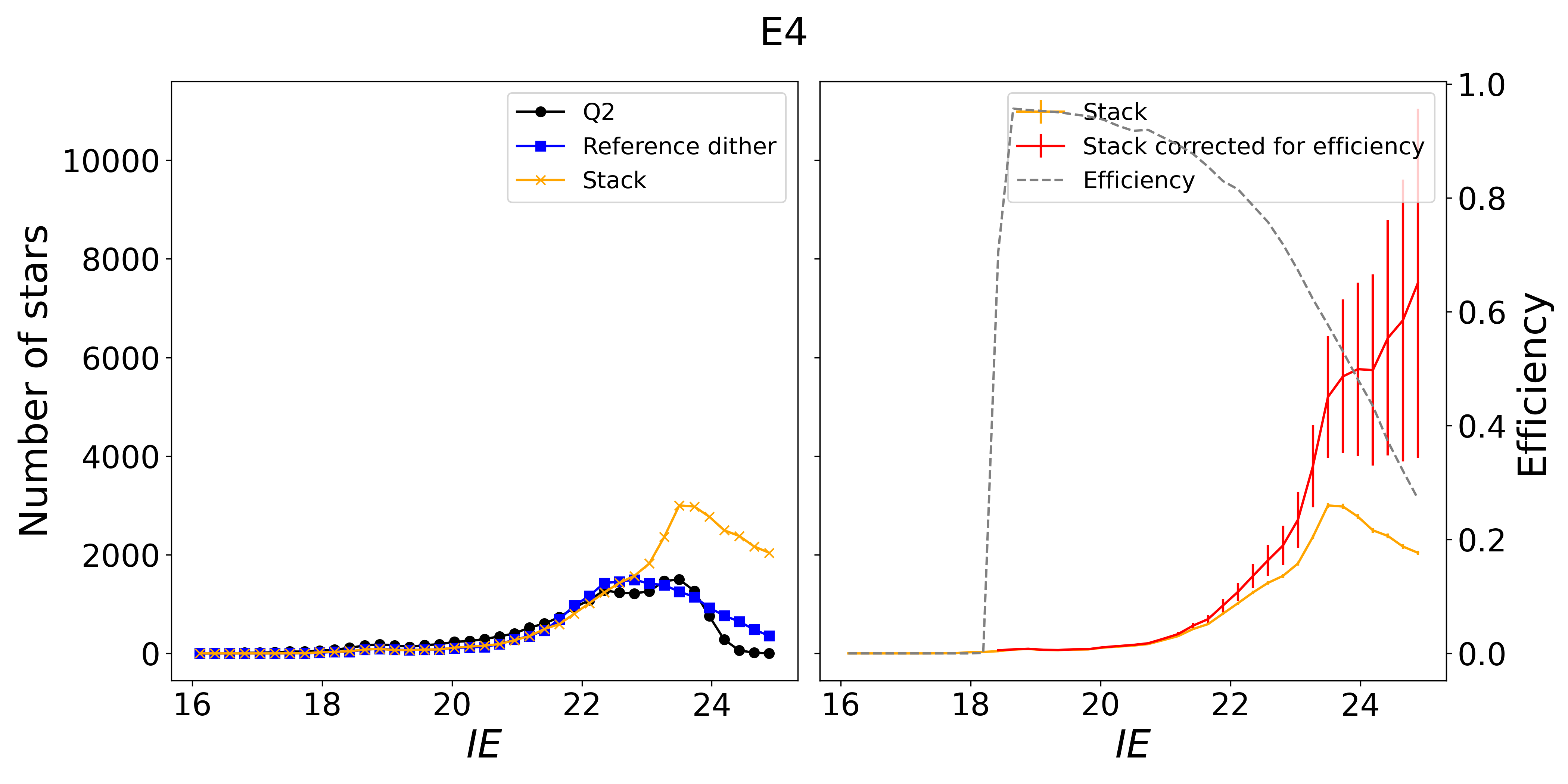}
\includegraphics[width=9cm]{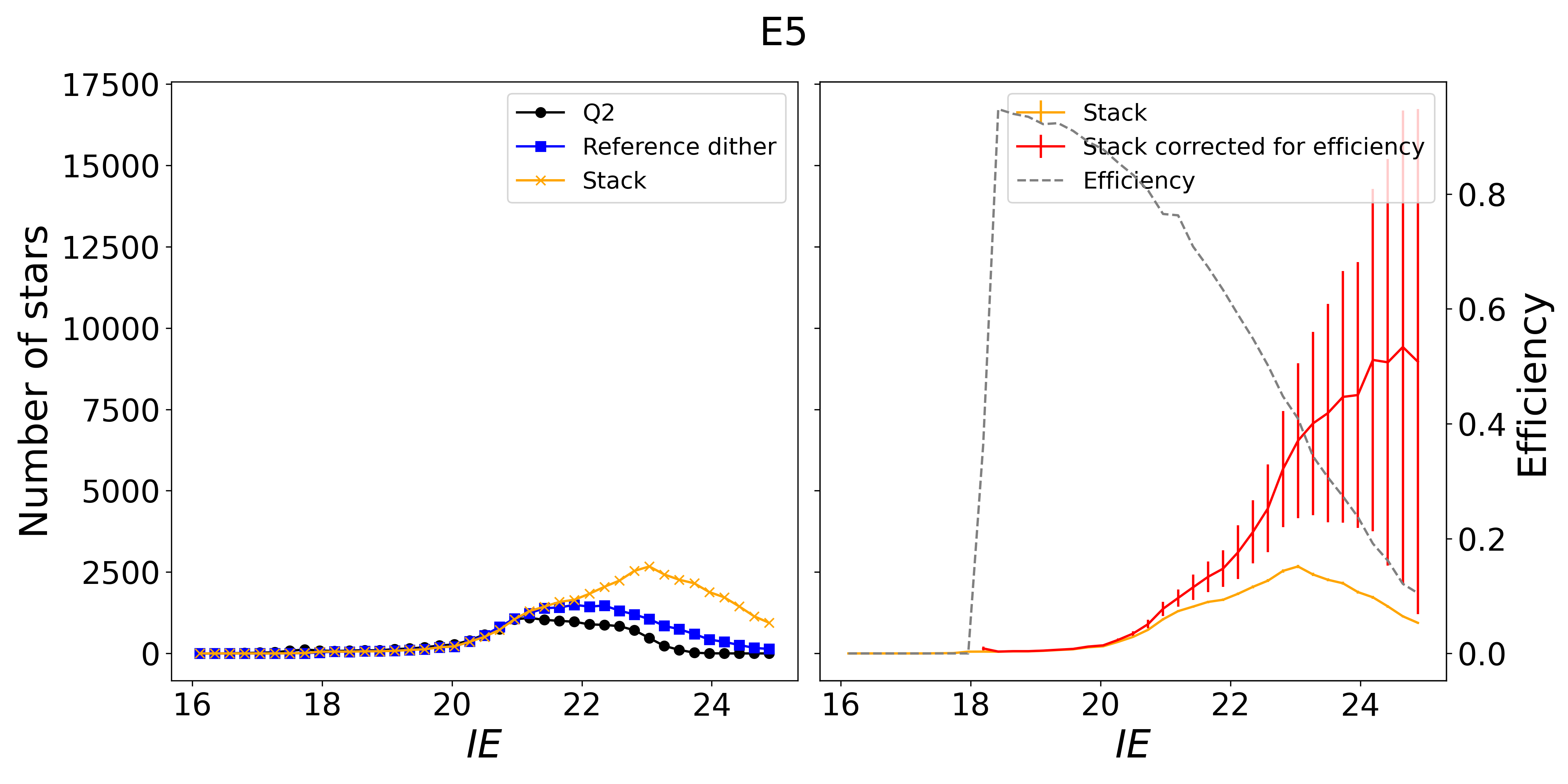}
\includegraphics[width=9cm]{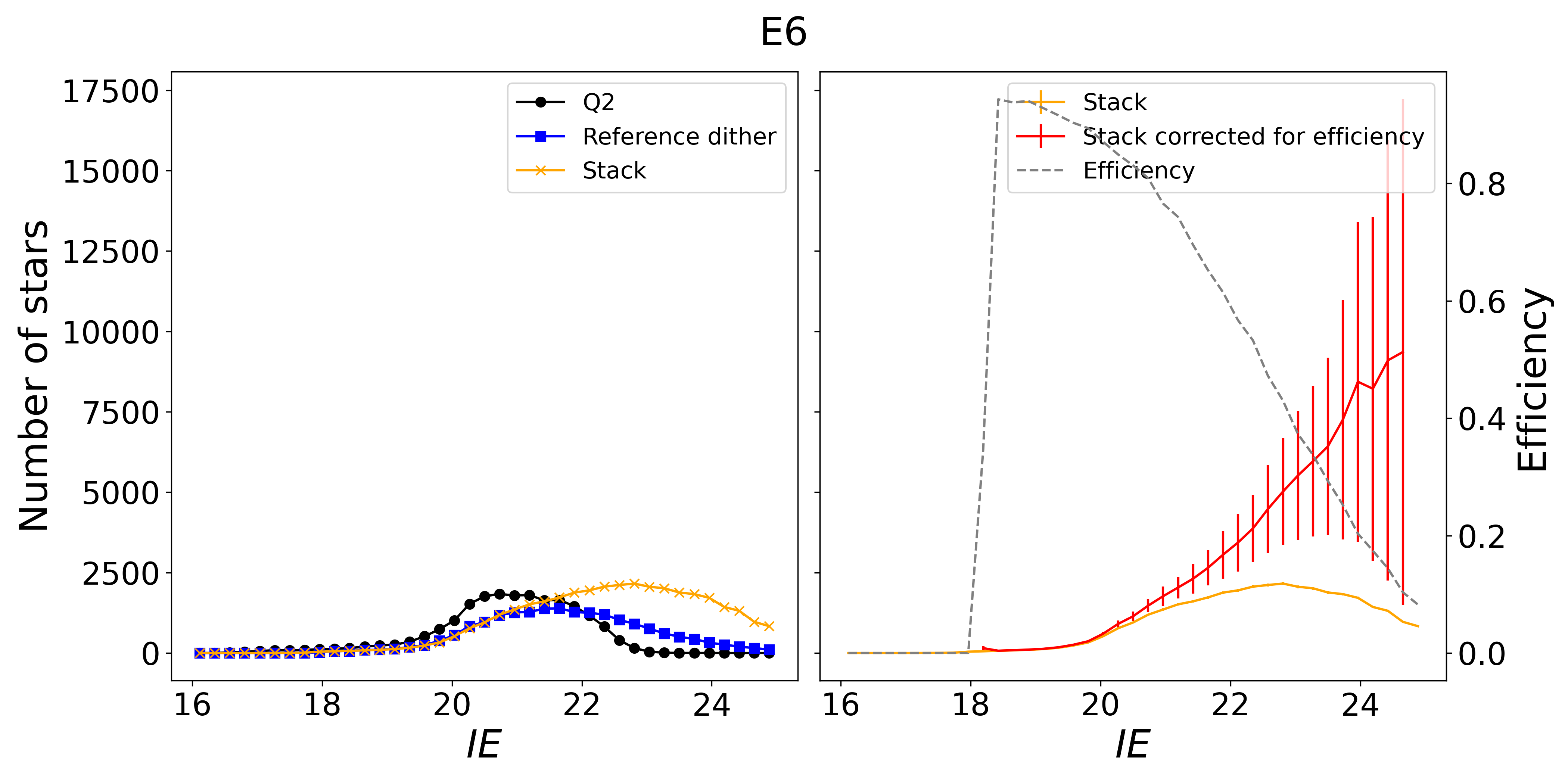}
\includegraphics[width=9cm]{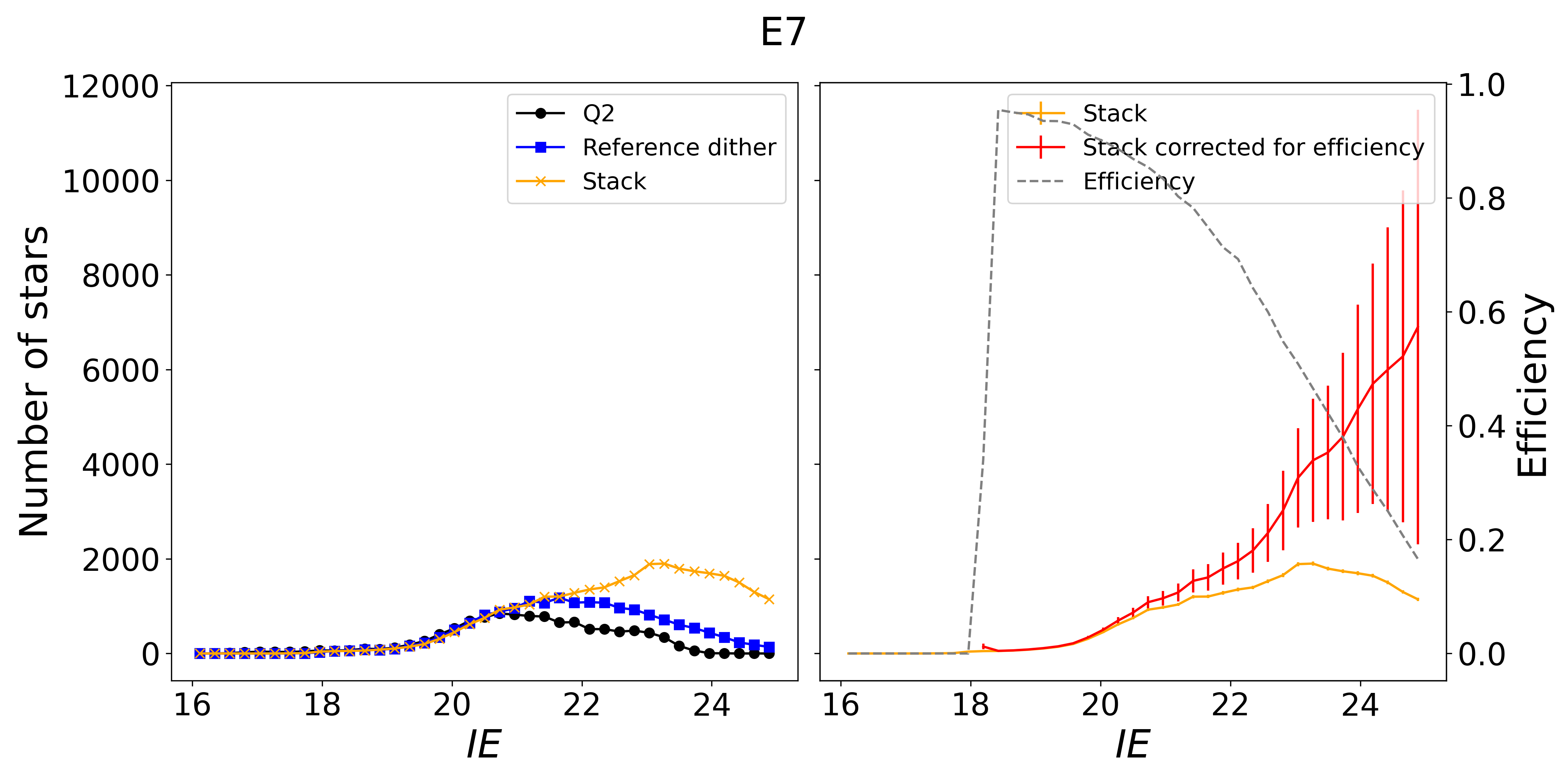}
\includegraphics[width=9cm]{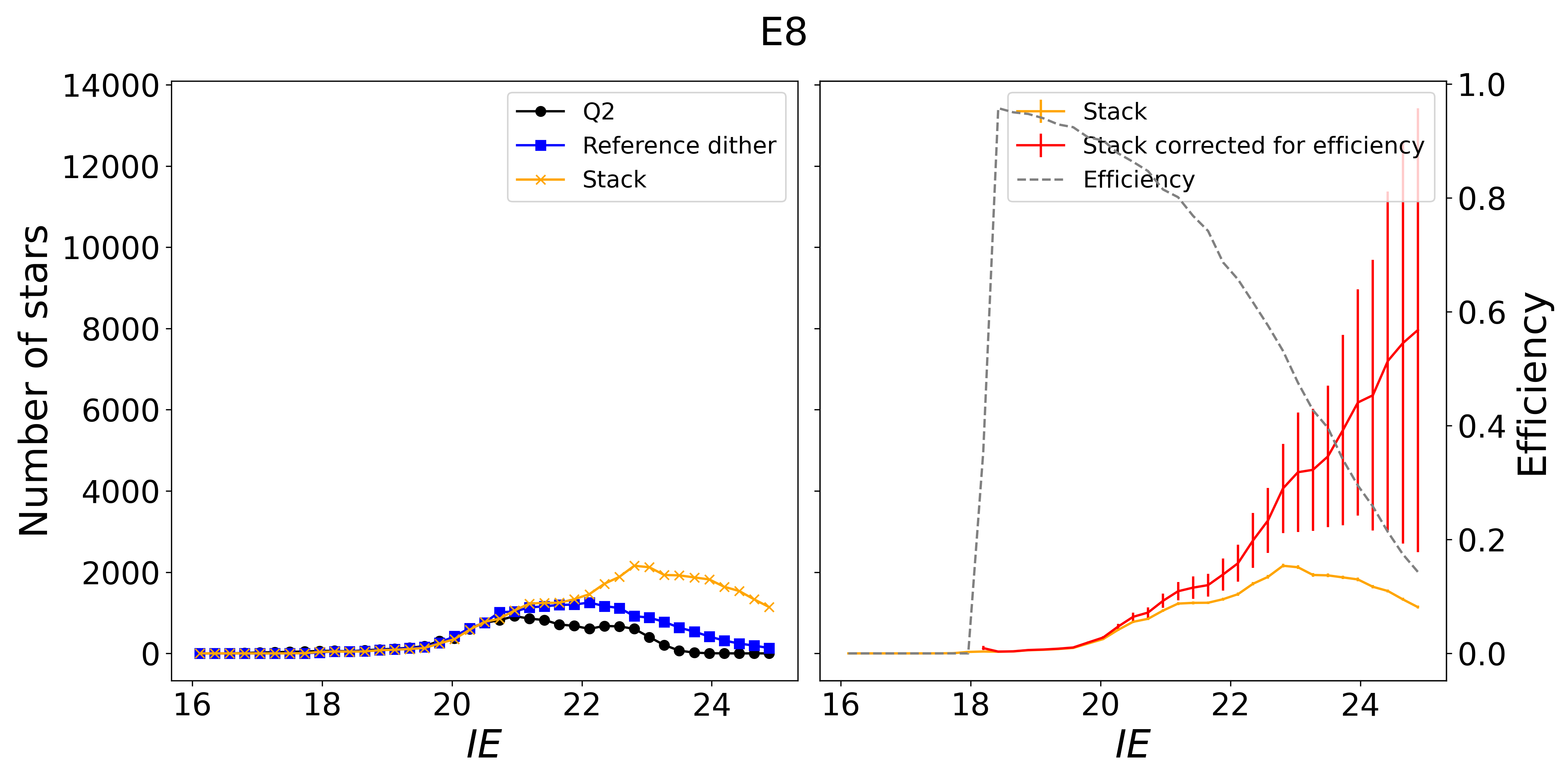}
\includegraphics[width=9cm]{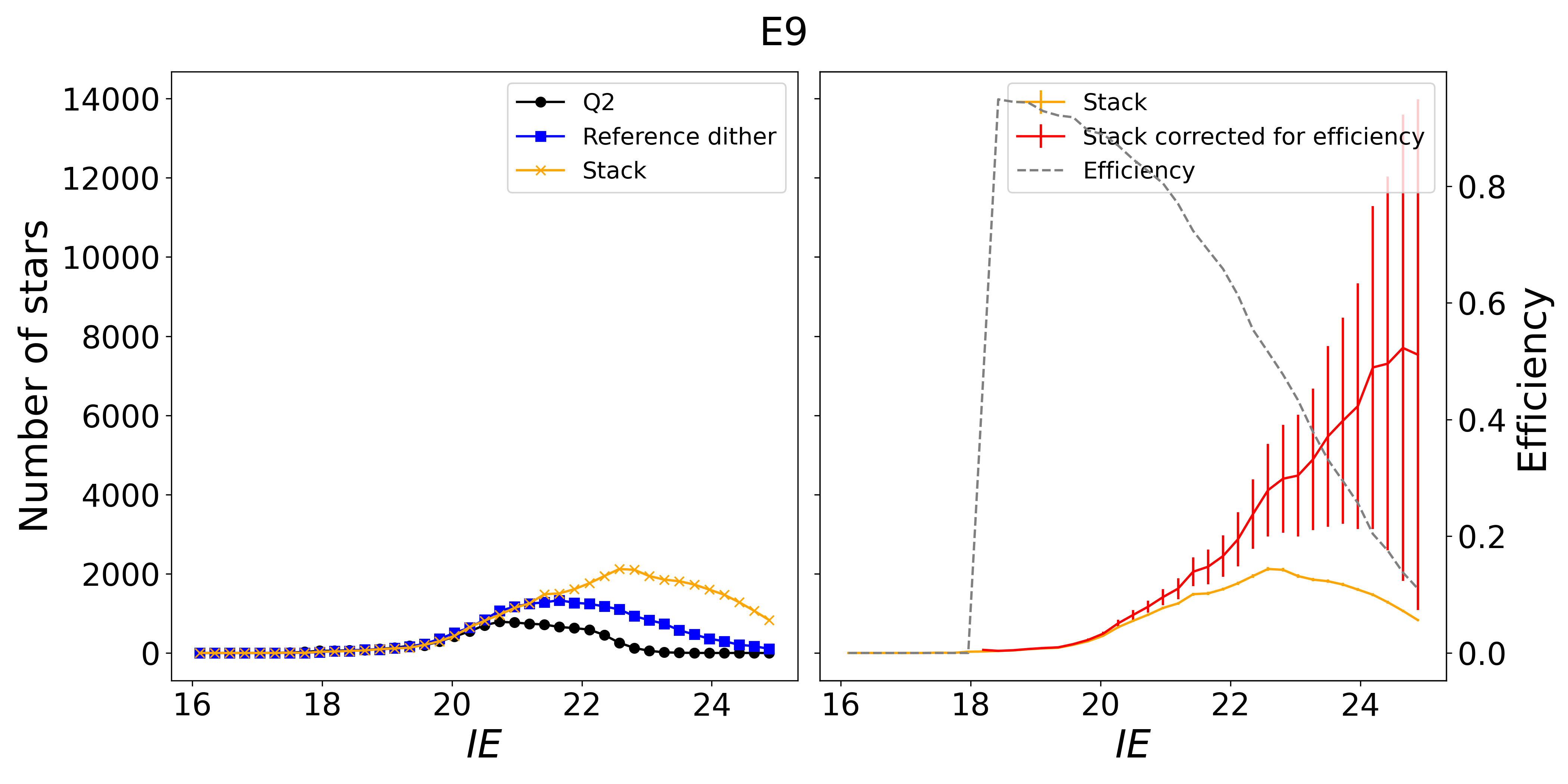}
\caption{%
As Fig.~\ref{fig:ob05390_completness}, but for $100\arcsecf5\times 100\arcsecf5$ cutouts centred at each of the nine EGBS pointings. The observed differences are the result of the interplay between crowding, background fitting, and threshold determinations by the VIS-PF compared to another processing.}
\label{fig:E1-E9_completness}
\end{figure*}

\end{appendix}

\end{document}